\documentclass{article}%
\usepackage{amssymb}
\usepackage{amsmath}
\usepackage{amsfonts}
\usepackage{graphicx}
\usepackage{hyperref}%
\providecommand{\U}[1]{\protect\rule{.1in}{.1in}}
\def\be{\begin{equation}}
\def\ee{\end{equation}}
\def\bea{\begin{eqnarray*}}
\def\eea{\end{eqnarray*}}

\begin{document}

\title{A Galileon Primer}
\author{Thomas Curtright$^{\S }${\Large , }David Fairlie$^{\boxtimes}$, and Hassan
Alshal$^{\S ,\triangle}${\small \bigskip}\\$^{\S }\,${\small Department of Physics, University of Miami}\\{\small Coral Gables, Florida 33124-8046, USA}\\$^{\boxtimes}\,${\small Department of Mathematical Sciences, University of
Durham}\\{\small Durham, DH1 3LE, UK}\\$^{\triangle}$\ {\small Department of Physics, Faculty of Science, Cairo
University}\\{\small Giza, 12613, Egypt\bigskip}\\{\small curtright@physics.miami.edu\ \ \ \ \ dma6dbf@durham.ac.uk\ \ \ \ \ halshal@sci.cu.edu.eg}%
}
\date{}
\maketitle

\begin{abstract}
Elementary features of galileon models are discussed at an introductory level.
\ Following a simple example, a general formalism leading to a hierarchy of
field equations and Lagrangians is developed for flat spacetimes. \ Legendre
duality is discussed. \ Implicit and explicit solutions are then constructed
and analyzed in some detail. \ Galileon shock fronts are conjectured to exist.
\ Finally, some interesting general relativistic effects are studied for
galileons coupled minimally to gravity. \ Spherically symmetric galileon and
metric solutions with naked curvature singularities are obtained and are shown
to be separated from solutions which exhibit event horizons by a critical
curve in the space of boundary data.\newpage

\end{abstract}
\tableofcontents

\newpage

\section{Introduction}

Galileon theories are a class of models for hypothetical scalar fields whose
Lagrangians involve multilinears of first and second derivatives, but whose
nonlinear field equations are still only second order. \ They may be important
for the description of large-scale features in astrophysics as well as for
elementary particle theory \cite{deRham,F}. \ Hierarchies of galileon
Lagrangians were discussed mathematically for flat spacetime in
\cite{FG1,FG2,FGM}, independently of an earlier systematic survey of
second-order scalar-tensor field equations in curved 4D spacetime \cite{H}.
\ The simplest example involves a single scalar field, $\phi$. \ This galileon
field may be coupled \textquotedblleft universally\textquotedblright\ to the
trace of the energy-momentum tensor, $\Theta$, and upon so doing, it is
gravitation-like by virtue of the similarity between this universal coupling
and that of the metric $g_{\mu\nu} $ to $\Theta_{\mu\nu}$ in general
relativity. \ As might be expected from this similarity and the ubiquitous
generation of scalar fields by the process of dimensional reduction, it is
possible to obtain some galileon models from limits of higher dimensional
gravitation theories. \ Indeed, galileon models were discovered yet again by
this process \cite{DGPetc.}.

\section{A simple example}

Although higher derivative actions \emph{usually} lead to higher derivative
field equations, as is well-known, nonetheless it is \emph{possible} to
accommodate higher derivatives in the action while retaining second-order
field equations if the Lagrangian is not quadratic in the fields. \ The price
to be paid is that the second-order field equations are always
\emph{nonlinear}. \ This is the basic ingredient that underlies all
galileon\ models. \ 

This point has already been well-appreciated in the literature, of course, but
for purposes of illustration, consider the \textquotedblleft
simplest\textquotedblright\ cubic, fourth-order Lagrangian density,
\begin{equation}
L=\phi^{2}~\phi_{\alpha\alpha\beta\beta}\ , \label{SimplestL}%
\end{equation}
where $\phi_{\alpha}=\partial\phi/\partial x^{\alpha}$, etc., and repeated
indices are summed using either Lorentzian or Euclidean signatures. \ The
corresponding action $A=\int L$ has local variation,%
\begin{equation}
\frac{\delta A}{\delta\phi}=2\phi~\phi_{\alpha\alpha\beta\beta}+\left(
\phi^{2}\right)  _{\alpha\alpha\beta\beta}\ .
\end{equation}
Surface terms, irrelevant for the field equations in the bulk, have been
discarded. \ That is to say,%
\begin{equation}
\frac{\delta A}{\delta\phi}=4\phi~\phi_{\alpha\alpha\beta\beta}+8\phi_{\alpha
}~\phi_{\alpha\beta\beta}+2\phi_{\alpha\alpha}~\phi_{\beta\beta}+4\phi
_{\alpha\beta}~\phi_{\alpha\beta}\ ,
\end{equation}
and the field equation, $\delta A/\delta\phi=0$, is both nonlinear and
fourth-order. \ 

However, the order of the field equation may be reduced by adding to the
Lagrangian a judicious amount of the \textquotedblleft
next-to-simplest\textquotedblright\ cubic term that involves first, second, or
third derivatives. \ So far as the field equations are concerned, there is
actually only \emph{one} other term that can be added, namely, $\phi
~\phi_{\alpha\alpha}~\phi_{\beta\beta}$. \ Superficially different terms, e.g.
$\phi~\phi_{\alpha}~\phi_{\alpha\beta\beta}$, $\phi~\phi_{\alpha\beta}%
~\phi_{\alpha\beta}$, $\phi_{\alpha}~\phi_{\alpha}~\phi_{\beta\beta}$, and
$\phi_{\alpha}~\phi_{\beta}~\phi_{\alpha\beta}$,\ do \emph{not} give
independent contributions to the local variation of the action in the bulk,
although they differ in their surface contributions. \ In particular,%
\begin{equation}
2~\phi~\phi_{\alpha\alpha}~\phi_{\beta\beta}+4~\phi~\phi_{\alpha\beta}%
~\phi_{\alpha\beta}-3~\phi^{2}~\phi_{\alpha\alpha\beta\beta}=\left(
4~\phi~\phi_{\alpha}~\phi_{\alpha\beta}+2~\phi~\phi_{\beta}~\phi_{\alpha
\alpha}-3~\phi^{2}~\phi_{\alpha\alpha\beta}-2~\phi_{\alpha}~\phi_{\alpha}%
~\phi_{\beta}\right)  _{\beta}\ ,
\end{equation}
i.e. a total divergence. \ Thus it is sufficient to include in the action any
two of the three terms on the LHS, with an arbitrary relative coefficient.

So, rather than (\ref{SimplestL}), consider instead the Lagrangian density%
\begin{equation}
L=\phi^{2}~\phi_{\alpha\alpha\beta\beta}-\lambda~\phi~\phi_{\alpha\alpha}%
~\phi_{\beta\beta}\ , \label{1ParameterL}%
\end{equation}
with constant $\lambda$. \ The variation of the action obtained from
(\ref{1ParameterL}) is%
\begin{align}
\frac{\delta A}{\delta\phi}  &  =2\phi~\phi_{\alpha\alpha\beta\beta}+\left(
\phi^{2}\right)  _{\alpha\alpha\beta\beta}-\lambda\left(  \phi_{\alpha\alpha
}~\phi_{\beta\beta}+2\left(  \phi~\phi_{\alpha\alpha}\right)  _{\beta\beta
}\right) \nonumber\\
&  =\left(  4-2\lambda\right)  \left(  \phi~\phi_{\alpha\alpha\beta\beta
}+2\phi_{\alpha}~\phi_{\alpha\beta\beta}\right)  +\left(  2-3\lambda\right)
\phi_{\alpha\alpha}~\phi_{\beta\beta}+4\phi_{\alpha\beta}~\phi_{\alpha\beta
}\ .
\end{align}
Thus $\lambda=2$ uniquely eliminates from the variation all derivatives higher
than the second, leaving just%
\begin{equation}
\left.  \frac{\delta A}{\delta\phi}\right\vert _{\lambda=2}=-4\left(
\phi_{\alpha\alpha}~\phi_{\beta\beta}-\phi_{\alpha\beta}~\phi_{\alpha\beta
}\right)  \ .
\end{equation}
While this equation is still nonlinear, it is now only second-order.

Moreover, the action for the $\lambda=2$ model can be rewritten in various
ways upon integrating by parts. \ Perhaps the most compact and memorable of
these is%
\begin{equation}
A_{2}=\int\phi_{\alpha}~\phi_{\alpha}~\phi_{\beta\beta}~d^{n}x\ .
\label{Action2}%
\end{equation}
This differs from the previous $\left.  A\right\vert _{\lambda=2}$ by a factor
of $2$ and a boundary term,
\[
A_{2}=\frac{1}{2}\int\left(  \phi^{2}~\phi_{\alpha\alpha\beta\beta}%
-2~\phi~\phi_{\alpha\alpha}~\phi_{\beta\beta}\right)  ~d^{n}x-\frac{1}{2}%
\int\partial_{\alpha}B_{\alpha}~d^{n}x\ ,
\]
involving the current
\begin{equation}
B_{\alpha}=\phi^{2}~\overleftrightarrow{\partial_{\alpha}}~\phi_{\beta\beta
}=\phi^{2}~\phi_{\alpha\beta\beta}-2~\phi~\phi_{\alpha}~\phi_{\beta\beta}\ .
\end{equation}
Indeed, most discussions of this model are developed around $A_{2}$, after
defining the system's Lagrangian to be%
\begin{equation}
L_{2}=\phi_{\alpha}~\phi_{\alpha}~\phi_{\beta\beta}\ . \label{L2}%
\end{equation}

To complete our discussion of this elementary case, consider the
energy-momentum density arising from $L_{2}$. \ The canonical result is
straightforwardly obtained, even though the Lagrangian involves higher
derivatives, but the resulting density is not a symmetric tensor. \ However,
minimal coupling to gravity is guaranteed to yield a symmetric tensor, so we
take that route. \ Generally covariant forms of (\ref{L2}) and (\ref{Action2})
are obtained through the replacements $\phi_{\alpha}\phi_{\alpha}%
~d^{n}x\rightarrow g^{\alpha\beta}\phi_{\alpha}\phi_{\beta}\sqrt{-g}~d^{n}x$
and $\phi_{\beta\beta}\rightarrow\frac{1}{\sqrt{-g}}~\partial_{\mu}\left(
\sqrt{-g}g^{\mu\nu}\phi_{\nu}\right)  $. \ Thus an invariant action is
\begin{equation}
\left.  A_{2}\right\vert _{\text{curved space}}=\int g^{\alpha\beta}%
\phi_{\alpha}\phi_{\beta}~\partial_{\mu}\left(  \sqrt{-g}g^{\mu\nu}\phi_{\nu
}\right)  ~d^{n}x\ . \label{InvariantA2}%
\end{equation}
Varying the metric gives $\Theta_{\alpha\beta}$. \ In the flat-space limit,
the result is%
\begin{equation}
\Theta_{\mu\nu}=\phi_{\mu}\phi_{\nu}\phi_{\alpha\alpha}-\phi_{\alpha}%
\phi_{\alpha\nu}\phi_{\mu}-\phi_{\alpha}\phi_{\alpha\mu}\phi_{\nu}+\delta
_{\mu\nu}\phi_{\alpha}\phi_{\beta}\phi_{\alpha\beta}\ .
\end{equation}
This is seen to be conserved
\begin{equation}
\partial_{\mu}\Theta_{\mu\nu}=\phi_{\nu}~\mathcal{E}_{2}\ ,
\end{equation}
upon using the field equation, $\mathcal{E}_{2}=0$, where%
\begin{equation}
\mathcal{E}_{2}\equiv\phi_{\alpha\alpha}\phi_{\beta\beta}-\phi_{\alpha\beta
}\phi_{\alpha\beta}\ . \label{FE2}%
\end{equation}
The justification for the name \textquotedblleft galileon\textquotedblright%
\ is now apparent. \ Any shift of the field by a constant, or by a term linear
in \ $x$, as is reminiscent of a galilean transformation in classical
mechanics, will leave the action (\ref{Action2}) invariant, up to surface
terms, and therefore not falsify a solution of the field equation.

An interesting wrinkle now appears: $\ \Theta_{\mu\nu}$ is not traceless
on-shell. \ Consequently, the usual form of the scale current, $x_{\alpha
}\Theta_{\alpha\mu}$, is not conserved. \ On the other hand, the action
(\ref{Action2}) is homogeneous in $\phi$ and its derivatives, and is clearly
invariant under the scale transformations $x\rightarrow sx$ and $\phi\left(
x\right)  \rightarrow s^{\left(  4-n\right)  /3}\phi\left(  sx\right)  $.
\ Hence the corresponding Noether current must be conserved. \ This current is
easily found, at least in four dimensions, since the trace is obviously a
total divergence in that case:%
\begin{equation}
\left.  \Theta_{\mu\mu}\right\vert _{n=4}=\phi_{\alpha\alpha}\phi_{\beta}%
\phi_{\beta}+2\phi_{\alpha}\phi_{\beta}\phi_{\alpha\beta}=\partial_{\alpha
}\left(  \phi_{\alpha}\phi_{\beta}\phi_{\beta}\right)  \ .
\end{equation}
That is to say, for $n=4$ the virial is the trilinear $V_{\alpha}=\phi
_{\alpha}\phi_{\beta}\phi_{\beta}$. \ So a conserved scale current is given by
the combination,%
\begin{equation}
\left.  S_{\mu}\right\vert _{n=4}=x_{\alpha}\Theta_{\alpha\mu}-\phi_{\alpha
}\phi_{\alpha}\phi_{\mu}\ .
\end{equation}
However, the virial here is not a divergence modulo a conserved current. \ So
the theory is \emph{not} conformally invariant despite being scale invariant
\cite{J}.

Some additional algebra is needed for $n\neq4$, but eventually one finds:%
\begin{align}
S_{\mu}  &  =x_{\alpha}\Theta_{\alpha\mu}-V_{\mu}\ ,\label{SCurrentAnyDim}\\
V_{\mu}  &  =\frac{n-1}{3}~\phi_{\alpha}\phi_{\alpha}\phi_{\mu}+\frac{4-n}%
{3}~\phi J_{\mu}\ ,\label{VirialAnyDim}\\
\partial_{\mu}S_{\mu}  &  =\left[  x_{\alpha}\phi_{\alpha}\mathcal{+}%
\frac{n-4}{3}~\phi\right]  \mathcal{E}_{2}\ . \label{SConservationAnyDim}%
\end{align}
The last term in\ the virial $V_{\mu}$ for $n\neq4$ involves a bilinear
current which is conserved merely as a restatement of the field equation:%
\begin{equation}
J_{\mu}=\phi_{\mu}\overleftrightarrow{\partial_{\alpha}}\phi_{\alpha
}\ ,\ \ \ \mathcal{E}_{2}=\partial_{\mu}J_{\mu}\ .
\end{equation}
In fact, this current is itself a total divergence,%
\begin{equation}
J_{\mu}=\partial_{\nu}\left(  \delta_{\mu\nu}\phi\phi_{\alpha\alpha}-\phi
~\phi_{\mu\nu}\right)  \ ,
\end{equation}
so the field equation for the model is a double divergence for any $n$. \ But
once again, although the model is scale invariant in any number of dimensions,
it is not conformally invariant for $n>2$.

Since (\ref{L2}) has the form of the conventional free field Lagrangian
density $\times$ the Klein-Gordon equation, for a massless scalar field, it
immediately suggests a generalization to a hierarchy of such systems, where
the Lagrangian density for the $k$th system is just a product of the free
field Lagrangian density and the equation of motion for the $\left(
k-1\right)  $st system. \ In fact, this simple generalization is easy to
formulate in explicit detail. \ A systematic theory for the hierarchy is
elegantly expressed using determinants.

\section{General formalism}

\textit{This section may be skipped by anyone with a phobia for determinants,
and definitely should be passed over by anyone under a doctor's orders to cut
back on tensor index shuffling. \ Later sections of the paper rarely invoke
results obtained in this section. \ However, the material presented here may
be helpful for applications beyond those considered in this primer.
\ Accordingly, the last part (\S 3.9) provides an encapsulation of the results
in terms of determinants and standard Kronecker symbols. \ }

\subsection{Determinant and trace identities}

For any $n\times n$ matrix $M$, consider the expansion%
\begin{equation}
\det\left(  \boldsymbol{1}+\lambda M\right)  =\sum_{k=0}^{n}\frac{\lambda^{k}%
}{k!}~\mathcal{E}_{k}\left(  M\right)  \ , \label{DetExpansion}%
\end{equation}
where $\mathcal{E}_{0}\equiv1$. $\ $Elementary cases are $\mathcal{E}%
_{1}=\mathrm{Tr}\left(  M\right)  $ and $\mathcal{E}_{n}=n!\det\left(
M\right)  $. \ Other cases may not be so familiar. \ However, from the
identity%
\begin{equation}
\det\left(  \boldsymbol{1}+\lambda M\right)  =\exp\left(  \mathrm{Tr}%
\ln\left(  \boldsymbol{1}+\lambda M\right)  \right)  \ ,
\end{equation}
it follows that the $\mathcal{E}_{k}$ obey a recursion relation for any $M$,%
\begin{equation}
\frac{1}{k!}~\mathcal{E}_{k}=\sum_{\ell=0}^{k-1}\frac{\left(  -1\right)
^{k-1-\ell}}{\ell!}~\mathcal{T}_{k-\ell}\mathcal{E}_{\ell}\ ,\text{ \ \ for
}k\leq n\text{ , \ \ where \ \ }\mathcal{T}_{m}\equiv\mathrm{Tr}\left(
M^{m}\right)  \ . \label{EofMRecursion}%
\end{equation}
The solution of this recursion for all $k\leq n$ can be expressed in terms of
\emph{another} set of determinants,
\begin{equation}
\mathcal{E}_{k}=\det\left(  \mathbb{T}_{k}\right)  \ , \label{DetEofM}%
\end{equation}
where $\mathbb{T}_{k}$ is an auxiliary $k\times k$ matrix containing the
various traces:%
\begin{equation}
\mathbb{T}_{k}=\left(
\begin{array}
[c]{ccccccc}%
\mathcal{T}_{1} & k-1 & 0 & \cdots & 0 & 0 & 0\\
\mathcal{T}_{2} & \mathcal{T}_{1} & k-2 & \cdots & 0 & 0 & 0\\
\mathcal{T}_{3} & \mathcal{T}_{2} & \mathcal{T}_{1} & \cdots & 0 & 0 & 0\\
\vdots & \vdots & \vdots & \ddots & \vdots & \vdots & \vdots\\
\mathcal{T}_{k-2} & \mathcal{T}_{k-3} & \mathcal{T}_{k-4} & \cdots &
\mathcal{T}_{1} & 2 & 0\\
\mathcal{T}_{k-1} & \mathcal{T}_{k-2} & \mathcal{T}_{k-3} & \cdots &
\mathcal{T}_{2} & \mathcal{T}_{1} & 1\\
\mathcal{T}_{k} & \mathcal{T}_{k-1} & \mathcal{T}_{k-2} & \cdots &
\mathcal{T}_{3} & \mathcal{T}_{2} & \mathcal{T}_{1}%
\end{array}
\right)  \ . \label{DetEofMT}%
\end{equation}
The recursion relation (\ref{EofMRecursion}) is recovered by expanding
$\det\left(  \mathbb{T}_{k}\right)  $ in the minors of the first column. \ In
addition to (\ref{DetEofM}) we also note the identity%
\begin{equation}
k\mathcal{E}_{k-1}=\mathrm{Tr}\left(  \mathrm{adj}\mathbb{T}_{k}\right)  \ .
\label{TrEofM}%
\end{equation}
In this last trace relation, we use the \emph{adjugate} (a.k.a. the
\emph{classical} \emph{adjoint}) matrix notation, $\mathrm{adj}\left(
\mathbb{T}\right)  =$ $\left(  \det\mathbb{T}\right)  ~\mathbb{T}^{-1}$. \ 

For example,%
\begin{align}
\mathcal{E}_{1}  &  =\det\mathbb{T}_{1}=\mathcal{T}_{1}\ ,\\
\mathcal{E}_{2}  &  =\det\mathbb{T}_{2}=\det\left(
\begin{array}
[c]{cc}%
\mathcal{T}_{1} & 1\\
\mathcal{T}_{2} & \mathcal{T}_{1}%
\end{array}
\right)  =\mathcal{T}_{1}^{2}-\mathcal{T}_{2}\ ,\\
\mathcal{E}_{3}  &  =\det\mathbb{T}_{3}=\det\left(
\begin{array}
[c]{ccc}%
\mathcal{T}_{1} & 2 & 0\\
\mathcal{T}_{2} & \mathcal{T}_{1} & 1\\
\mathcal{T}_{3} & \mathcal{T}_{2} & \mathcal{T}_{1}%
\end{array}
\right)  =\mathcal{T}_{1}^{3}-3\mathcal{T}_{1}\mathcal{T}_{2}+2\mathcal{T}%
_{3}\ ,\\
\mathcal{E}_{4}  &  =\det\mathbb{T}_{4}=\det\left(
\begin{array}
[c]{cccc}%
\mathcal{T}_{1} & 3 & 0 & 0\\
\mathcal{T}_{2} & \mathcal{T}_{1} & 2 & 0\\
\mathcal{T}_{3} & \mathcal{T}_{2} & \mathcal{T}_{1} & 1\\
\mathcal{T}_{4} & \mathcal{T}_{3} & \mathcal{T}_{2} & \mathcal{T}_{1}%
\end{array}
\right)  =\mathcal{T}_{1}^{4}-6\mathcal{T}_{1}^{2}\mathcal{T}_{2}%
+8\mathcal{T}_{1}\mathcal{T}_{3}+3\mathcal{T}_{2}^{2}-6\mathcal{T}_{4}\ ,
\end{align}
etc. \ Actually, for an $n\times n$ matrix $M$, explicit computation of the
traces inserted into the expression (\ref{DetEofMT}) gives a \emph{null}
determinant for $k>n$. \ That is, $\mathcal{E}_{k>n}=0$, as would be expected
from the expansion of $\det\left(  \boldsymbol{1}+\lambda M\right)  $.

Moreover, a slight modification of the auxiliary matrix in (\ref{DetEofMT})
gives directly the characteristic polynomial for any $n\times n$ matrix $M$,%
\begin{equation}
\det\left(  M-\lambda~\boldsymbol{1}\right)  =\frac{1}{n!}~\det\left(
\begin{array}
[c]{ccccccc}%
1 & n & 0 & \cdots & 0 & 0 & 0\\
\lambda & \mathcal{T}_{1} & n-1 & \cdots & 0 & 0 & 0\\
\lambda^{2} & \mathcal{T}_{2} & \mathcal{T}_{1} & \cdots & 0 & 0 & 0\\
\vdots & \vdots & \vdots & \ddots & \vdots & \vdots & \vdots\\
\lambda^{n-2} & \mathcal{T}_{n-2} & \mathcal{T}_{n-3} & \cdots &
\mathcal{T}_{1} & 2 & 0\\
\lambda^{n-1} & \mathcal{T}_{n-1} & \mathcal{T}_{n-2} & \cdots &
\mathcal{T}_{2} & \mathcal{T}_{1} & 1\\
\lambda^{n} & \mathcal{T}_{n} & \mathcal{T}_{n-1} & \cdots & \mathcal{T}_{3} &
\mathcal{T}_{2} & \mathcal{T}_{1}%
\end{array}
\right)  \ . \label{Characteristic}%
\end{equation}
This follows immediately from expanding in the minors of the first column,
using (\ref{DetExpansion}) and (\ref{EofMRecursion}).

From the elementary identity $\det\left(  AB\right)  =\left(  \det A\right)
\left(  \det B\right)  $ we have%
\begin{equation}
\det\left(  \boldsymbol{1}+\lambda M\right)  =\lambda^{n}\left(  \det
M\right)  \det\left(  \boldsymbol{1}+\lambda^{-1}M^{-1}\right)  \ ,
\end{equation}
for an $n\times n$ nonsingular $M$. \ It follows for $0\leq k\leq n$\ that%
\begin{equation}
\frac{1}{k!}~\mathcal{E}_{k}\left(  M\right)  =\left(  \det M\right)
~\frac{1}{\left(  n-k\right)  !}~\mathcal{E}_{n-k}\left(  M^{-1}\right)  \ .
\label{EofMDuality}%
\end{equation}
For example,%
\begin{gather}
\mathcal{E}_{n}\left(  M\right)  =n!\left(  \det M\right)  \ ,\\
\mathcal{E}_{n-1}\left(  M\right)  =\left(  n-1\right)  !\left(  \det
M\right)  \ \mathrm{Tr}\left(  M^{-1}\right)  =\left(  n-1\right)
!~\mathcal{E}_{1}\left(  \mathrm{adj}M\right)  \ ,
\end{gather}
etc. \ Or, to rewrite (\ref{EofMDuality}) more symmetrically, for $n\times n$
nonsingular $M$,
\begin{equation}
\frac{1}{\sqrt{\det\left(  M\right)  }}~\frac{1}{k!}~\mathcal{E}_{k}\left(
M\right)  =\frac{1}{\sqrt{\det\left(  M^{-1}\right)  }}~\frac{1}{\left(
n-k\right)  !}~\mathcal{E}_{n-k}\left(  M^{-1}\right)  \ .
\label{SymmetricEofMDuality}%
\end{equation}

\subsection{Field equations}

Take $M$ to be $\mathcal{H}=\partial\partial\phi$, the Hessian matrix of
second partial derivatives of $\phi\left(  x_{1},\cdots,x_{n}\right)  $, then
\begin{equation}
\det\left(  1+\lambda\partial\partial\phi\right)  =\sum_{k=0}^{n}\frac
{\lambda^{k}}{k!}~\mathcal{E}_{k}\left(  \partial\partial\phi\right)  \ .
\end{equation}
For $k\geq1$ also define \textquotedblleft the equation of motion at level $k
$\textquotedblright\ as $\mathcal{E}_{k}\left(  \partial\partial\phi\right)
=0 $. \ For example, with $\phi_{\alpha\beta}=\partial_{\alpha}\partial
_{\beta}\phi$,
\begin{align}
\mathcal{E}_{1}\left(  \partial\partial\phi\right)   &  =\mathrm{Tr}\left(
\partial\partial\phi\right)  =\phi_{\alpha\alpha}\ ,\\
\mathcal{E}_{2}\left(  \partial\partial\phi\right)   &  =\left(
\mathrm{Tr}\left(  \partial\partial\phi\right)  \right)  ^{2}-\mathrm{Tr}%
\left(  \left(  \partial\partial\phi\right)  ^{2}\right)  =\phi_{\alpha\alpha
}\phi_{\beta\beta}-\phi_{\alpha\beta}\phi_{\alpha\beta}\ ,
\end{align}
etc., while at the highest levels, in $n$ dimensions,
\begin{gather}
\mathcal{E}_{n}\left(  \partial\partial\phi\right)  =n!~\det\left(
\partial\partial\phi\right)  \ ,\\
\mathcal{E}_{n-1}\left(  \partial\partial\phi\right)  =\left(  n-1\right)
!~\det\left(  \partial\partial\phi\right)  ~\mathrm{Tr}\left(  \left(
\partial\partial\phi\right)  ^{-1}\right)  =\left(  n-1\right)  !~\mathcal{E}%
_{1}\left(  \mathrm{adj}\left(  \partial\partial\phi\right)  \right)  \ ,
\label{TraceAdjEofM}%
\end{gather}
etc. \ We shall refer to the $\mathcal{E}_{n}\left(  \partial\partial
\phi\right)  =0$\ case as the \textquotedblleft maximal\textquotedblright%
\ galileon field equations.

\subsection{Lagrangians}

These may be defined recursively and yield the above equations of motion after
varying $\phi$ and integrating by parts:%
\begin{equation}
\mathcal{L}_{k}=\phi_{\alpha}\phi_{\alpha}~\mathcal{E}_{k-1}\left(
\partial\partial\phi\right)  \ ,\ \ \ \delta\int\mathcal{L}_{k}~d^{n}%
x=-2\int\mathcal{E}_{k}\left(  \partial\partial\phi\right)  ~\delta\phi
~d^{n}x\ . \label{kLevelSystem}%
\end{equation}
Thus, stationarity of the action for $\mathcal{L}_{k}$\ implies the equation
of motion:%
\begin{equation}
0=\mathcal{E}_{k}\left(  \partial\partial\phi\right)  \ .
\end{equation}

A systematic method to obtain this recursion is to work out the variation of%
\begin{equation}
A=\int\phi_{\alpha}\phi_{\alpha}\det\left(  1+\lambda\partial\partial
\phi\right)  ~d^{n}x\ ,
\end{equation}
and then use (\ref{DetExpansion}) to single out the action for $\mathcal{L}%
_{k}$. \ More generally, with%
\begin{equation}
A\left[  F\right]  =\int F\left(  \phi_{\alpha}\phi_{\alpha}\right)
\det\left(  \boldsymbol{1}+\lambda\partial\partial\phi\right)  ~d^{n}x\ ,
\end{equation}
we find%
\begin{equation}
\delta A\left[  F\right]  =-2\int\mathcal{E}\left[  F\right]  ~\delta
\phi~d^{n}x\ ,
\end{equation}
where%
\begin{equation}
\mathcal{E}\left[  F\right]  \mathcal{=}\det\left(  \boldsymbol{1}%
+\lambda\partial\partial\phi\right)  ~\left\{  \left(  \phi_{\alpha\alpha
}-\lambda\left(  \boldsymbol{1}+\lambda\partial\partial\phi\right)  _{\mu\nu
}^{-1}~\phi_{\mu\alpha}\phi_{\nu\alpha}\right)  F^{\prime}\right.  \left.
+\left(  \phi_{\mu}~\left(  \boldsymbol{1}+\lambda\partial\partial\phi\right)
_{\mu\nu}^{-1}~\partial_{\nu}\left(  \phi_{\alpha}\phi_{\alpha}\right)
\right)  F^{\prime\prime}\right\}  \ . \label{E[F]}%
\end{equation}
Setting $F\left(  \phi_{\alpha}\phi_{\alpha}\right)  =\phi_{\alpha}%
\phi_{\alpha}$, i.e. $F^{\prime}=1$ and $F^{\prime\prime}=0$, and expanding
the RHS of (\ref{E[F]}) in powers of $\lambda$ leads to (\ref{kLevelSystem}).

\subsection{Universal field equations}

Given in $n$ dimensions an arbitrary Lagrangian dependent only upon first
derivatives of the field, $\phi$, and homogeneous of weight one, there is an
iterative procedure for calculating a sequence of equations of motion which
always terminates with the same final equation, $\mathcal{U}_{n}=0$,
independent of the starting Lagrangian \cite{FGM,FG0}. \ This final equation
has therefore been called a \textquotedblleft universal field
equation\textquotedblright\ (UFE).\footnote{In two dimensions, the UFE is just
\emph{the Bateman equation}, $\phi_{xx}\phi_{t}^{2}-2\phi_{xt}\phi_{x}\phi
_{t}+\phi_{tt}\phi_{t}^{2}=0$.} \ It involves only first and second
derivatives of $\phi$. \ Here we describe the relation between $\mathcal{U}%
_{n}$ and the galileon Lagrangian $\mathcal{L}_{n}$\ in $n$ dimensions.

The functional form appearing in the UFE in $n$ dimensions can be expressed as
a \textquotedblleft bordered determinant\textquotedblright\ \cite{F,Chaundy},%
\begin{equation}
\mathcal{U}_{n}\left[  \phi\right]  =\det\left(
\begin{array}
[c]{cc}%
0 & \partial\phi\\
\partial\phi & \partial\partial\phi
\end{array}
\right)  \ , \label{Un}%
\end{equation}
where the entries in the top row, and in the left column, are $0$ and
$\phi_{\alpha}$ for $\alpha=1,\cdots,n$, and the Hessian matrix occupies the
$n\times n$ block on the lower right. \ However, unlike $\mathcal{E}%
_{k}\left(  \partial\partial\phi\right)  $ that appears in the galileon field
equations, $\mathcal{U}_{n}$ is \textbf{not }identical to a total divergence,
so the integral $\int\mathcal{U}_{n}\left[  \phi\right]  d^{n}x$ can serve to
specify nontrivial dynamics in the bulk.

For example, in $n=2$ Euclidean dimensions,
\begin{align}
\mathcal{U}_{2}  &  =\det\left(
\begin{array}
[c]{ccc}%
0 & \phi_{1} & \phi_{2}\\
\phi_{1} & \phi_{11} & \phi_{12}\\
\phi_{2} & \phi_{21} & \phi_{22}%
\end{array}
\right) \nonumber\\
&  =-\phi_{1}\left(  \phi_{1}\phi_{22}-\phi_{12}\phi_{2}\right)  +\phi
_{2}\left(  \phi_{21}\phi_{1}-\phi_{2}\phi_{11}\right) \nonumber\\
&  =-\phi_{\alpha}\phi_{\alpha}\phi_{\beta\beta}+\phi_{\alpha}\phi
_{\alpha\beta}\phi_{\beta}\ .
\end{align}
But then $\phi_{\alpha}\phi_{\alpha\beta}\phi_{\beta}=\partial_{\beta}\left(
\phi_{\alpha}\phi_{\alpha}\phi_{\beta}\right)  -\phi_{\alpha\beta}\phi
_{\alpha}\phi_{\beta}-\phi_{\alpha}\phi_{\alpha}\phi_{\beta\beta}$ so
$\phi_{\alpha}\phi_{\alpha\beta}\phi_{\beta}=\frac{1}{2}\partial_{\beta
}\left(  \phi_{\alpha}\phi_{\alpha}\phi_{\beta}\right)  -\frac{1}{2}%
\phi_{\alpha}\phi_{\alpha}\phi_{\beta\beta}$. \ Thus $\mathcal{U}_{2}%
=-\frac{3}{2}\phi_{\alpha}\phi_{\alpha}\phi_{\beta\beta}+\frac{1}{2}%
\partial_{\beta}\left(  \phi_{\alpha}\phi_{\alpha}\phi_{\beta}\right)  $.
\ That is to say,%
\begin{equation}
\mathcal{U}_{2}=-\frac{3}{2}\mathcal{L}_{2}+\frac{1}{2}\partial_{\beta}\left(
\phi_{\alpha}\phi_{\alpha}\phi_{\beta}\right)  \ . \label{U2}%
\end{equation}
The same result applies in spaces with Lorentz signature, when repeated
indices are summed with the Lorentz metric.

Similarly, in $n=3$ dimensions, $\mathcal{U}_{3}$ differs from a constant
multiple of $\mathcal{L}_{3}$ just by a divergence,%
\begin{equation}
\mathcal{U}_{3}=-\mathcal{L}_{3}+\frac{1}{2}\partial_{\gamma}\left(
\phi_{\alpha\alpha}\phi_{\beta}\phi_{\beta}\phi_{\gamma}-\phi_{\alpha}%
\phi_{\alpha}\phi_{\beta}\phi_{\beta\gamma}\right)  \ . \label{U3}%
\end{equation}
Indeed, it turns out that in any $n$ dimensions, $\mathcal{U}_{n}$ is always
proportional to the maximal galileon Lagrangian $\mathcal{L}_{n}$ modulo a
divergence, or boundary term, $\mathcal{B}_{n}$,\footnote{Note this would
still be true if the zero in the upper left corner of $\mathcal{U}_{n}$ were
replaced by any constant $c$, for then $\det\left(
\begin{array}
[c]{cc}%
c & \partial\phi\\
\partial\phi & \partial\partial\phi
\end{array}
\right)  =\mathcal{U}_{n}+\frac{c}{n!}~\mathcal{E}_{n}$ and the last term is
again a total divergence.} \
\begin{equation}
\left(  n-1\right)  !~\mathcal{U}_{n}=-\frac{1}{2}\left(  n+1\right)
\mathcal{L}_{n}+\frac{1}{2}\left(  n-1\right)  \mathcal{B}_{n}\ . \label{UnLn}%
\end{equation}
The relative coefficient between $\mathcal{U}_{n}$ and $\mathcal{L}_{n}$ is
worked out explicitly in the next subsection, where an explicit form for
$\mathcal{B}_{n}$ is also given. \ The upshot is that the action for maximal
galileon fields that vanish on the spacetime boundary is obtained just by
integrating the functional form appearing in the UFE, $\int\mathcal{U}%
_{n}\left[  \phi\right]  d^{n}x$.

As someone well-schooled in determinants might guess, especially in light of
the discussion following (\ref{DetExpansion}), there is another way to express
the UFE\ in $n$ dimensions, in terms of traces. \ This is given by%
\begin{equation}
\mathcal{V}_{n}\equiv\det\left(
\begin{array}
[c]{ccccccc}%
\mathcal{S}_{0} & n-1 & 0 & \cdots & 0 & 0 & 0\\
\mathcal{S}_{1} & \mathcal{T}_{1} & n-2 & \cdots & 0 & 0 & 0\\
\mathcal{S}_{2} & \mathcal{T}_{2} & \mathcal{T}_{1} & \cdots & 0 & 0 & 0\\
\vdots & \vdots & \vdots & \ddots & \vdots & \vdots & \vdots\\
\mathcal{S}_{n-3} & \mathcal{T}_{n-3} & \mathcal{T}_{n-4} & \cdots &
\mathcal{T}_{1} & 2 & 0\\
\mathcal{S}_{n-2} & \mathcal{T}_{n-2} & \mathcal{T}_{n-3} & \cdots &
\mathcal{T}_{2} & \mathcal{T}_{1} & 1\\
\mathcal{S}_{n-1} & \mathcal{T}_{n-1} & \mathcal{T}_{n-2} & \cdots &
\mathcal{T}_{3} & \mathcal{T}_{2} & \mathcal{T}_{1}%
\end{array}
\right)  =-\left(  n-1\right)  !~\mathcal{U}_{n}\ , \label{UnTraces}%
\end{equation}
where we have defined%
\begin{equation}
\mathcal{T}_{k}=\mathrm{Tr}\left[  \left(  \partial\partial\phi\right)
^{k}\right]  \ ,\ \ \ \mathcal{S}_{k}=\mathrm{Tr}\left[  \left(  \partial
\phi\partial\phi\right)  \left(  \partial\partial\phi\right)  ^{k}\right]  \ .
\label{T&S}%
\end{equation}
The special results in (\ref{U2}) and (\ref{U3}) may be confirmed from
(\ref{UnTraces}).

Perhaps the most elegant proof that the determinants in (\ref{Un}) and
(\ref{UnTraces}) are proportional is to make use of an orthogonal
transformation at each point $x$ to find local frames such that the symmetric
Hessian matrix is diagonal,
\begin{equation}
\left(  \partial\partial\phi\right)  =\mathrm{diag}\left(  \lambda_{1}%
,\cdots,\lambda_{n}\right)  \ .
\end{equation}
(We do not diagonalize the full $\left(  n+1\right)  \times\left(  n+1\right)
$ matrix appearing in (\ref{Un})\ because we wish to keep track of the first
derivatives, $\partial\phi$.) \ In such frames it is straightforward to show,
from either (\ref{Un}) or (\ref{UnTraces}), that%
\begin{equation}
\mathcal{U}_{n}=-\sum_{\alpha=1}^{n}\phi_{\alpha}^{2}~(%
{\displaystyle\prod\limits_{\substack{\beta=1 \\\beta\neq\alpha}}^{n}}
\lambda_{\beta})\ .
\end{equation}
The signs here are for $n$ Euclidean dimensions. \ For spaces with Lorentz
signature, the result holds upon making the usual sign changes.

The determinant in (\ref{UnTraces}) should be compared to that in
(\ref{DetEofM}). \ This suggests that we consider more generally, for $0\leq
k\leq n$, the determinant%
\begin{equation}
\mathcal{V}_{k}=\det\left(
\begin{array}
[c]{ccccccc}%
\mathcal{S}_{0} & k-1 & 0 & \cdots & 0 & 0 & 0\\
\mathcal{S}_{1} & \mathcal{T}_{1} & k-2 & \cdots & 0 & 0 & 0\\
\mathcal{S}_{2} & \mathcal{T}_{2} & \mathcal{T}_{1} & \cdots & 0 & 0 & 0\\
\vdots & \vdots & \vdots & \ddots & \vdots & \vdots & \vdots\\
\mathcal{S}_{k-3} & \mathcal{T}_{k-3} & \mathcal{T}_{k-4} & \cdots &
\mathcal{T}_{1} & 2 & 0\\
\mathcal{S}_{k-2} & \mathcal{T}_{k-2} & \mathcal{T}_{k-3} & \cdots &
\mathcal{T}_{2} & \mathcal{T}_{1} & 1\\
\mathcal{S}_{k-1} & \mathcal{T}_{k-1} & \mathcal{T}_{k-2} & \cdots &
\mathcal{T}_{3} & \mathcal{T}_{2} & \mathcal{T}_{1}%
\end{array}
\right)  \ . \label{VkDet}%
\end{equation}
Expand this determinant in the minors of the first column to obtain%
\begin{equation}
\mathcal{V}_{k}=\mathcal{S}_{0}\mathcal{E}_{k-1}-\left(  k-1\right)
\mathcal{S}_{1}\mathcal{E}_{k-2}+\left(  k-1\right)  \left(  k-2\right)
\mathcal{S}_{2}\mathcal{E}_{k-3}-+\cdots+\left(  -1\right)  ^{k-1}\left(
k-1\right)  !\mathcal{S}_{k-1}\mathcal{E}_{0}\ . \label{Vk}%
\end{equation}
This in turn suggests a more direct --- but somewhat brute force ---
derivation of the relation between (\ref{Un}) and (\ref{UnTraces}), and the
result in (\ref{UnLn}). \ This is given in the next section.

\subsection{More on determinant identities}

We have defined $\mathcal{E}_{k}\left(  M\right)  $ by the expansion
(\ref{DetExpansion}). \ On the other hand, we also have the identities%
\begin{align}
&  \det\left(  \boldsymbol{1}+\lambda M\right) \nonumber\\
&  =\frac{1}{n!}~\varepsilon_{\alpha_{1}\cdots\alpha_{n}}\varepsilon
_{\beta_{1}\cdots\beta_{n}}~\left(  \delta_{\alpha_{1}\beta_{1}}+\lambda
M_{\alpha_{1}\beta_{1}}\right)  \cdots\left(  \delta_{\alpha_{n}\beta_{n}%
}+\lambda M_{\alpha_{n}\beta_{n}}\right) \nonumber\\
&  =\frac{1}{n!}~\delta_{\beta_{1}\cdots\beta_{n}}^{\alpha_{1}\cdots\alpha
_{n}}~\left(  \delta_{\alpha_{1}\beta_{1}}+\lambda M_{\alpha_{1}\beta_{1}%
}\right)  \cdots\left(  \delta_{\alpha_{n}\beta_{n}}+\lambda M_{\alpha
_{n}\beta_{n}}\right) \nonumber\\
&  =\frac{1}{n!}\left(
\begin{array}
[c]{c}%
\delta_{\alpha_{1}\cdots\alpha_{n}}^{\alpha_{1}\cdots\alpha_{n}}+\delta
_{\beta_{1}\alpha_{2}\cdots\alpha_{n}}^{\alpha_{1}\alpha_{2}\cdots\alpha_{n}%
}\times n~\lambda M_{\alpha_{1}\beta_{1}}+\delta_{\beta_{1}\beta_{2}\alpha
_{3}\cdots\alpha_{n}}^{\alpha_{1}\alpha_{2}\alpha_{3}\cdots\alpha_{n}}%
\times\frac{n\left(  n-1\right)  }{2}~\lambda^{2}M_{\alpha_{1}\beta_{1}%
}M_{\alpha_{2}\beta_{2}}+\cdots\\
+\delta_{\beta_{1}\cdots\beta_{n}}^{\alpha_{1}\cdots\alpha_{n}}\times\frac
{n!}{n!}~\lambda^{n}M_{\alpha_{1}\beta_{1}}\cdots M_{\alpha_{n}\beta_{n}}%
\end{array}
\right) \nonumber\\
&  =\frac{1}{n!}\left(
\begin{array}
[c]{c}%
n!+\left(  n-1\right)  !\delta_{\beta_{1}}^{\alpha_{1}}\times n~\lambda
M_{\alpha_{1}\beta_{1}}+\left(  n-2\right)  !\delta_{\beta_{1}\beta_{2}%
}^{\alpha_{1}\alpha_{2}}\times\frac{n\left(  n-1\right)  }{2}~\lambda
^{2}M_{\alpha_{1}\beta_{1}}M_{\alpha_{2}\beta_{2}}+\cdots\\
+\delta_{\beta_{1}\cdots\beta_{n}}^{\alpha_{1}\cdots\alpha_{n}}\times\frac
{n!}{n!}~\lambda^{n}M_{\alpha_{1}\beta_{1}}\cdots M_{\alpha_{n}\beta_{n}}%
\end{array}
\right) \nonumber\\
&  =1+\lambda~\delta_{\beta_{1}}^{\alpha_{1}}\times M_{\alpha_{1}\beta_{1}%
}+\frac{\lambda^{2}}{2!}~\delta_{\beta_{1}\beta_{2}}^{\alpha_{1}\alpha_{2}%
}\times M_{\alpha_{1}\beta_{1}}M_{\alpha_{2}\beta_{2}}+\cdots+\frac
{\lambda^{n}}{n!}~\delta_{\beta_{1}\cdots\beta_{n}}^{\alpha_{1}\cdots
\alpha_{n}}\times M_{\alpha_{1}\beta_{1}}\cdots M_{\alpha_{n}\beta_{n}}\ .
\label{DetExpansionKronecker}%
\end{align}
where the generalized Kronecker symbols are defined by\footnote{There is a
difference here between Euclidean and Minkoski metrics. \ For Euclidean space,%
\[
\varepsilon_{\alpha_{1}\cdots\alpha_{n}}\varepsilon^{\beta_{1}\cdots\beta_{n}%
}=\delta_{\alpha_{1}\cdots\alpha_{n}}^{\beta_{1}\cdots\beta_{n}}%
\]
is true for any $n$, but the corresponding identity in Minkowski space is%
\[
\varepsilon_{\alpha_{1}\cdots\alpha_{n}}\varepsilon^{\beta_{1}\cdots\beta_{n}%
}=\left(  -1\right)  ^{n-1}\delta_{\alpha_{1}\cdots\alpha_{n}}^{\beta
_{1}\cdots\beta_{n}}\ .
\]
The remaining discussion in this subsection will be given for the Euclidean
case.}%
\begin{align}
\delta_{\beta_{1}\cdots\beta_{k}}^{\alpha_{1}\cdots\alpha_{k}}  &  =\frac
{1}{\left(  n-k\right)  !}~\delta_{\beta_{1}\cdots\beta_{k}\underline{\alpha
_{k+1}\cdots\alpha_{n}}}^{\alpha_{1}\cdots\alpha_{k}\underline{\alpha
_{k+1}\cdots\alpha_{n}}}\\
&  =\frac{1}{\left(  n-k\right)  !}~\varepsilon_{\alpha_{1}\cdots\alpha
_{k}\underline{\alpha_{k+1}\cdots\alpha_{n}}}\varepsilon_{\beta_{1}\cdots
\beta_{k}\underline{\alpha_{k+1}\cdots\alpha_{n}}}\nonumber\\
&  =\delta_{\beta_{1}}^{\alpha_{1}}\delta_{\beta_{2}}^{\alpha_{2}}\cdots
\delta_{\beta_{k-1}}^{\alpha_{k-1}}\delta_{\beta_{k}}^{\alpha_{k}}%
\pm\text{permutations of }\alpha\text{s or }\beta\text{s, but not
both.}\nonumber
\end{align}
For emphasis, on the RHS we have underlined the repeated indices that are
implicitly summed. \ Thus the $\mathcal{E}_{k}\left(  M\right)  $ can be
expressed in terms of generalized Kronecker symbols:%
\begin{equation}
\mathcal{E}_{k}\left(  M\right)  =\delta_{\beta_{1}\beta_{2}\cdots\beta_{k}%
}^{\alpha_{1}\alpha_{2}\cdots\alpha_{k}}\times M_{\alpha_{1}\beta_{1}}\cdots
M_{\alpha_{k}\beta_{k}}\ . \label{KroneckerEk}%
\end{equation}
Note the last term in the expansion (\ref{DetExpansionKronecker}) is the
familiar
\begin{equation}
\mathcal{E}_{n}\left(  M\right)  =\delta_{\beta_{1}\beta_{2}\cdots\beta_{n}%
}^{\alpha_{1}\alpha_{2}\cdots\alpha_{n}}\times M_{\alpha_{1}\beta_{1}}\cdots
M_{\alpha_{n}\beta_{n}}=n!\det M\ .
\end{equation}

Applying generalized Kronecker symbol methods to bordered determinants, and
making use of (\ref{KroneckerEk}), leads to relations that may be usefully
applied to the UFE. \ Consider
\begin{equation}
\mathcal{M}=\left(
\begin{array}
[c]{cc}%
0 & \widetilde{v}\\
v & M
\end{array}
\right)  \ ,
\end{equation}
where $M$ is any symmetric $n\times n$ matrix, $v$ is an arbitrary $n\times1$
column matrix, and its transpose $\widetilde{v}$ is a $1\times n$ row matrix.
\ Clearly $\det\mathcal{M}$ is bilinear in the components of $v$. \ For
convenience, we index the rows and columns of $\mathcal{M}$\ from $0$ to $n$.
\ Since $\mathcal{M}_{00}=0$ it follows that
\begin{align}
\det\mathcal{M}  &  \mathcal{=}\frac{1}{\left(  n+1\right)  !}~\delta
_{\beta_{0}\beta_{1}\cdots\beta_{n}}^{\alpha_{0}\alpha_{1}\cdots\alpha_{n}%
}\times\mathcal{M}_{\alpha_{0}\beta_{0}}\cdots\mathcal{M}_{\alpha_{n}\beta
_{n}}\nonumber\\
&  =\frac{1}{\left(  n+1\right)  !}~\left(  n+1\right)  ~\delta_{\beta
_{0}\beta_{1}\beta_{2}\cdots\beta_{n-1}\beta_{n}}^{0\alpha_{1}\alpha_{2}%
\cdots\alpha_{n-1}\alpha_{n}}\times\mathcal{M}_{0\beta_{0}}\mathcal{M}%
_{\alpha_{1}\beta_{1}}\cdots\mathcal{M}_{\alpha_{n}\beta_{n}}\nonumber\\
&  =\frac{1}{\left(  n+1\right)  !}~\left(  n+1\right)  n~\delta_{\beta
_{0}\beta_{1}\beta_{2}\cdots\beta_{n-1}0}^{0\alpha_{1}\alpha_{2}\cdots
\alpha_{n-1}\alpha_{n}}\times\mathcal{M}_{0\beta_{0}}\mathcal{M}_{\alpha
_{1}\beta_{1}}\cdots\mathcal{M}_{\alpha_{n-1}\beta_{n-1}}\mathcal{M}%
_{\alpha_{n}0}\ ,
\end{align}
where other $0$ subscripts can \emph{not} appear because of the antisymmetry
of the Kronecker delta. \ So, substituting $v$ and $M$ for the components of
$\mathcal{M}$,
\begin{align}
\det\mathcal{M}  &  =\frac{-1}{\left(  n-1\right)  !}~\delta_{\beta_{1}%
\beta_{2}\cdots\beta_{n-1}\beta_{0}}^{\alpha_{1}\alpha_{2}\cdots\alpha
_{n-1}\alpha_{n}}\times v_{\beta_{0}}v_{\alpha_{n}}M_{\alpha_{1}\beta_{1}%
}\cdots M_{\alpha_{n-1}\beta_{n-1}}\label{detMasKroneckers}\\
&  =\frac{-1}{\left(  n-1\right)  !}~\left[  _{\ }\left.  \delta_{\beta
_{1}\beta_{2}\cdots\beta_{n-1}}^{\alpha_{1}\alpha_{2}\cdots\alpha_{n-1}}%
\delta_{\beta_{0}}^{\alpha_{n}}-\left(  n-1\right)  \delta_{\beta_{1}\beta
_{2}\cdots\beta_{n-2}\beta_{0}}^{\alpha_{1}\alpha_{2}\cdots\alpha_{n-2}%
\alpha_{n-1}}\delta_{\beta_{n-1}}^{\alpha_{n}}\right.  _{\ }\right]  \times
v_{\beta_{0}}v_{\alpha_{n}}M_{\alpha_{1}\beta_{1}}\cdots M_{\alpha_{n-2}%
\beta_{n-2}}M_{\alpha_{n-1}\beta_{n-1}}\nonumber\\
&  =\frac{-1}{\left(  n-1\right)  !}~\left[  _{\ }\left.  v_{\alpha}v_{\alpha
}\mathcal{E}_{n-1}\left(  M\right)  -\left(  n-1\right)  \delta_{\beta
_{1}\beta_{2}\cdots\beta_{n-2}\beta_{0}}^{\alpha_{1}\alpha_{2}\cdots
\alpha_{n-2}\alpha_{n-1}}\times v_{\beta_{0}}\left(  Mv\right)  _{\alpha
_{n-1}}M_{\alpha_{1}\beta_{1}}\cdots M_{\alpha_{n-2}\beta_{n-2}}\right.
_{\ }\right]  \ .\nonumber
\end{align}
Repeating the steps of the last three lines to reduce the remaining Kronecker
symbol gives the series%
\begin{align}
\det\mathcal{M}  &  =\frac{-1}{\left(  n-1\right)  !}~\left[
\begin{array}
[c]{c}%
\left(  \widetilde{v}v\right)  \mathcal{E}_{n-1}\left(  M\right)  -\left(
n-1\right)  \left(  \widetilde{v}Mv\right)  \mathcal{E}_{n-2}\left(  M\right)
+\left(  n-1\right)  \left(  n-2\right)  \left(  \widetilde{v}M^{2}v\right)
\mathcal{E}_{n-2}\left(  M\right) \\
-\cdots+\left(  -1\right)  ^{n-1}\left(  n-1\right)  !\left(  \widetilde{v}%
M^{n-1}v\right)  \mathcal{E}_{0}\left(  M\right)
\end{array}
\right] \nonumber\\
&  =\frac{-1}{\left(  n-1\right)  !}\sum_{j=0}^{n-1}\left(  -1\right)
^{j}\frac{\left(  n-1\right)  !}{\left(  n-1-j\right)  !}\left(
\widetilde{v}M^{j}v\right)  \mathcal{E}_{n-1-j}\left(  M\right)
\end{align}
where $\left(  \widetilde{v}M^{k}v\right)  =\mathrm{Tr}\left(  v\widetilde{v}%
M^{k}\right)  =v_{\alpha}\left(  M^{k}\right)  _{\alpha\beta}v_{\beta}$, and
of course, $\mathcal{E}_{0}\left(  M\right)  =1$.

In particular, setting $v=\partial\phi$ and $M=\partial\partial\phi$, and
recalling the second definition in (\ref{T&S}), this last result leads
directly from (\ref{Un}) to%
\begin{equation}
\mathcal{U}_{n}=\frac{-1}{\left(  n-1\right)  !}\sum_{j=0}^{n-1}\left(
-1\right)  ^{j}\frac{\left(  n-1\right)  !}{\left(  n-1-j\right)
!}~\mathcal{S}_{j}~\mathcal{E}_{n-1-j}\left[  \partial\partial\phi\right]  \ .
\end{equation}
But the sum on the right-hand side here is just another way to write
(\ref{VkDet}) for $k=n$, as expressed in (\ref{Vk}). \ Thus we establish again
the relation between $\mathcal{U}_{n}$ and $\mathcal{V}_{n}$, (\ref{UnTraces}%
), only this time without invoking a local frame to diagonalize the Hessian matrix.

Next, we consider the local variation of $\int\mathcal{U}_{n}\left[
\phi\right]  d^{n}x$ using (\ref{detMasKroneckers}). \ Thus
\begin{equation}
\int\mathcal{U}_{n}\left[  \phi\right]  d^{n}x=\frac{-1}{\left(  n-1\right)
!}\int\delta_{\beta_{1}\beta_{2}\cdots\beta_{n-1}\beta}^{\alpha_{1}\alpha
_{2}\cdots\alpha_{n-1}\alpha}\times\phi_{\beta}\phi_{\alpha}\times\phi
_{\alpha_{1}\beta_{1}}\cdots\phi_{\alpha_{n-1}\beta_{n-1}}d^{n}x\ ,
\end{equation}
and dropping surface terms in the variation,%
\begin{align}
&  \delta\int\mathcal{U}_{n}\left[  \phi\right]  d^{n}x\nonumber\\
&  =\frac{-1}{\left(  n-1\right)  !}\int\delta_{\beta_{1}\beta_{2}\cdots
\beta_{n-1}\beta}^{\alpha_{1}\alpha_{2}\cdots\alpha_{n-1}\alpha}\times\left(
_{\ }\left.  2\left(  \delta\phi_{\beta}\right)  \phi_{\alpha}\times
\phi_{\alpha_{1}\beta_{1}}\cdots\phi_{\alpha_{n-1}\beta_{n-1}}+\left(
n-1\right)  \phi_{\beta}\phi_{\alpha}\times\left(  \delta\phi_{\alpha_{1}%
\beta_{1}}\right)  \phi_{\alpha_{2}\beta_{2}}\cdots\phi_{\alpha_{n-1}%
\beta_{n-1}}\right.  _{\ }\right)  d^{n}x\nonumber\\
&  =\frac{-1}{\left(  n-1\right)  !}\int\delta\phi\times\delta_{\beta_{1}%
\beta_{2}\cdots\beta_{n-1}\beta}^{\alpha_{1}\alpha_{2}\cdots\alpha_{n-1}%
\alpha}\times\left(  _{\ }\left.  -2\phi_{\alpha\beta}\times\phi_{\alpha
_{1}\beta_{1}}\cdots\phi_{\alpha_{n-1}\beta_{n-1}}+\left(  n-1\right)  \left(
\phi_{\beta}\phi_{\alpha}\right)  _{\alpha_{1}\beta_{1}}\times\phi_{\alpha
_{2}\beta_{2}}\cdots\phi_{\alpha_{n-1}\beta_{n-1}}\right.  _{\ }\right)
d^{n}x\nonumber\\
&  =\frac{-1}{\left(  n-1\right)  !}\int\delta\phi\times\delta_{\beta_{1}%
\beta_{2}\cdots\beta_{n-1}\beta}^{\alpha_{1}\alpha_{2}\cdots\alpha_{n-1}%
\alpha}\times\left(  _{\ }\left.  -2\phi_{\alpha\beta}\times\phi_{\alpha
_{1}\beta_{1}}\cdots\phi_{\alpha_{n-1}\beta_{n-1}}+\left(  n-1\right)
\phi_{\alpha_{1}\beta}\phi_{\alpha\beta_{1}}\times\phi_{\alpha_{2}\beta_{2}%
}\cdots\phi_{\alpha_{n-1}\beta_{n-1}}\right.  _{\ }\right)  d^{n}x\nonumber\\
&  =\frac{n+1}{\left(  n-1\right)  !}\int\delta\phi\times\delta_{\beta
_{1}\beta_{2}\cdots\beta_{n-1}\beta}^{\alpha_{1}\alpha_{2}\cdots\alpha
_{n-1}\alpha}\times\left(  _{\ }\left.  \phi_{\alpha_{1}\beta_{1}}\phi
_{\alpha_{2}\beta_{2}}\cdots\phi_{\alpha_{n-1}\beta_{n-1}}\phi_{\alpha
_{n}\beta_{n}}\right.  _{\ }\right)  d^{n}x \label{deltaUasKroneckers}%
\end{align}
Now use $\mathcal{E}_{n}\left(  \partial\partial\phi\right)  $ as given by
(\ref{KroneckerEk}) to obtain the result,%
\begin{equation}
\delta\int\mathcal{U}_{n}\left[  \phi\right]  d^{n}x=\frac{n+1}{\left(
n-1\right)  !}\int\mathcal{E}_{n}\left(  \partial\partial\phi\right)
~\delta\phi~d^{n}x\ .
\end{equation}
Compare this to $\delta\int\mathcal{L}_{n}~d^{n}x=-2\int\mathcal{E}_{n}\left(
\partial\partial\phi\right)  ~\delta\phi~d^{n}x,$ as given in
(\ref{kLevelSystem}) for $k=n$, to conclude%
\begin{equation}
\delta\int\mathcal{U}_{n}\left[  \phi\right]  d^{n}x=\delta\int\left(
-\frac{n+1}{2\left(  n-1\right)  !}\mathcal{L}_{n}\left[  \phi\right]
\right)  d^{n}x\ ,
\end{equation}
where spacetime boundary terms have been dropped. \ Thus the unvaried
integrands can only differ by a divergence. \ This establishes (\ref{UnLn}).
\ An explicit form for the divergence $\mathcal{B}_{n}$, as normalized in
(\ref{UnLn}), can be found by keeping track of the discarded boundary terms
produced by integrating by parts in (\ref{kLevelSystem}) and in
(\ref{deltaUasKroneckers}). \ This is left as an exercise. \ The result is%
\begin{equation}
\mathcal{B}_{n}=\partial_{\sigma}\left(  _{\ }\left.  \delta_{\rho\beta
_{1}\beta_{2}\cdots\beta_{n-2}}^{\sigma\alpha_{1}\alpha_{2}\cdots\alpha_{n-2}%
}\times\phi_{\alpha_{1}\beta_{1}}\cdots\phi_{\alpha_{n-2}\beta_{n-2}}%
\times\phi_{\rho}\phi_{\gamma}\phi_{\gamma}\right.  _{\ }\right)  \ .
\label{nDivergence}%
\end{equation}
For example, this reduces to the divergence terms in (\ref{U2}) and
(\ref{U3}), for $n=2$ and $n=3$, respectively.

To complete our discussion of the UFE, we consider the effects of a quadratic
constraint on $\phi_{\alpha}$ for a field in $n+1$ dimensions. \ The
constraint will effectively reduce the number of dimensions to be $n$. \ For
convenience, we let indices $\alpha,\beta=0,1,\cdots,n,$ while we let
$\lambda,\mu,\nu=1,\cdots,n$. \ Then the constraint of interest to us is
\begin{equation}
\phi_{\alpha}\phi_{\alpha}=0\ .
\end{equation}
Nontrivial solutions would of course require complex fields in the Euclidean
case, but for the time being, let us not be deterred by this.. \ Solve for
$\phi_{0}^{2}$ and differentiate to obtain%
\begin{equation}
\phi_{0}^{2}=-\phi_{\mu}\phi_{\mu}\ ,\ \ \ \phi_{0\mu}=-\frac{\phi_{\nu}%
\phi_{\nu\mu}}{\phi_{0}}\ ,\ \ \ \phi_{00}=\frac{\phi_{\nu}\phi_{\nu\mu}%
\phi_{\mu}}{\phi_{0}^{2}}\ .
\end{equation}
Now compute $\mathcal{E}_{k}\left(  \partial\partial\phi\right)  $ subject to
the constraint, specifically displaying the occurrences of $\phi_{0\mu}$\ and
$\phi_{00}$. \ Thus%
\begin{align}
\left.  \mathcal{E}_{k}\left(  \partial\partial\phi\right)  \right\vert
_{\substack{n+1\text{ dimensions} \\\text{with }\phi_{\alpha}\phi_{\alpha}%
=0}}  &  =\delta_{\beta_{1}\beta_{2}\cdots\beta_{k}}^{\alpha_{1}\alpha
_{2}\cdots\alpha_{k}}\times\phi_{\alpha_{1}\beta_{1}}\cdots\phi_{\alpha
_{k}\beta_{k}}\ \nonumber\\
&  =k\delta_{\nu_{1}\nu_{2}\cdots\nu_{k-1}}^{\mu_{1}\mu_{2}\cdots\mu_{k-1}%
}\times\phi_{\mu_{1}\nu_{1}}\cdots\phi_{\mu_{k-1}\nu_{k-1}}\times\phi
_{00}\nonumber\\
&  -k\left(  k-1\right)  \delta_{\nu_{1}\nu_{2}\cdots\nu_{k-2}}^{\mu_{1}%
\mu_{2}\cdots\mu_{k-2}}\times\phi_{\mu_{1}\nu_{1}}\cdots\phi_{\mu_{k-2}%
\nu_{k-2}}\times\phi_{0\lambda}\phi_{0\lambda}\nonumber\\
&  +\delta_{\nu_{1}\nu_{2}\cdots\nu_{k}}^{\mu_{1}\mu_{2}\cdots\mu_{k}}%
\times\phi_{\mu_{1}\nu_{1}}\cdots\phi_{\mu_{k}\nu_{k}}\nonumber\\
&  =k\delta_{\nu_{1}\nu_{2}\cdots\nu_{k-1}}^{\mu_{1}\mu_{2}\cdots\mu_{k-1}%
}\times\phi_{\mu_{1}\nu_{1}}\cdots\phi_{\mu_{k-1}\nu_{k-1}}\times\phi_{\nu
}\phi_{\nu\mu}\phi_{\mu}/\phi_{0}^{2}\nonumber\\
&  -k\left(  k-1\right)  \delta_{\nu_{1}\nu_{2}\cdots\nu_{k-2}}^{\mu_{1}%
\mu_{2}\cdots\mu_{k-2}}\times\phi_{\mu_{1}\nu_{1}}\cdots\phi_{\mu_{k-2}%
\nu_{k-2}}\times\phi_{\nu}\phi_{\nu\mu}\phi_{\mu\lambda}\phi_{\lambda}%
/\phi_{0}^{2}\nonumber\\
&  +\delta_{\nu_{1}\nu_{2}\cdots\nu_{k}}^{\mu_{1}\mu_{2}\cdots\mu_{k}}%
\times\phi_{\mu_{1}\nu_{1}}\cdots\phi_{\mu_{k}\nu_{k}}%
\end{align}
Factoring out $\phi_{0}^{2}=-\phi_{\lambda}\phi_{\lambda}$ we arrive at%
\begin{equation}
\left.  \mathcal{E}_{k}\left(  \partial\partial\phi\right)  \right\vert
_{\substack{n+1\text{ dimensions} \\\text{with }\phi_{\alpha}\phi_{\alpha}%
=0}}=\left.  \frac{1}{\phi_{\lambda}\phi_{\lambda}}~\delta_{\nu_{1}\nu
_{2}\cdots\nu_{k+1}}^{\mu_{1}\mu_{2}\cdots\mu_{k+1}}\times\phi_{\mu_{1}\nu
_{1}}\cdots\phi_{\mu_{k-1}\nu_{k-1}}\times\phi_{\mu_{k+1}}\phi_{\nu_{k+1}%
}\right\vert _{n\text{ dimensions}}\ .
\end{equation}
That is to say%
\begin{equation}
\left.  \mathcal{E}_{k}\left(  \partial\partial\phi\right)  \right\vert
_{\substack{n+1\text{ dimensions} \\\text{with }\phi_{\alpha}\phi_{\alpha}%
=0}}=\left.  \frac{1}{\phi_{\lambda}\phi_{\lambda}}~\mathcal{V}_{k+1}%
\right\vert _{\substack{n\text{ dimensions with} \\\text{varying }%
\phi_{\lambda}\phi_{\lambda}}}. \label{ConstraintRelation}%
\end{equation}

\subsection{Legendre transformations}

The standard form for a Legendre transformation $\phi,x\longleftrightarrow
\Phi,X$ is given by%
\begin{gather}
\phi\left(  x\right)  +\Phi\left(  X\right)  =\sum_{\alpha=1}^{n}x_{\alpha
}X_{\alpha}\ ,\label{LegendreTransf}\\
X_{\alpha}\left(  x\right)  =\frac{\partial\phi\left(  x\right)  }{\partial
x_{\alpha}}\equiv\partial_{\alpha}\phi\ ,\ \ \ x_{\alpha}\left(  X\right)
=\frac{\partial\Phi\left(  X\right)  }{\partial X_{\alpha}}\equiv
\nabla_{\alpha}\Phi\ .
\end{gather}
It follows that the Hessian matrices for $\phi$ and $\Phi$ are related by%
\begin{equation}
\left(  \partial\partial\phi\right)  ^{-1}=\left(  \nabla\nabla\Phi\right)
\ .
\end{equation}
From this and the previous matrix identity (\ref{SymmetricEofMDuality}) it
follows in $n$ dimensions that%
\begin{equation}
\frac{1}{\sqrt{\det\left(  \partial\partial\phi\right)  }}~\frac{1}%
{k!}~\mathcal{E}_{k}\left(  \partial\partial\phi\right)  =\frac{1}{\sqrt
{\det\left(  \nabla\nabla\Phi\right)  }}~\frac{1}{\left(  n-k\right)
!}~\mathcal{E}_{n-k}\left(  \nabla\nabla\Phi\right)  \ . \label{DualFieldEqns}%
\end{equation}
That is to say, field equations for $\phi$ and $\Phi$ are related by the
Legendre tranform, and so are their solutions. \ The transformation gives a
one-to-one local map between solutions of the nonlinear equations
$\mathcal{E}_{k}\left(  \partial\partial\phi\right)  =0$ and $\mathcal{E}%
_{n-k}\left(  \nabla\nabla\Phi\right)  =0$, valid for all $x$ or $X$ such that
the corresponding Hessian matrices are nonsingular, i.e. for all $x$ or $X$
such that $\det\left(  \partial\partial\phi\right)  \neq0\neq\det\left(
\nabla\nabla\Phi\right)  $.

This then is a general, implicit procedure for the construction of solutions
to the equation $\mathcal{E}_{k}=0$ given solutions to $\mathcal{E}_{n-k}=0$.
\ In practice it is challenging to find tractable examples where the procedure
can be fully realized. \ We will say more about solutions in \S 4.

\subsection{Legendre self-dual models}

The basic self-(anti)dual action consists of a pair of terms,%
\begin{equation}
A_{\pm}=\frac{1}{k!}~A_{k}\pm\frac{1}{\left(  n-k\right)  !}~A_{n-k}=\int%
\phi_{\alpha}\phi_{\alpha}~\left(  \frac{1}{k!}~\mathcal{E}_{k-1}\left(
\partial\partial\phi\right)  \pm\frac{1}{\left(  n-k\right)  !}~\mathcal{E}%
_{n-k-1}\left(  \partial\partial\phi\right)  \right)  \ .
\end{equation}
Thus from (\ref{kLevelSystem}) the first variation is
\begin{equation}
\delta A_{\pm}=-2\int\delta\phi~\left(  \frac{1}{k!}~\mathcal{E}_{k}\left(
\partial\partial\phi\right)  \pm\frac{1}{\left(  n-k\right)  !}~\mathcal{E}%
_{n-k}\left(  \partial\partial\phi\right)  \right)  \ .
\end{equation}
This exhibits a \emph{classical} self-(anti)duality for the resulting field
equations, and their solutions,\ under the Legendre transformation
(\ref{LegendreTransf}). \ Again in $n$ dimensions,%
\begin{align}
&  \frac{1}{\sqrt{\det\left(  \partial\partial\phi\right)  }}~\left(  \frac
{1}{k!}~\mathcal{E}_{k}\left(  \partial\partial\phi\right)  \pm\frac
{1}{\left(  n-k\right)  !}~\mathcal{E}_{n-k}\left(  \partial\partial
\phi\right)  \right) \nonumber\\
&  =\frac{\pm1}{\sqrt{\det\left(  \nabla\nabla\Phi\right)  }}~\left(  \frac
{1}{k!}~\mathcal{E}_{k}\left(  \nabla\nabla\Phi\right)  \pm\frac{1}{\left(
n-k\right)  !}~\mathcal{E}_{n-k}\left(  \nabla\nabla\Phi\right)  \right)  \ .
\end{align}

In particular, for $k=1$ this becomes%
\begin{align}
&  \frac{1}{\sqrt{\det\left(  \partial\partial\phi\right)  }}~\left(
\mathcal{E}_{1}\left(  \partial\partial\phi\right)  \pm\mathcal{E}_{1}\left(
\mathrm{adj}\left(  \partial\partial\phi\right)  \right)  \right) \nonumber\\
&  =\frac{\pm1}{\sqrt{\det\left(  \nabla\nabla\Phi\right)  }}~\left(
\mathcal{E}_{1}\left(  \nabla\nabla\Phi\right)  \pm\mathcal{E}_{1}\left(
\mathrm{adj}\left(  \nabla\nabla\Phi\right)  \right)  \right)  \ ,
\end{align}
where we have made use of (\ref{TraceAdjEofM}). \ 

The consequences of this duality for \emph{quantized} systems requires
consideration of how the Legendre transformation directly affects the actions,
$A_{\pm}$, and not just the field equations. \ We consider this next.

\subsection{Legendre transformations of the action}

When $M$ is taken to be the Hessian matrix, say, $M=\partial\partial\phi$,
then every $\mathcal{E}_{k}\left(  \partial\partial\phi\right)  $ is actually
a \emph{double} divergence,%
\begin{equation}
\mathcal{E}_{k}\left(  \partial\partial\phi\right)  =\delta_{\beta_{1}%
\beta_{2}\cdots\beta_{k}}^{\alpha_{1}\alpha_{2}\cdots\alpha_{k}}\times
\phi_{\alpha_{1}\beta_{1}}\phi_{\alpha_{2}\beta_{2}}\cdots\phi_{\alpha
_{k}\beta_{k}}=\partial_{\alpha_{1}}\partial_{\beta_{1}}\left(  \delta
_{\beta_{1}\beta_{2}\cdots\beta_{k}}^{\alpha_{1}\alpha_{2}\cdots\alpha_{k}%
}\times\phi~\phi_{\alpha_{2}\beta_{2}}\cdots\phi_{\alpha_{k}\beta_{k}}\right)
\ .
\end{equation}
Recall the Legendre transformation result (\ref{DualFieldEqns}),%
\begin{equation}
\mathcal{E}_{k}\left(  \partial\partial\phi\right)  =\frac{k!}{\left(
n-k\right)  !}\frac{1}{\det\left(  \nabla\nabla\Phi\right)  }\mathcal{E}%
_{n-k}\left(  \nabla\nabla\Phi\right)  \ .
\end{equation}
Thus, using $\det\left(  \frac{\partial x}{\partial X}\right)  =\det\left(
\nabla\nabla\Phi\right)  $, we also have%
\begin{align}
\int\phi_{\mu}\left(  x\right)  \phi_{\mu}\left(  x\right)  ~\mathcal{E}%
_{k}\left(  \partial\partial\phi\right)  ~d^{n}x  &  =\int X_{\mu}X_{\mu
}~\frac{k!}{\left(  n-k\right)  !}\frac{1}{\det\left(  \nabla\nabla
\Phi\right)  }\mathcal{E}_{n-k}\left(  \nabla\nabla\Phi\right)  \det\left(
\frac{\partial x}{\partial X}\right)  d^{n}X\nonumber\\
&  =\frac{k!}{\left(  n-k\right)  !}\int X_{\mu}X_{\mu}~\mathcal{E}%
_{n-k}\left(  \nabla\nabla\Phi\right)  d^{n}X\nonumber\\
&  =\frac{k!}{\left(  n-k\right)  !}\int X_{\mu}X_{\mu}~\nabla_{\alpha_{1}%
}\nabla_{\beta_{1}}\left(  \delta_{\beta_{1}\beta_{2}\cdots\beta_{n-k}%
}^{\alpha_{1}\alpha_{2}\cdots\alpha_{n-k}}\times\Phi~\Phi_{\alpha_{2}\beta
_{2}}\cdots\Phi_{\alpha_{n-k}\beta_{n-k}}\right)  d^{n}X\nonumber\\
&  =\frac{k!}{\left(  n-k\right)  !}~2\int\delta_{\mu\beta_{2}\cdots
\beta_{n-k}}^{\mu\alpha_{2}\cdots\alpha_{n-k}}\times\Phi~\Phi_{\alpha_{2}%
\beta_{2}}\cdots\Phi_{\alpha_{n-k}\beta_{n-k}}d^{n}X\nonumber\\
&  =\frac{k!}{\left(  n-k\right)  !}~2\left(  k+1\right)  \int\delta
_{\beta_{2}\cdots\beta_{n-k}}^{\alpha_{2}\cdots\alpha_{n-k}}\times\Phi
~\Phi_{\alpha_{2}\beta_{2}}\cdots\Phi_{\alpha_{n-k}\beta_{n-k}}d^{n}X\ ,
\end{align}
where in the last step we used%
\begin{equation}
\delta_{\mu\lambda_{1}\cdots\lambda_{m}}^{\mu\nu_{1}\cdots\nu_{m}}=\left(
n-m\right)  \delta_{\lambda_{1}\cdots\lambda_{m}}^{\nu_{1}\cdots\nu_{m}}\ .
\end{equation}
The effect of the Legendre transformation is therefore%
\begin{align}
\mathcal{A}_{k+1}  &  =\int\phi_{\mu}\left(  x\right)  \phi_{\mu}\left(
x\right)  ~\mathcal{E}_{k}\left(  \partial\partial\phi\right)  ~d^{n}%
x\nonumber\\
&  =\frac{\left(  k+1\right)  !}{\left(  n-k\right)  !}~2\int\delta_{\beta
_{2}\cdots\beta_{n-k}}^{\alpha_{2}\cdots\alpha_{n-k}}\times\Phi~\Phi
_{\alpha_{2}\beta_{2}}\cdots\Phi_{\alpha_{n-k}\beta_{n-k}}~d^{n}X\nonumber\\
&  =\frac{\left(  k+1\right)  !}{\left(  n-k\right)  !}~2\int\Phi
~\mathcal{E}_{n-k-1}\left(  \nabla\nabla\Phi\right)  ~d^{n}X
\end{align}
After the usual integrations by parts, the latter Lagrangian has variation%
\[
\delta\int\delta_{\beta_{2}\cdots\beta_{n-k}}^{\alpha_{2}\cdots\alpha_{n-k}%
}\times\Phi~\Phi_{\alpha_{2}\beta_{2}}\cdots\Phi_{\alpha_{n-k}\beta_{n-k}%
}d^{n}X=\left(  n-k\right)  \int\left(  \delta\Phi\right)  \times\delta
_{\beta_{2}\cdots\beta_{n-k}}^{\alpha_{2}\cdots\alpha_{n-k}}~\Phi_{\alpha
_{2}\beta_{2}}\cdots\Phi_{\alpha_{n-k}\beta_{n-k}}~d^{n}X\ ,
\]
giving the expected equation of motion
\begin{equation}
0=\mathcal{E}_{n-k-1}\left(  \nabla\nabla\Phi\right)  =\delta_{\beta_{2}%
\cdots\beta_{n-k}}^{\alpha_{2}\cdots\alpha_{n-k}}~\Phi_{\alpha_{2}\beta_{2}%
}\cdots\Phi_{\alpha_{n-k}\beta_{n-k}}\ .
\end{equation}
It should be possible, therefore, to express the transformed Lagrangian in the
standard form, upon integrating by parts:%
\begin{align}
&  \int\delta_{\lambda_{1}\cdots\lambda_{m}}^{\nu_{1}\cdots\nu_{m}}\times
\Phi~\Phi_{\nu_{1}\lambda_{1}}\cdots\Phi_{\nu_{m}\lambda_{m}}d^{n}X\\
&  =-\int\delta_{\lambda_{1}\cdots\lambda_{m}}^{\nu_{1}\cdots\nu_{m}}%
\times\Phi_{\lambda_{1}}\Phi_{\nu_{1}}~\Phi_{\nu_{2}\lambda_{2}}\cdots
\Phi_{\nu_{m}\lambda_{m}}~d^{n}X\text{ \ \ (}\lambda_{1}\text{ integrated by
parts)}\nonumber\\
&  =-\int\left(  \delta_{\lambda_{1}}^{\nu_{1}}\delta_{\lambda_{2}%
\cdots\lambda_{m}}^{\nu_{2}\cdots\nu_{m}}-\left(  m-1\right)  \delta
_{\lambda_{2}}^{\nu_{1}}\delta_{\lambda_{1}\lambda_{3}\cdots\lambda_{m}}%
^{\nu_{2}\nu_{3}\cdots\nu_{m}}\right)  \times\left(  \Phi_{\lambda_{1}}%
\Phi_{\nu_{1}}\Phi_{\nu_{2}\lambda_{2}}\right)  \Phi_{\nu_{3}\lambda_{3}%
}\cdots\Phi_{\nu_{m}\lambda_{m}}~d^{n}X\nonumber\\
&  =-\int\left(  \Phi_{\lambda}\Phi_{\lambda}\right)  \left(  \delta
_{\lambda_{2}\cdots\lambda_{m}}^{\nu_{2}\cdots\nu_{m}}\times\Phi_{\nu
_{2}\lambda_{2}}\Phi_{\nu_{3}\lambda_{3}}\cdots\Phi_{\nu_{m}\lambda_{m}%
}\right)  ~d^{n}X\nonumber\\
&  +\frac{1}{2}\left(  m-1\right)  \int\delta_{\lambda_{1}\lambda_{3}%
\cdots\lambda_{m}}^{\nu_{2}\nu_{3}\cdots\nu_{m}}\times\left(  \Phi
_{\lambda_{1}}\partial_{\nu_{2}}\left(  \Phi_{\lambda}\Phi_{\lambda}\right)
\right)  \Phi_{\nu_{3}\lambda_{3}}\cdots\Phi_{\nu_{m}\lambda_{m}}%
~d^{n}X\nonumber\\
&  =-\int\left(  \Phi_{\lambda}\Phi_{\lambda}\right)  \left(  \delta
_{\lambda_{2}\cdots\lambda_{m}}^{\nu_{2}\cdots\nu_{m}}\times\Phi_{\nu
_{2}\lambda_{2}}\Phi_{\nu_{3}\lambda_{3}}\cdots\Phi_{\nu_{m}\lambda_{m}%
}\right)  ~d^{n}X\nonumber\\
&  -\frac{1}{2}\left(  m-1\right)  \int\delta_{\lambda_{1}\lambda_{3}%
\cdots\lambda_{m}}^{\nu_{2}\nu_{3}\cdots\nu_{m}}\times\left(  \Phi_{\lambda
}\Phi_{\lambda}\right)  \Phi_{\lambda_{1}\nu_{2}}\Phi_{\nu_{3}\lambda_{3}%
}\cdots\Phi_{\nu_{m}\lambda_{m}}~d^{n}X\text{ \ \ (}\nu_{2}\text{ integrated
by parts)}\nonumber
\end{align}
So the re-expressed Lagrangian is simply%
\begin{equation}
\int\delta_{\lambda_{1}\cdots\lambda_{m}}^{\nu_{1}\cdots\nu_{m}}\times
\Phi~\Phi_{\nu_{1}\lambda_{1}}\cdots\Phi_{\nu_{m}\lambda_{m}}~d^{n}X=\int%
\Phi~\mathcal{E}_{m}\left(  \nabla\nabla\Phi\right)  ~d^{n}X=-\frac{1}%
{2}\left(  m+1\right)  \int\left(  \Phi_{\lambda}\Phi_{\lambda}\right)
\mathcal{E}_{m-1}\left(  \nabla\nabla\Phi\right)  ~d^{n}X\ .
\end{equation}
OK then, we have%
\begin{align}
&  \mathcal{A}_{k+1}\left[  \phi\right] \nonumber\\
&  =\int\phi_{\mu}\left(  x\right)  \phi_{\mu}\left(  x\right)  ~\mathcal{E}%
_{k}\left(  \partial\partial\phi\right)  ~d^{n}x=\frac{\left(  k+1\right)
!}{\left(  n-k\right)  !}~2\left(  -\frac{1}{2}\right)  \left(
n-k-1+1\right)  \int\left(  \Phi_{\lambda}\Phi_{\lambda}\right)
\mathcal{E}_{n-k-1-1}\left(  \nabla\nabla\Phi\right)  ~d^{n}X\nonumber\\
&  =-\frac{\left(  k+1\right)  !}{\left(  n-k-1\right)  !}\int\left(
\Phi_{\lambda}\Phi_{\lambda}\right)  \mathcal{E}_{n-k-1-1}\left(  \nabla
\nabla\Phi\right)  ~d^{n}X\nonumber\\
&  =-\frac{\left(  k+1\right)  !}{\left(  n-k-1\right)  !}~\mathcal{A}%
_{n-k-1}\left[  \Phi\right]  \ .
\end{align}
If we shift the index, this may be written more symmetrically. \ 

Thus we have established that the Legendre transform (\ref{LegendreTransf})
gives directly a relation between the actions for the two theories:%
\begin{equation}
\frac{1}{k!}~\mathcal{A}_{k}\left[  \phi\right]  =\frac{\left(  -1\right)
}{\left(  n-k\right)  !}~\mathcal{A}_{n-k}\left[  \Phi\right]
\ \ \ \text{\text{Euclidean},} \label{EuclideanRelatedActions}%
\end{equation}
provided boundary terms from integrating by parts may be discarded, where%
\begin{align}
\mathcal{A}_{k}\left[  \phi\right]   &  =\int\phi_{\mu}\left(  x\right)
\phi_{\mu}\left(  x\right)  ~\mathcal{E}_{k-1}\left(  \partial\partial
\phi\right)  ~d^{n}x=\frac{-2}{k+1}\int\phi\left(  x\right)  ~\mathcal{E}%
_{k}\left(  \partial\partial\phi\right)  ~d^{n}x\ ,\\
\mathcal{A}_{n-k}\left[  \Phi\right]   &  =\int\Phi_{\mu}\left(  X\right)
\Phi_{\mu}\left(  X\right)  ~\mathcal{E}_{n-k-1}\left(  \nabla\nabla
\Phi\right)  ~d^{n}X=\frac{-2}{n-k+1}\int\Phi\left(  X\right)  ~\mathcal{E}%
_{n-k}\left(  \nabla\nabla\Phi\right)  ~d^{n}X\ .
\end{align}

But alas, the sign in (\ref{EuclideanRelatedActions}) disagrees with the
result from the explicit calculation of the $n=2$, $k=1$ case with Lorentz
signature! \ Namely,%
\begin{equation}
\int\partial_{\alpha}\phi~\partial^{\alpha}\phi~d^{2}x=\int\nabla_{\alpha}%
\Phi~\nabla^{\alpha}\Phi~d^{2}X\ .
\end{equation}
However, this discrepancy is due to the difference between the Euclidean and
Lorentzian space identities for the product of two Levi-Civita symbols,
namely, in $n$ dimensions with Lorentz metric sign conventions $\left(
+,-,-,-,\cdots\right)  $,%
\begin{align}
\varepsilon_{\alpha_{1}\cdots\alpha_{n}}\varepsilon^{\beta_{1}\cdots\beta
_{n}}  &  =\delta_{\alpha_{1}\cdots\alpha_{n}}^{\beta_{1}\cdots\beta_{n}%
}\text{ \ \ Euclidean,}\\
\varepsilon_{\alpha_{1}\cdots\alpha_{n}}\varepsilon^{\beta_{1}\cdots\beta
_{n}}  &  =\left(  -1\right)  ^{n-1}\delta_{\alpha_{1}\cdots\alpha_{n}}%
^{\beta_{1}\cdots\beta_{n}}\text{ \ \ Lorentzian.}%
\end{align}
In Lorentz space then, the appropriate identity in $n$ dimensions is%
\begin{equation}
\frac{1}{k!}~\mathcal{A}_{k}\left[  \phi\right]  =\frac{\left(  -1\right)
^{n}}{\left(  n-k\right)  !}~\mathcal{A}_{n-k}\left[  \Phi\right]
\ \ \ \text{\text{Lorentzian}.} \label{LorentzianRelatedActions}%
\end{equation}

\subsection{Hidden symmetry}

Hinterbichler and Joyce (HJ) \cite{HJ}\ have pointed out that Legendre
self-(anti)dual combinations in four dimensions%
\begin{equation}
\mathcal{A}_{\pm}\mathcal{\equiv A}_{1}\left[  \phi\right]  \pm\mathcal{A}%
_{3}\left[  \phi\right]
\end{equation}
realize (nonlinearly) a surprising amount of symmetry, namely, the
\emph{semi}direct sum of the Heisenberg and special linear algebras: $h\left(
4\right)  \oplus_{s}sl\left(  4\right)  $, although HJ do not identify the
algebra by these standard names. \ Moreover, HJ show that additional
symmetries are also present if particular linear combinations of Legendre dual
galileon Lagrangians are considered in $D$ spacetime dimensions. \ 

The most succinct \emph{verbal} description is just to say the HJ galileon
symmetry algebra is\ isomorphic to a \emph{semi}direct sum of the Heisenberg
algebra $h\left(  D\right)  $ and $sl\left(  D\right)  $. \ Thus,%
\begin{equation}
h\left(  D\right)  \oplus_{s}sl\left(  D\right)
\end{equation}
Recall that $h\left(  D\right)  $ is realized by $\left\{  x_{a}%
,p_{b},C\right\}  $ where $C$ is the central charge appearing in $\left[
x_{a},p_{b}\right]  =\delta_{ab}C$, and $sl\left(  D\right)  $ is realized by
$\left\{  x_{a}p_{b}-\frac{1}{D}x_{c}p_{c}\delta_{ab}\right\}  $.\newpage

\subsection{Summary of results for Euclidean metrics}

\noindent Local relations:\ \ 

\begin{center}
\fbox{$%
\begin{array}
[c]{ccc}%
\mathcal{L}_{k}=\phi_{\alpha}\phi_{\alpha}~\mathcal{E}_{k-1} &  &
\mathcal{E}_{k-1}=\delta_{\beta_{1}\beta_{2}\cdots\beta_{k-1}}^{\alpha
_{1}\alpha_{2}\cdots\alpha_{k-1}}\times\phi_{\alpha_{1}\beta_{1}}\phi
_{\alpha_{2}\beta_{2}}\cdots\phi_{\alpha_{k-1}\beta_{k-1}}%
\end{array}
$}\medskip
\end{center}

\noindent$\hspace{-0.25in}\fbox{$%
\begin{array}
[c]{lll}%
\mathcal{E}_{k}=\partial_{\rho}\partial_{\sigma}\left(  \delta_{\sigma
\beta_{2}\cdots\beta_{k}}^{\rho\alpha_{2}\cdots\alpha_{k}}\times\phi
_{\alpha_{2}\beta_{2}}\cdots\phi_{\alpha_{k}\beta_{k}}\times\phi\right)  &  &
\mathcal{B}_{k}=\partial_{\sigma}\left(  \delta_{\rho\beta_{1}\beta_{2}%
\cdots\beta_{k-2}}^{\sigma\alpha_{1}\alpha_{2}\cdots\alpha_{k-2}}\times
\phi_{\alpha_{1}\beta_{1}}\cdots\phi_{\alpha_{k-2}\beta_{k-2}}\times\phi
_{\rho}\phi_{\tau}\phi_{\tau}\right)  \smallskip\\
\mathcal{D}_{k}=\partial_{\rho}\partial_{\sigma}\left(  \delta_{\sigma
\beta_{2}\cdots\beta_{k-1}\beta_{k}}^{\rho\alpha_{2}\cdots\alpha_{k-1}%
\alpha_{k}}\times\phi_{\alpha_{2}\beta_{2}}\cdots\phi_{\alpha_{k}\beta_{k}%
}\times\tfrac{1}{2}\phi^{2}\right)  &  & \mathcal{V}_{k}=\delta_{\beta
_{1}\beta_{2}\cdots\beta_{k-1}\beta_{k}}^{\alpha_{1}\alpha_{2}\cdots
\alpha_{k-1}\alpha_{k}}\times\phi_{\alpha_{1}\beta_{1}}\cdots\phi
_{\alpha_{k-1}\beta_{k-1}}\times\phi_{\alpha_{k}}\phi_{\beta_{k}}%
\end{array}
$}$

\begin{center}
{\small Note that }$\mathcal{E}_{k}${\small \ and }$\mathcal{D}_{k}%
${\small \ are double divergences.}\bigskip

\fbox{$%
\begin{array}
[c]{lll}%
\mathcal{V}_{k}=\frac{1}{2}\left(  1-k\right)  \mathcal{B}_{k}+\frac{1}%
{2}\left(  1+k\right)  \mathcal{L}_{k} &  & \mathcal{V}_{k}=\mathcal{D}%
_{k}-\phi\mathcal{E}_{k}%
\end{array}
$}
\end{center}

\noindent Integrated relations: \ 

\begin{center}
\fbox{$\mathcal{A}_{k}=\int\mathcal{L}_{k}~d^{n}x$}\medskip

$\fbox{$%
\begin{array}
[c]{ll}%
\mathcal{A}_{k}=\frac{2}{k+1}\int\mathcal{V}_{k}~d^{n}x & \text{{\small upon
setting} }\int\mathcal{B}_{k}~d^{n}x=0\smallskip\\
\int\mathcal{V}_{k}~d^{n}x=-\int\phi~\mathcal{E}_{k}~d^{n}x &
\text{{\small upon setting} }\int\mathcal{D}_{k}~d^{n}x=0\smallskip\\
\mathcal{A}_{k}=\frac{-2}{k+1}\int\phi~\mathcal{E}_{k}~d^{n}x &
\text{{\small upon discarding both of the above boundary terms}}%
\end{array}
$}$\medskip
\end{center}

\noindent Constraint relation:

\begin{center}
\fbox{$\left.  \mathcal{E}_{k}\left(  \partial\partial\phi\right)  \right\vert
_{\substack{n+1\text{ dimensions} \\\text{with }\phi_{\alpha}\phi_{\alpha}%
=0}}=\left.  \frac{1}{\phi_{\lambda}\phi_{\lambda}}~\mathcal{V}_{k+1}%
\right\vert _{\substack{n\text{ dimensions with} \\\text{varying }%
\phi_{\lambda}\phi_{\lambda}}}$}
\end{center}

\noindent Legendre relations: \ 

\begin{center}
\fbox{$%
\begin{array}
[c]{c}%
\phi\left(  x\right)  +\Phi\left(  X\right)  =\sum_{\alpha=1}^{n}x_{\alpha
}X_{\alpha}\smallskip\\
X_{\alpha}\left(  x\right)  =\frac{\partial\phi\left(  x\right)  }{\partial
x_{\alpha}}\equiv\partial_{\alpha}\phi\ ,\ \ \ x_{\alpha}\left(  X\right)
=\frac{\partial\Phi\left(  X\right)  }{\partial X_{\alpha}}\equiv
\nabla_{\alpha}\Phi\smallskip\\
\frac{1}{\sqrt{\det\left(  \partial\partial\phi\right)  }}~\frac{1}%
{k!}~\mathcal{E}_{k}\left(  \partial\partial\phi\right)  =\frac{1}{\sqrt
{\det\left(  \nabla\nabla\Phi\right)  }}~\frac{1}{\left(  n-k\right)
!}~\mathcal{E}_{n-k}\left(  \nabla\nabla\Phi\right)  \smallskip\\
\frac{1}{k!}\mathcal{A}_{k}\left[  \phi\right]  =-\frac{1}{\left(  n-k\right)
!}~\mathcal{A}_{n-k}\left[  \Phi\right]  \smallskip
\end{array}
$}\medskip
\end{center}

\noindent Trace relations: \ 

\begin{center}
\fbox{$%
\begin{array}
[c]{ccc}%
\mathcal{T}_{k}=\mathrm{Tr}\left[  \left(  \partial\partial\phi\right)
^{k}\right]  &  & \mathcal{S}_{k}=\mathrm{Tr}\left[  \left(  \partial
\phi\partial\phi\right)  \left(  \partial\partial\phi\right)  ^{k}\right]
\end{array}
$}\medskip

\fbox{$%
\begin{array}
[c]{c}%
\mathcal{E}_{k}=\det\left(
\begin{array}
[c]{ccccccc}%
\mathcal{T}_{1} & k-1 & 0 & \cdots & 0 & 0 & 0\\
\mathcal{T}_{2} & \mathcal{T}_{1} & k-2 & \cdots & 0 & 0 & 0\\
\mathcal{T}_{3} & \mathcal{T}_{2} & \mathcal{T}_{1} & \cdots & 0 & 0 & 0\\
\vdots & \vdots & \vdots & \ddots & \vdots & \vdots & \vdots\\
\mathcal{T}_{k-2} & \mathcal{T}_{k-3} & \mathcal{T}_{k-4} & \cdots &
\mathcal{T}_{1} & 2 & 0\\
\mathcal{T}_{k-1} & \mathcal{T}_{k-2} & \mathcal{T}_{k-3} & \cdots &
\mathcal{T}_{2} & \mathcal{T}_{1} & 1\\
\mathcal{T}_{k} & \mathcal{T}_{k-1} & \mathcal{T}_{k-2} & \cdots &
\mathcal{T}_{3} & \mathcal{T}_{2} & \mathcal{T}_{1}%
\end{array}
\right)
\end{array}
$}\medskip

\fbox{$%
\begin{array}
[c]{c}%
\mathcal{V}_{k}=\det\left(
\begin{array}
[c]{ccccccc}%
\mathcal{S}_{0} & k-1 & 0 & \cdots & 0 & 0 & 0\\
\mathcal{S}_{1} & \mathcal{T}_{1} & k-2 & \cdots & 0 & 0 & 0\\
\mathcal{S}_{2} & \mathcal{T}_{2} & \mathcal{T}_{1} & \cdots & 0 & 0 & 0\\
\vdots & \vdots & \vdots & \ddots & \vdots & \vdots & \vdots\\
\mathcal{S}_{k-3} & \mathcal{T}_{k-3} & \mathcal{T}_{k-4} & \cdots &
\mathcal{T}_{1} & 2 & 0\\
\mathcal{S}_{k-2} & \mathcal{T}_{k-2} & \mathcal{T}_{k-3} & \cdots &
\mathcal{T}_{2} & \mathcal{T}_{1} & 1\\
\mathcal{S}_{k-1} & \mathcal{T}_{k-1} & \mathcal{T}_{k-2} & \cdots &
\mathcal{T}_{3} & \mathcal{T}_{2} & \mathcal{T}_{1}%
\end{array}
\right)
\end{array}
$}\medskip
\end{center}

\noindent Lorentz metric results depend on sign conventions, i.e.
$(+,-,-,\cdots)$ or $(-,+,+,\cdots)$, and are left to the reader to
determine.\newpage

\section{Classical solutions}

\textit{In this section we consider classical solutions of the field equations
stemming from individual }$A_{k}$\textit{. \ We present and illustrate a
variety of methods to find solutions of }$\mathcal{E}_{k}\left(
\partial\partial\phi\right)  =0$\textit{.}

As a warm-up, consider the first example beyond the standard free massless
field, namely, $\mathcal{E}_{2}\left(  \partial\partial\phi\right)  =0$. \ For
an extremely simple case, a \textquotedblleft spherically
symmetric\textquotedblright\ solution is valid almost everywhere in $n$
dimensions:%
\begin{equation}
\phi\left(  x\right)  =\left\{
\begin{array}
[c]{c}%
\left(  x_{\alpha}x_{\alpha}\right)  ^{1-\frac{n}{4}}\text{ if }n\neq4\\
\ln\left(  x_{\alpha}x_{\alpha}\right)  \text{ if }n=4
\end{array}
\right.  \ ,
\end{equation}
where $\alpha=1,\cdots,n$ is summed. \ However, in $n$-dimensional Minkowski
space (except for $n=8,\ 12,\ \cdots$) this solution obviously has an
imaginary part outside the light-cone. \ So, in general, such functions with
branch points give \emph{real} solutions only on subspaces. \ We may think of
the branch points as defining boundaries for the subspaces, with some
particular boundary conditions. \ 

At the opposite extreme, another solution is%
\begin{equation}
\phi\left(  x\right)  =\sqrt{x_{1}x_{2}\cdots x_{n}}\ .
\end{equation}
It is straightforward to check the equation of motion $\mathcal{E}_{2}=0$ is
satisfied. \ This solution has branch points for both Euclidean and Minkowski space.

Next consider $\mathcal{E}_{k}=0$ for general $k$. \ Although $\mathcal{E}%
_{k>1}$ is nonlinear in $\phi$, it is nevertheless still true that some plane
waves are exact solutions. \ For \textquotedblleft light-ray\textquotedblright%
\ plane waves,
\begin{equation}
\mathcal{E}_{k}\left[  A\exp\left(  iq_{\alpha}x_{\alpha}\right)  \right]  =0
\end{equation}
for constant $A$ and $q_{\alpha}$, if $q_{\alpha}q_{\alpha}=0$ with $A$
arbitrary. \ In this case, each of the terms in\ $\mathcal{E}_{k}$ vanish
separately. \ In fact, light-ray plane waves are only one among many possible
solutions for which\ both $\phi_{\alpha\alpha}=0$ and $\phi_{\beta}\phi
_{\beta}=0$. \ That is to say, the general galileon equation $\mathcal{E}%
_{k}=0$ possesses a class of solutions given by the simultaneous solution of%
\begin{equation}
\frac{\partial^{2}\phi}{\partial x^{\alpha}\partial x^{\alpha}}=0\text{
\ \ and \ \ }\left(  \frac{\partial\phi}{\partial x^{\alpha}}\right)  \left(
\frac{\partial\phi}{\partial x^{\alpha}}\right)  =0\ . \label{TwoConditions}%
\end{equation}
The proof of this statement is elementary for $\mathcal{E}_{2}=0$, while for
higher $k$, the hierarchical construction in (\ref{kLevelSystem}) and the
nature of the variation procedure guarantees that $\mathcal{E}_{k}=0$ will
hold if both $\phi_{\alpha}\phi_{\alpha}=0$ and $\mathcal{E}_{k-1}=0$, thereby
establishing the general result.\footnote{It is interesting that this class of
solutions is not given by the method of Legendre transformations, described
above in \S 3.6\ and to be discussed further below. \ The Legendre method
fails for this class because the second equation in (\ref{TwoConditions})
implies the existence of a functional relation among the Legendre transformed
variables so that $\det\left(  \partial X_{\alpha}/\partial x_{\beta}\right)
=0$.}

In \emph{three} dimensions it is only necessary to take the single constraint
$\phi_{\alpha}\phi_{\alpha}=0$ since a consequence of taking additional
derivatives of this one constraint is $\phi_{\alpha\alpha}=0$. \ This is not
true in higher dimensions, but it does suggest another method for $n=3$.

\subsection{Implicit solutions}

In three dimensions, choose four arbitrary functions
$f(u,v),\ g(u,v),\ h(u,v),\ k(u,v)$ constrained by the three relations
\begin{align}
&  xf(u,v)+yg(u,v)+zh(u,v)+k(u,v)=\phi(x,y,z),\\
&  x{f(u,v)_{u}}+y{g_{u}(u,v)}+zh_{u}(u,v)+k_{u}(u,v)=0,\\
&  x{f_{v}(u,v)}+y{g_{v}(u,v)}+zh_{v}(u,v)+k_{v}(u,v)=0.
\end{align}
Here subscripts denote partial differentiation with respect to $u,\ v$. \ Then
the implicit solution of these equations for $\phi(x,y,z)$ is a solution to
the Monge-Ampere equation $\det\left(  \phi_{\mu\nu}\right)  =0$ in 3
dimensions. \ Here
\begin{equation}
\phi_{x}=f(u,v),\ \phi_{y}=g(u,v),\ \phi_{z}=g(u,v),
\end{equation}
so the solution implies that there exists a functional relationship amongst
$(\phi_{x},\ \phi_{y},\ \phi_{z})$. \ This remark is enough to guarantee that
the elimination of $(u,\ v)$ will give a solution to Monge Ampere. \ To solve
the galileon equation some further constraints are necessary. \ Suppose
\begin{equation}
\phi_{z}=Q(\phi_{x},\phi_{y})=Q(\alpha,\beta)\
\end{equation}
defining $\alpha=\phi_{x},\ \beta=\phi_{y}$. \ Then
\begin{subequations}
\begin{align}
\phi_{xz}  &  =Q_{\alpha}\phi_{xx}+Q_{\beta}\phi_{xy}\ ,\\
\phi_{xz}  &  =Q_{\alpha}\phi_{xy}+Q_{\beta}\phi_{yy}\ ,\\
\phi_{zz}  &  =Q_{\alpha}\phi_{xz}+Q_{\beta}\phi_{yz}\ ,\\
&  =(Q_{\alpha})^{2}\phi_{xx}+2Q_{\alpha}Q_{\beta}\phi_{xy}+(Q_{\beta})^{2}\ .
\end{align}
Combining these relations we obtain
\end{subequations}
\begin{equation}
\left(  \phi_{xx}\phi_{yy}-(\phi_{xy})^{2}\right)  \left(  1+Q_{\alpha}%
^{2}+Q_{\beta}^{2}\right)  \ ,
\end{equation}
so either the constraint is
\begin{equation}
1+Q_{\alpha}^{2}+Q_{\beta}^{2}=0 \label{quod}%
\end{equation}
or
\begin{equation}
\phi_{xx}\phi_{yy}-(\phi_{xy})^{2}=0\ . \label{quad2}%
\end{equation}
In the case (\ref{quod}) this places a constraint upon the functional
dependence of the derivatives. \ For example, if ${Q=i\sqrt{\phi_{x}^{2}%
+\phi_{y}^{2}}}$ then the constraint is satisfied automatically. \ Then
$\phi_{x}^{2}+\phi_{y}^{2}+\phi_{z}^{2}=0$, a constraint encountered earlier.
\ In turn, this constrains the functions $\left(
f(u,v),\ g(u,v),\ h(u,v)\right)  $ to satisfy $f^{2}+g^{2}+h^{2}=0.$ \ Another
possibility is $Q=i\sin(\phi_{x})+i\cos(\phi_{y})$.

In the other case where (\ref{quad2}) holds it is not sufficient to impose
just this to solve the galileon equations, but we also require that $\phi
_{z}=Q(\phi_{x},\phi_{y})$, or else demand that all leading subdeterminants
vanish, i.e.
\begin{equation}
\phi_{xx}\phi_{yy}-(\phi_{xy})^{2}=0,\ \ \phi_{yy}\phi_{zz}-(\phi_{yz}%
)^{2}=0,\ \ \phi_{zz}\phi_{xx}-(\phi_{zx})^{2}=0\ .
\end{equation}
A solution of this type is given by $\phi(x,y,z)=\sqrt{xyz}$.

\subsection{Envelope method}

The \emph{envelope method} also gives solutions. \ To be explicit, in 3D, the
method is to take%
\begin{align}
xf\left(  u\right)  +yg\left(  u\right)  +zh\left(  u\right)   &  =\phi\left(
x,y,z\right)  \ ,\\
xf^{\prime}\left(  u\right)  +yg^{\prime}\left(  u\right)  +zh^{\prime}\left(
u\right)   &  =0\ ,
\end{align}
and then choose various functions $f$,\ $g$, and $h$ to determine $u\left(
x,y,z\right)  $. \ For example, inserting $f\left(  u\right)  =u$, $g\left(
u\right)  =\frac{1}{2}u^{2}$, and $h\left(  u\right)  =\frac{1}{3}u^{3}$ into
the second of these two equations gives \ $x+yu+zu^{2}=0$, whose solutions are%
\begin{equation}
u\left(  x,y,z\right)  =\frac{1}{2z}\left(  -y\pm\sqrt{y^{2}-4xz}\right)  \ .
\end{equation}
Therefore the corresponding solutions for $\phi$ are%
\begin{equation}
\phi\left(  x,y,z\right)  =-\frac{1}{12z^{2}}\left(  -y^{3}+6xyz\pm\left(
y^{2}-4xz\right)  \sqrt{y^{2}-4xz}\right)  \ .
\end{equation}
Again, these functions have a branch point, as well as a pole, so they are
not\emph{\ real} solutions \emph{unless} $y^{2}\geq4xz$ and $z\neq0$. \ For
these examples, it is again straightforward to check the equation of motion is satisfied.

This procedure can be extended if all three derivatives are functionally
related, and the auxiliary functions depend only upon one function:
\begin{align}
&  xf(u)+yg(u)+zh(u)+\omega(u)=\phi(x,y,z)\ ,\\
&  x{f_{u}(u)}+y{g_{u}(u)}+z{h_{u}(u)}+\omega_{u}(u)=0.
\end{align}
The first of these equations possesses a tantalizing similarity to the
Legendre transform. \ But here the procedure is to solve the second equation
for $u$ in terms of $x,\ y,\ z$ which results in a solution of $\mathcal{E}%
_{2}=0$ when inserted into the first equation. \ However, having obtained this
solution for $u$, a huge class of solutions may be obtained by replacing
$f(u),\ g(u),\ h(u),\ $and $\omega(u)$ by
\begin{align}
F(u)  &  =q(u)f(u)-\int q^{\prime}(u)f(u)du\ ,\\
G(u)  &  =q(u)g(u)-\int q^{\prime}(u)g(u)du\ ,\\
H(u)  &  =q(u)h(u)-\int q^{\prime}(u)h(u)du\ ,\\
\Omega(u)  &  =q(u)\omega(u)-\int q^{\prime}(u)\omega(u)du\ .
\end{align}
The equation to determine $u$ is still the same as before, so a new solution
is generated. \ This may be checked on the specific example, taking $q(u)=u$.

\subsubsection*{Another example}

Take $f(u)=u,\ g(u)=u^{2},\ h(u)=u^{3}$. \ Solving the second of these
equations for $u$,
\begin{equation}
u=1/3\,{\frac{-y+\sqrt{{y}^{2}-3\,zx}}{z}\ ,}%
\end{equation}
and
\begin{equation}
\phi=1/27\,{\frac{\left(  -y+\sqrt{{y}^{2}-3\,zx}\right)  \left(
6\,zx-{\ y}^{2}+y\sqrt{{y}^{2}-3\,zx}\right)  }{{z}^{2}}\ .}%
\end{equation}
This is a solution to the $n=3$ case. \ Note this is of weight one, so it is
also a solution of the Monge-Ampe\'{r}e equation.

\subsection{ Power law solutions in other dimensions and Legendre
equivalences}

In general, the equation $\mathcal{E}_{k}\left(  \partial\partial\phi\right)
=0$ is homogeneous in $\phi$, of degree $k$, and therefore the overall
normalization of any solution is not determined. \ Consider again spherically
symmetric solutions as given by a power ansatz:%
\begin{equation}
\phi\left(  x\right)  =\left(  x_{\alpha}x_{\alpha}\right)  ^{p}\ .
\label{SphericalAnsatz}%
\end{equation}
In $n$ dimensions, for this ansatz, the products and traces of $\partial
\partial\phi$ are:%
\begin{align}
\left(  \partial\partial\phi\right)  _{\mu\nu}^{k}  &  =\left(  2p\right)
^{k}\left(  x_{\alpha}x_{\alpha}\delta_{\mu\nu}+\left(  \left(  2p-1\right)
^{k}-1\right)  x_{\mu}x_{\nu}\right)  \left(  x_{\alpha}x_{\alpha}\right)
^{kp-k-1}\ ,\nonumber\\
\left(  \partial\partial\phi\right)  _{\mu\mu}^{k}  &  =\left(  2p\right)
^{k}\left(  n-1+\left(  2p-1\right)  ^{k}\right)  \left(  x_{\alpha}x_{\alpha
}\right)  ^{kp-k}\ .
\end{align}
Inserting these traces into (\ref{DetEofMT}) and evaluating (\ref{DetEofM}) we
find, for example,%
\begin{align}
\mathcal{E}_{1}  &  =2p\left(  n-2+2p\right)  \left(  x_{\alpha}x_{\alpha
}\right)  ^{p-1}=0\Longrightarrow p=1-n/2\ ,\\
\mathcal{E}_{2}  &  =4p^{2}\left(  n-1\right)  \left(  n+4p-4\right)  \left(
x_{\alpha}x_{\alpha}\right)  ^{2p-2}=0\Longrightarrow p=1-n/4\ ,\\
\mathcal{E}_{3}  &  =8p^{3}\left(  n-1\right)  \left(  n-2\right)  \left(
n+6p-6\right)  \left(  x_{\alpha}x_{\alpha}\right)  ^{3p-3}=0\Longrightarrow
p=1-n/6\ ,
\end{align}
Thus the power $p$ required for a solution is determined, as indicated. \ 

For other levels in the hierarchy,%
\begin{equation}
\left.  \mathcal{E}_{k}\left(  \partial\partial\phi\right)  \right\vert
_{\phi\left(  x\right)  =\left(  x_{\alpha}x_{\alpha}\right)  ^{p}}=\left(
x_{\alpha}x_{\alpha}\right)  ^{kp-k}\left(  n+2kp-2k\right)  \left(
2p\right)  ^{k}~\frac{\left(  n+1-k\right)  !}{\left(  n-2\right)  !}\ .
\end{equation}
The condition on $p$ for the ansatz (\ref{SphericalAnsatz}) to be a solution
of $\mathcal{E}_{k}=0$\ is therefore%
\begin{equation}
p=1-\frac{n}{2k}\ . \label{PowerCondition}%
\end{equation}
By taking a limit, a nontrivial solution for $n=2k$ is easily found to be
$\ln\left(  x_{a}x_{\alpha}\right)  $. \ 

So, in $n$ dimensions the $k$th equation of motion of the hierarchy is solved
by
\begin{equation}
\phi\left(  x\right)  =\left\{
\begin{array}
[c]{c}%
\left(  x_{\alpha}x_{\alpha}\right)  ^{1-\frac{n}{2k}}\text{ if }n\neq2k\\
\\
\ln\left(  x_{\alpha}x_{\alpha}\right)  \text{ if }n=2k
\end{array}
\right.  \ . \label{SphericalPowerSolutions}%
\end{equation}
Moreover, under the Legendre tranformation, the ansatz solution for level $k$
is mapped into the ansatz solution for level $n-k$. \ 

In particular, when $n=k+1$ the ansatz is mapped into an harmonic function by
the Legendre transformation.\footnote{Two such examples for solutions of
$\mathcal{E}_{2}\left(  \partial\partial\phi\right)  =0$ and $\mathcal{E}%
_{1}\left(  \nabla\nabla\Phi\right)  =0$ in three dimensions are
\begin{equation}
\Phi=XYZ\ ,\ \ \ \phi=\sqrt{xyz}\ ,
\end{equation}
and
\begin{equation}
\Phi=\frac{1}{\sqrt{X^{2}+Y^{2}+Z^{2}}}\ ,\ \ \ \phi=\left(  x^{2}+y^{2}%
+z^{2}\right)  ^{1/4}\ .
\end{equation}
\ For another, take $\Phi=Z\left(  Z^{2}+Y^{2}-4X^{2}\right)  $, etc. \ We
leave it as an exercise for the reader to find the corresponding $\phi$.} \ In
$n$ dimensions the spherically symmetric harmonic solution is given by
$0=\nabla_{\beta}\nabla_{\beta}\left(  X_{\alpha}X_{\alpha}\right)
^{\frac{2-n}{2}}$, so this last statement is equivalent to%
\begin{equation}
\left(  x_{\alpha}x_{\alpha}\right)  ^{\frac{n-2}{2\left(  n-1\right)  }%
}=\left(  x_{\alpha}x_{\alpha}\right)  ^{\frac{k-1}{2k}}%
\underset{\text{Legendre for }n=k+1}{\sim}\left(  X_{\alpha}X_{\alpha}\right)
^{\frac{1-k}{2}}=\left(  X_{\alpha}X_{\alpha}\right)  ^{\frac{2-n}{2}}%
\end{equation}
Note the effect of the transformation is just to replace $k\longmapsto\frac
{1}{k}$ in the exponent, thereby changing the scaling properties of the
solution:%
\begin{align}
\left(  X_{\beta}\nabla_{\beta}-1\right)  \left(  X_{\alpha}X_{\alpha}\right)
^{\frac{2-n}{2}}  &  =\left(  1-n\right)  \left(  X_{\alpha}X_{\alpha}\right)
^{\frac{2-n}{2}}\ ,\\
\left(  x_{\beta}\partial_{\beta}-1\right)  \left(  x_{\alpha}x_{\alpha
}\right)  ^{\frac{n-2}{2\left(  n-1\right)  }}  &  =\left(  \frac{1}%
{1-n}\right)  \left(  x_{\alpha}x_{\alpha}\right)  ^{\frac{n-2}{2\left(
n-1\right)  }}\ .
\end{align}

Let's go through the details for the $n=k+1$ case. \ Under the Legendre
transformation:
\begin{subequations}
\begin{align}
x_{\beta}X_{\beta}  &  =\phi\left(  x\right)  +\Phi\left(  X\right)  \ ,\\
X_{\beta}  &  =\frac{\partial\phi\left(  x\right)  }{\partial x_{\beta}}%
=\frac{\partial}{\partial x_{\beta}}\left(  x_{\alpha}x_{\alpha}\right)
^{\frac{k-1}{2k}}=\tfrac{k-1}{k}x_{\beta}\left(  x_{\alpha}x_{\alpha}\right)
^{\frac{-k-1}{2k}}\ ,\\
x_{\beta}X_{\beta}  &  =\tfrac{k-1}{k}\left(  x_{\alpha}x_{\alpha}\right)
^{\frac{k-1}{2k}}=\tfrac{k-1}{k}~\phi\left(  x\right)  \text{\ },\\
\Phi\left(  X\right)   &  =-\tfrac{1}{k}~\phi\left(  x\right)  \ ,\\
X_{\beta}X_{\beta}  &  =\left(  \tfrac{k-1}{k}\right)  ^{2}\left(  x_{\alpha
}x_{\alpha}\right)  ^{\frac{-1}{k}}\ ,\ \ \ x_{\alpha}x_{\alpha}=\left(
\tfrac{k-1}{k}\right)  ^{2k}\left(  X_{\beta}X_{\beta}\right)  ^{-k}\ ,\\
\left(  x_{\alpha}x_{\alpha}\right)  ^{\frac{k-1}{2k}}  &  =\left(
\tfrac{k-1}{k}\right)  ^{k-1}\left(  X_{\beta}X_{\beta}\right)  ^{\frac
{1-k}{2}}\ .
\end{align}
So then, an harmonic function of $X$\ is indeed the result of transforming
$\left(  x_{\alpha}x_{\alpha}\right)  ^{\frac{k-1}{2k}}$, for dimension
$n=k+1$. \ Including a convenient normalization,
\end{subequations}
\begin{equation}
\phi\left(  x\right)  =k^{k}\left(  x_{\alpha}x_{\alpha}\right)  ^{\frac
{k-1}{2k}}\underset{\text{Legendre for }n=k+1}{\longmapsto}\Phi\left(
X\right)  =-\left(  k-1\right)  ^{k-1}\left(  X_{\beta}X_{\beta}\right)
^{\frac{1-k}{2}}\ .
\end{equation}
Note that this procedure could be reversed, starting from the harmonic
solution. \ Thus there is a local one-to-one map between harmonic functions
$\Phi\left(  X\right)  $ and solutions of the nonlinear equation
$\mathcal{E}_{n-1}\left(  \partial\partial\phi\right)  =0$, in $n$ dimensions,
valid so long as $0<\left\vert \det\left(  \nabla\nabla\Phi\right)
\right\vert <\infty$.

\subsection{Self-Dual Solutions}

Another approach to the solution provides a class of self dual solutions; i.e
solutions both to the original equation, and to the Legendre transformed
equation (in the same variables). \ Suppose we impose the ansatz
\begin{equation}
x_{\mu}V_{\mu}(\phi)=1.
\end{equation}
(Here the subscript $\mu$ is a vector index, not a derivative.) \ Then
\begin{align}
\frac{\partial\phi}{\partial x_{\mu}}  &  =-\frac{V_{\mu}}{\sum{x_{\mu}V_{\mu
}^{\prime}}}\ ,\\
\frac{\partial^{2}\phi}{\partial x_{\mu}\partial x_{\nu}}  &  =-\frac{(V_{\mu
}V_{\nu}^{\prime}+V_{\nu}V_{\mu}^{\prime})}{(\sum{x_{\alpha}V_{\alpha}%
^{\prime}})^{2})}+\frac{V_{\mu}V_{\nu}(\sum{x_{\beta}V_{\beta}^{\prime\prime
})}}{(\sum{x_{\alpha}V_{\alpha}^{\prime}})^{3})}\ .
\end{align}
Inserting this expression into $\mathcal{E}_{2}$ we obtain
\begin{equation}
\frac{1}{(\sum{x_{\alpha}V_{\alpha}^{\prime})^{4}}}\left(  (\sum{V_{\mu}%
V_{\mu}})(\sum{V_{\nu}^{\prime}V_{\nu}^{\prime}})-(\sum{V_{\mu}V_{\mu}%
^{\prime}})^{2}\right)  \ .
\end{equation}
By construction solutions of this type will inevitably be complex in Euclidean
space, but may be real in Minkowski space. \ This is clearly zero if the
constraint $\sum{V_{\mu}V_{\mu}}=0$ is imposed. \ Moreover, it is easy to see
that this solution also solves $\nabla^{2}\phi=0$, the Legendre transform of
$\mathcal{E}_{2}$ in 3 dimensions. \ Another remarkable property \cite{F2} of
this solution is that if you replace $\phi$ by any function of $\phi$, it
remains a solution of these equations! Indeed, this class of solutions is
universal, as it solves \textit{all} equations $\mathcal{E}_{k}=0$ in the
appropriate dimension, as every term contains at least one factor of the form
$\sum{V_{\mu}V_{\mu}}$ or $\sum{V_{\mu}V_{\mu}^{\prime}}$, which both vanish
by the constraint. \ Furthermore this extends to covariant equations which
also include first, as well as second derivatives; thus the Lagrangians
themselves vanish on this class of solutions.

\subsubsection*{Example}

As a simple example consider the equation
\begin{equation}
x\phi+iy\sqrt{1+{\phi}^{2}}+z=1
\end{equation}
whose coefficients satisfy the constraint. \ Solving for $\phi$ we obtain
\begin{equation}
\phi={\frac{\pm(x-zx)+\sqrt{-{y}^{4}-{y}^{2}{z}^{2}+2\,{y}^{2}z-{y}^{2}%
-{x}^{2}{y}^{2}}}{{y}^{2}+{x}^{2}}\ .}%
\end{equation}
These solutions may be verified to satisfy $\mathcal{E}_{1}=0,\ \mathcal{E}%
_{2}=0,\ $and $\mathcal{E}_{3}=0.$

\section{Mixtures}

\textit{In this section we consider solutions of the field equations for
systems governed by linear combinations of the }$A_{k}$\textit{\ for different
}$k$\textit{.}

\subsection{Symmetric spacetime solutions}

Consider the modified field equation that follows from $\mathcal{A}_{1}%
-\kappa\mathcal{A}_{2}$:
\begin{equation}
\phi_{\lambda\lambda}=\kappa\left(  \phi_{\nu\nu}\phi_{\lambda\lambda}%
-\phi_{\lambda\nu}\phi_{\lambda\nu}\right)  \ . \label{MixedFieldEqn}%
\end{equation}
Make a spherically symmetric spacetime ansatz:%
\begin{equation}
\phi=f\left(  \sigma\right)  \ ,\ \ \ \sigma\equiv x_{\lambda}x_{\lambda}\ .
\end{equation}
Then
\begin{equation}
\phi_{\lambda\lambda}=\partial_{\lambda}\left(  2x_{\lambda}f^{\prime}\right)
=2nf^{\prime}+4x_{\lambda}x_{\lambda}f^{\prime\prime}\ ,\ \ \ \phi_{\nu
\lambda}=\partial_{\nu}\left(  2x_{\lambda}f^{\prime}\right)  =2\delta
_{\nu\lambda}f^{\prime}+4x_{\nu}x_{\lambda}f^{\prime\prime}\ ,
\end{equation}%
\begin{align}
\phi_{\lambda\lambda}\phi_{\nu\nu}-\phi_{\nu\lambda}\phi_{\nu\lambda}  &
=\left(  2nf^{\prime}+4x_{\lambda}x_{\lambda}f^{\prime\prime}\right)
^{2}-\left(  2\delta_{\nu\lambda}f^{\prime}+4x_{\nu}x_{\lambda}f^{\prime
\prime}\right)  \left(  2\delta_{\nu\lambda}f^{\prime}+4x_{\nu}x_{\lambda
}f^{\prime\prime}\right) \nonumber\\
&  =4n\left(  n-1\right)  \left(  f^{\prime}\right)  ^{2}+16\left(
n-1\right)  x_{\lambda}x_{\lambda}f^{\prime}f^{\prime\prime}\ ,
\end{align}%
\begin{equation}
\phi_{\lambda\lambda}\phi_{\nu\nu}-\phi_{\nu\lambda}\phi_{\nu\lambda}=4\left(
n-1\right)  \left(  f^{\prime}\right)  \left(  nf^{\prime}+4\sigma
f^{\prime\prime}\right)  \ , \label{UnMod}%
\end{equation}
and the field equation (\ref{MixedFieldEqn}) becomes%
\begin{equation}
2nf^{\prime}+4\sigma f^{\prime\prime}=4\kappa\left(  n-1\right)  \left(
f^{\prime}\right)  \left(  nf^{\prime}+4\sigma f^{\prime\prime}\right)  \ .
\label{Mod}%
\end{equation}
This is again a first order differential equation for $g=f^{\prime}$: \
\begin{equation}
\frac{2}{n}\sigma g^{\prime}=-g+\frac{2\kappa\left(  n-1\right)  }{\left(
4\kappa\left(  n-1\right)  g-1\right)  }g^{2}\ .
\end{equation}
Therefore%
\begin{equation}
\ln\left(  g\left(  g+\frac{1}{2\kappa\left(  1-n\right)  }\right)  \right)
=-\frac{n}{2}\ln\sigma\ ,
\end{equation}
and we have%
\begin{equation}
g\left(  g+\frac{1}{2\kappa\left(  1-n\right)  }\right)  =C\sigma^{-n/2}\text{
\ \ where \ \ }C=g_{1}\left(  g_{1}+\frac{1}{2\kappa\left(  1-n\right)
}\right)  \ ,\ \ \ g_{1}\equiv g\left(  \sigma=1\right)  \ .
\end{equation}
That is to say,%
\begin{equation}
g\left(  \sigma\right)  =\frac{1}{4\kappa\left(  n-1\right)  }\left\{
\begin{array}
[c]{ccc}%
1-\sqrt{1+16\kappa^{2}\left(  n-1\right)  ^{2}C\sigma^{-n/2}} & \text{ \ \ if
\ \ } & 4\kappa\left(  n-1\right)  g_{1}<1\\
1+\sqrt{1+16\kappa^{2}\left(  n-1\right)  ^{2}C\sigma^{-n/2}} & \text{ \ \ if
\ \ } & 4\kappa\left(  n-1\right)  g_{1}>1
\end{array}
\right.  \ .
\end{equation}
One more integration gives $f=\int g$:%
\begin{equation}
f\left(  \sigma\right)  =f\left(  \sigma_{0}\right)  +\int_{\sigma_{0}%
}^{\sigma}g\left(  \rho\right)  d\rho\ .
\end{equation}
For convenience, let us take $\sigma_{0}=1$. \ The result of the integral is
then
\begin{equation}
\int\sqrt{1+16\kappa^{2}\left(  n-1\right)  ^{2}C\sigma^{-n/2}}d\sigma
=\sigma\left.  _{2}F_{1}\right.  \left(  -\frac{1}{2},-\frac{2}{n};1-\frac
{2}{n};-16\kappa^{2}\left(  n-1\right)  ^{2}C\sigma^{-\frac{1}{2}n}\right)
\ ,
\end{equation}
where $\left.  _{2}F_{1}\right.  $\ is the usual Gauss hypergeometric
function,%
\begin{equation}
\left.  _{2}F_{1}\right.  \left(  a,b;c;z\right)  =1+\frac{abz}{c}%
+\frac{a\left(  1+a\right)  b\left(  1+b\right)  z^{2}}{c\left(  1+c\right)
2!}+\frac{a(1+a)(2+a)b(1+b)(2+b)z^{3}}{c(1+c)(2+c)3!}+\cdots\ .
\end{equation}
$\allowbreak$The final result for the spherically symmetric spacetime solution
(recall $\sigma\equiv x_{\lambda}x_{\lambda}$) is then%
\begin{align}
f\left(  \sigma\right)   &  =f\left(  1\right)  +\frac{1}{4\kappa\left(
n-1\right)  }\left(  \sigma-1\right) \nonumber\\
&  +\frac{\pm1}{4\kappa\left(  n-1\right)  }\left(
\begin{array}
[c]{c}%
\left.  _{2}F_{1}\right.  \left(  -\frac{1}{2},-\frac{2}{n};1-\frac{2}%
{n};-16\kappa^{2}\left(  n-1\right)  ^{2}C\right) \\
-\sigma\left.  _{2}F_{1}\right.  \left(  -\frac{1}{2},-\frac{2}{n};1-\frac
{2}{n};-16\kappa^{2}\left(  n-1\right)  ^{2}C\sigma^{-\frac{1}{2}n}\right)
\end{array}
\right)  \ ,
\end{align}
where the $\pm1$ choice is made depending on whether $4\kappa\left(
n-1\right)  g_{1}\lessgtr1$, thereby giving various values for $-16\kappa
^{2}\left(  n-1\right)  ^{2}C=\left(  2-4\kappa\left(  n-1\right)
g_{1}\right)  \times4\kappa\left(  n-1\right)  g_{1}$. \ At the critical value
$4\kappa\left(  n-1\right)  g_{1}=1$ we have $-16\kappa^{2}\left(  n-1\right)
^{2}C=1$. \ (Note: \ The series expansion is not valid for the hypergeometric
function if $16\kappa^{2}\left(  n-1\right)  ^{2}C\sigma^{-\frac{1}{2}n}>1$.)
\ With $z=4\kappa\left(  n-1\right)  g_{1}$ we have $-16\kappa^{2}\left(
n-1\right)  ^{2}C=z\left(  2-z\right)  $, and we note for $z\left(
2-z\right)  =-1$, the solutions are: $z=1\pm\sqrt{2}$.

\subsection{Static, spherically symmetric solutions}

For example, for the free field in 4 spacetime dimensions, $\mathcal{E}_{1}=0
$, the static, spherically symmetric solutions of $\nabla^{2}\phi\left(
r\right)  =0$ in 3 space dimensions are of course
\begin{equation}
\phi_{1}\left(  r\right)  =C_{0}+\frac{C_{1}}{r}\ . \label{static1}%
\end{equation}
For the next step up in the hierarchy, $\mathcal{E}_{2}=0$, the static,
spherically symmetric solutions in 3 space dimensions are%
\begin{equation}
\phi_{2}\left(  r\right)  =C_{0}+C_{1}\sqrt{r}\ . \label{static2}%
\end{equation}
These two solutions are Legendre duals in 3D space (but not in 1+3 spacetime).
\ So, what happens if we mix them up? \ 

For example, take%
\begin{equation}
\mathcal{E}_{1}=\lambda\mathcal{E}_{2}\ .
\end{equation}
The static, spherically symmetric solutions in this case satisfy%
\begin{equation}
-\frac{1}{r^{2}}\partial_{r}\left(  r^{2}\partial_{r}\phi\right)
=\frac{2\lambda}{r^{2}}\partial_{r}\left(  r\left(  \partial_{r}\phi\right)
^{2}\right)  \ , \label{E1=E2}%
\end{equation}
which has an immediate first integral, and solution,%
\begin{align}
C_{1}  &  =r^{2}\partial_{r}\phi+2\lambda r\left(  \partial_{r}\phi\right)
^{2}\ ,\\
\partial_{r}\phi &  =\frac{r}{4\lambda}\left(  -1\pm\sqrt{1+\frac{8\lambda
C_{1}}{r^{3}}}\right)  \ .
\end{align}
Integrating this gives
\begin{equation}
\phi\left(  r\right)  =\phi\left(  0\right)  +\frac{R^{2}}{8\lambda}\left\{
-\frac{r^{2}}{R^{2}}\pm\sqrt{\frac{r}{R}}\left(  \sqrt{1+\frac{r^{3}}{R^{3}}%
}+3\left.  _{2}F_{1}\right.  \left(  \frac{1}{2},\frac{1}{6};\frac{7}%
{6};-\frac{r^{3}}{R^{3}}\right)  \right)  \right\}  \ , \label{mixed1&2}%
\end{equation}
where the length scale is related to the previous first integral by%
\begin{equation}
R=\left(  8\lambda C_{1}\right)  ^{1/3}\ .
\end{equation}

For small $r$ the solution (\ref{mixed1&2}) behaves like $\phi_{2}\left(
r\right)  $ in (\ref{static2}), and therefore it is not singular at the
origin. \ On the other hand, for large $r$, upon taking the upper $+$ sign in
(\ref{mixed1&2}), the solution behaves like $\phi_{1}\left(  r\right)  $ in
(\ref{static1}), while the lower $-$ sign choice in (\ref{mixed1&2}) gives a
solution that grows like $r^{2}$ for large $r$.

Taking the upper sign in (\ref{mixed1&2}),%
\begin{align}
&  \phi\left(  r\right)  \underset{r\ll R}{=}\phi\left(  0\right)
+\frac{R^{2}}{8\lambda}\left\{  4\sqrt{\frac{r}{R}}-\frac{r^{2}}{R^{2}%
}+O\left(  \left(  \frac{r}{R}\right)  ^{7/2}\right)  \right\}
\ ,\label{shortdistance}\\
&  \phi\left(  r\right)  \underset{r\gg R}{=}\phi\left(  0\right)
+\frac{R^{2}}{8\lambda}\left\{  \frac{3\Gamma\left(  1/3\right)  \Gamma\left(
7/6\right)  }{\sqrt{\pi}}-\frac{R}{r}+O\left(  \left(  \frac{R}{r}\right)
^{4}\right)  \right\}  \ , \label{largedistance}%
\end{align}
where $\frac{3\Gamma\left(  1/3\right)  \Gamma\left(  7/6\right)  }{\sqrt{\pi
}}=4.\,2065\ \cdots\ $. \ Taking the upper sign, the graph of $8\lambda\left(
\phi\left(  r\right)  -\phi\left(  0\right)  \right)  /R^{2}$ follows.%
\begin{center}
\includegraphics[
height=3.032in,
width=4.8227in
]%
{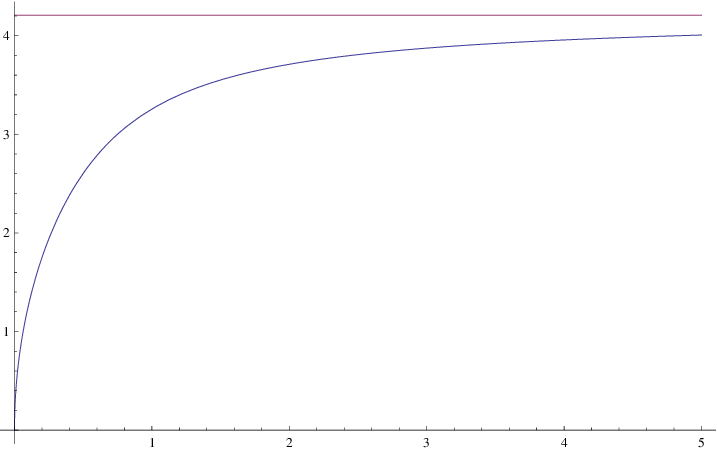}%
\\
$\frac{8\lambda}{R^{2}}\left(  \phi\left(  r\right)  -\phi\left(  0\right)
\right)  $ versus $r/R$, for $R>0$.
\end{center}
Another branch, obtained by taking the lower sign in (\ref{mixed1&2}), is not
so well-behaved for large $r$.\bigskip%

\begin{center}
\includegraphics[
trim=-0.016964in -0.008484in 0.016964in 0.008484in,
height=2.9664in,
width=4.8177in
]%
{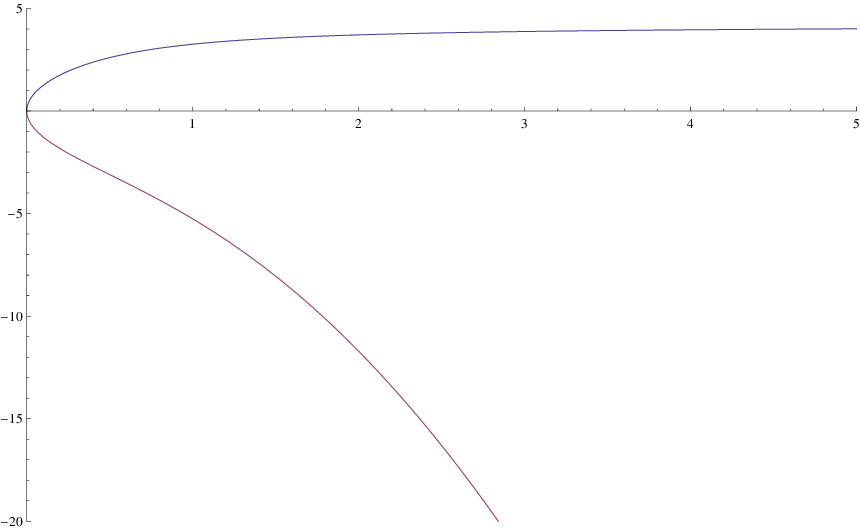}%
\\
Two branches of the mixed solution, $\frac{8\lambda}{R^{2}}\left(  \phi\left(
r\right)  -\phi\left(  0\right)  \right)  $, versus $r/R$, for $R>0$.
\end{center}

\subsection{Energy considerations}

A symmetric energy-momentum tensor for the mixed $\mathcal{E}_{1}$ ---
$\mathcal{E}_{2}$ model is%
\begin{align}
\Theta_{\mu\nu}  &  =\phi_{\mu}\phi_{\nu}-\tfrac{1}{2}\delta_{\mu\nu}%
\phi_{\alpha}\phi_{\alpha}-\lambda\left(  \phi_{\mu}\phi_{\nu}\phi
_{\alpha\alpha}-\phi_{\alpha}\phi_{\alpha\nu}\phi_{\mu}-\phi_{\alpha}%
\phi_{\alpha\mu}\phi_{\nu}+\delta_{\mu\nu}\phi_{\alpha}\phi_{\beta}%
\phi_{\alpha\beta}\right)  \ ,\\
\partial_{\mu}\Theta_{\mu\nu}  &  =\mathcal{E}\left[  \phi\right]
\mathcal{~}\phi_{\nu}\ ,\\
\mathcal{E}\left[  \phi\right]   &  =\phi_{\alpha\alpha}-\lambda\left(
\phi_{\alpha\alpha}\phi_{\beta\beta}-\phi_{\alpha\beta}\phi_{\alpha\beta
}\right)  \ ,
\end{align}
where the equation of motion is $\mathcal{E}\left[  \phi\right]  =0$. \ \ For
static, spherically symmetric $\phi$, the energy density is%
\begin{equation}
\Theta_{00}=\tfrac{1}{2}\left(  \partial_{r}\phi\left(  r\right)  \right)
^{2}-\frac{\lambda}{3}\partial_{r}\left(  \partial_{r}\phi\left(  r\right)
\right)  ^{3}\ ,
\end{equation}
and the total energy is
\begin{equation}
E=4\pi\int_{0}^{\infty}\Theta_{00}~r^{2}dr\ .
\end{equation}
This is finite for the bounded static solution that goes like $1/r$ for large
$r$, but it is not finite for the solution that goes like $r^{2}$. \ For the
finite case,%
\begin{equation}
\partial_{r}\phi=\frac{r}{4\lambda}\left(  -1+\sqrt{1+\frac{R^{3}}{r^{3}}%
}\right)  \ ,
\end{equation}
where $R^{3}=8\lambda C_{1}$ gives a length scale set by the first integral of
the static equation. \ After some playing around, we find%
\begin{equation}
E=\frac{\pi R^{5}}{6\lambda^{2}}\int_{0}^{\infty}s^{4}\left(  -1+\sqrt
{1+\frac{1}{s^{3}}}\right)  ^{2}ds=\frac{1}{90}\frac{2^{\frac{2}{3}}\pi^{3}%
}{\left(  \Gamma\left(  \frac{2}{3}\right)  \right)  ^{3}}\frac{R^{5}}%
{\lambda^{2}}=0.220\,25~\frac{R^{5}}{\lambda^{2}}\ .
\end{equation}
This is true for $R\geq0$, but actually, it is also of interest to consider
cases where $R<0$. \ 

The bounded solution for $R<0$ is real for $r\geq\left\vert R\right\vert $, as
obtained by integrating
\begin{equation}
\partial_{r}\phi=\frac{r}{4\lambda}\left(  -1+\sqrt{1-\frac{\left\vert
R\right\vert ^{3}}{r^{3}}}\right)  \ .
\end{equation}%
\begin{center}
\includegraphics[
height=2.9888in,
width=4.8177in
]%
{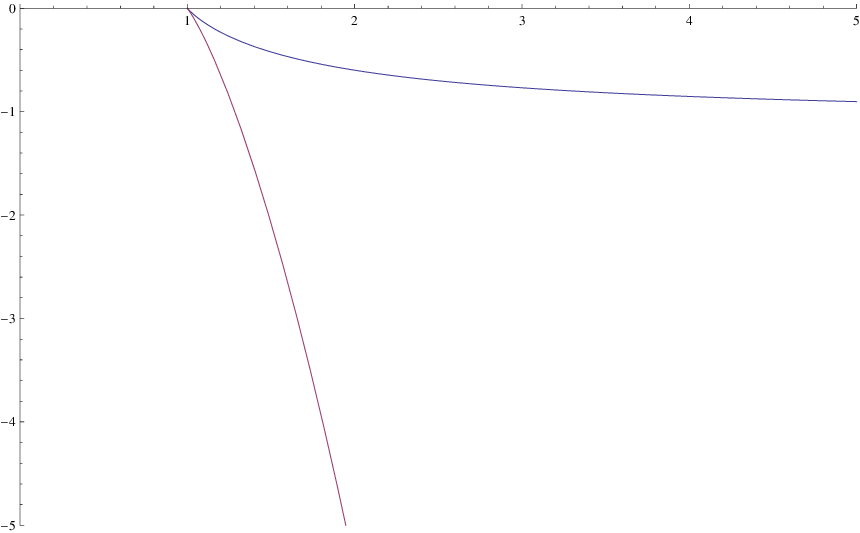}%
\\
Two branches of the mixed solution, $\frac{8\lambda}{R^{2}}\left(  \phi\left(
r\right)  -\phi\left(  \left\vert R\right\vert \right)  \right)  $, versus
$r/\left\vert R\right\vert $, for $R<0$.
\end{center}

The energy density has the same form as before, but now the total energy
\emph{outside} the singularity at $r=\left\vert R\right\vert $ is%
\begin{align}
E_{r\geq\left\vert R\right\vert }  &  =\frac{\pi\left\vert R\right\vert ^{5}%
}{6\lambda^{2}}\int_{1}^{\infty}s^{4}\left(  -1+\sqrt{1-\frac{1}{s^{3}}%
}\right)  ^{2}ds+\frac{4\pi\lambda}{3}\left.  r^{2}\left(  \partial_{r}%
\phi\right)  ^{3}\right\vert _{r=\left\vert R\right\vert }\nonumber\\
&  =\frac{1}{180}\left(  \frac{2^{\frac{2}{3}}\pi^{3}}{\left(  \Gamma\left(
\frac{2}{3}\right)  \right)  ^{3}}-\frac{3}{4}\pi\right)  \frac{\left\vert
R\right\vert ^{5}}{\lambda^{2}}=0.097\,037~\frac{\left\vert R\right\vert ^{5}%
}{\lambda^{2}}\ .
\end{align}

\subsection{Perturbative scattering}

Consider $p+q\longrightarrow p^{\prime}+q^{\prime}$\ for the $\mathcal{L}%
_{1}+\lambda\mathcal{L}_{2}$ model, perturbatively on-shell, i.e. $p^{2}%
=q^{2}=p^{\prime2}=q^{\prime2}=0$. \ The lowest-order scattering amplitude is%
\begin{equation}
M=\frac{1}{4}\lambda^{2}\left.  \left(  s^{3}+t^{3}+u^{3}\right)  \right\vert
_{u=-s-t}=\frac{3}{4}\lambda^{2}\left.  stu\right\vert _{u=-s-t}=\frac{-3}%
{4}\lambda^{2}st\left(  s+t\right)  \ .
\end{equation}
In the CM frame, in terms of the incident energy and scattering angle,
$p=\left(  E,\overrightarrow{p}\right)  $ and $\overrightarrow{p}%
\cdot\overrightarrow{p}^{\prime}=E^{2}\cos\theta$, we have $s=4E^{2}$,
$t=-2E^{2}\left(  1-\cos\theta\right)  $, and so \
\begin{equation}
M_{\text{CM}}=12\lambda^{2}E^{6}\sin^{2}\theta=12\lambda^{2}E^{6}\left(
1-\cos^{2}\theta\right)  =8\lambda^{2}E^{6}\left(  P_{0}\left(  \cos
\theta\right)  -P_{2}\left(  \cos\theta\right)  \right)  \ .\nonumber
\end{equation}
By the usual rules for the differential cross section in 4D, we then have%
\begin{equation}
\left.  \frac{d\sigma}{d\Omega}\right\vert _{\text{CM}}=\frac{1}{64\pi^{2}%
}\frac{1}{2!}\frac{\left\vert M_{\text{CM}}\right\vert ^{2}}{4E^{2}}%
=\frac{9\lambda^{4}E^{10}\sin^{4}\theta}{32\pi^{2}}\ ,
\end{equation}
and total cross section
\begin{equation}
\left.  \sigma\right\vert _{\text{CM}}=\frac{3\lambda^{4}E^{10}}{5\pi}\ .
\end{equation}
Note that $\left[  \lambda\right]  =1/m^{3}$ in 4D. \ This approximation for
$\sigma$ obviously exceeds the Froissart bound ($\propto\ln^{2}E$) as the
energy increases.\newpage

\section{Effects of $\phi~\Theta\lbrack\phi]$ self-couplings}

\textit{In this section, we consider galileon theories with an additional
self-coupling of the fields to the trace of their own energy-momentum tensor.
\ We explore the classical features of one such model, in flat 4D spacetime,
with emphasis on solutions that are scalar analogues of gravitational geons.
\ We discuss the stability of these scalar geons, and some of their possible
signatures, including shock fronts.}

For the simplest example, the galileon field is usually coupled to all
\emph{other} matter through the trace of the energy-momentum tensor,
$\Theta^{\text{(matter)}}$. \ But \emph{surely}, in a self-consistent theory
the galileon should also be coupled to its own energy-momentum trace, even in
the flat spacetime limit. \ Some consequences of this additional self-coupling
are considered in this section, based on work published in \cite{CF}.

Recall the action for the lowest non-trivial member of the galileon hierarchy,%
\begin{equation}
A_{2}=\tfrac{1}{2}\int\phi_{\alpha}\phi_{\alpha}\phi_{\beta\beta}~d^{n}x\ ,
\label{A2}%
\end{equation}
where $\phi$ is the scalar galileon field, $\phi_{\alpha}=\partial\phi\left(
x\right)  /\partial x^{\alpha}$, etc., and where repeated indices are summed
using the Lorentz metric $\delta_{\mu\nu}=\mathrm{diag}\left(  1,-1,-1,\cdots
\right)  $. \ 

\subsection{A non-vanishing trace}

As discussed above, including in $A_{2}$ a minimal coupling to a background
spacetime metric yields a symmetric energy-momentum tensor, which becomes in
the flat-space limit:%
\begin{equation}
\Theta_{\mu\nu}^{\left(  2\right)  }=\phi_{\mu}\phi_{\nu}\phi_{\alpha\alpha
}-\phi_{\alpha}\phi_{\alpha\nu}\phi_{\mu}-\phi_{\alpha}\phi_{\alpha\mu}%
\phi_{\nu}+\delta_{\mu\nu}\phi_{\alpha}\phi_{\beta}\phi_{\alpha\beta}\ .
\label{EnergyMomentum2}%
\end{equation}
This is seen to be conserved,%
\begin{equation}
\partial_{\mu}\Theta_{\mu\nu}^{\left(  2\right)  }=\phi_{\nu}~\mathcal{E}%
_{2}\left[  \phi\right]  \ ,
\end{equation}
upon using the field equation that follows from locally extremizing $A_{2}$,
$0=\delta A_{2}/\delta\phi=-\mathcal{E}_{2}\left[  \phi\right]  $, where%
\begin{equation}
\mathcal{E}_{2}\left[  \phi\right]  \equiv\phi_{\alpha\alpha}\phi_{\beta\beta
}-\phi_{\alpha\beta}\phi_{\alpha\beta}\ . \label{E2}%
\end{equation}
But, as previously noted, this$\ \Theta_{\mu\nu}^{\left(  2\right)  }$ is not
traceless. \ Consequently, the usual form of the scale current, $x_{\alpha
}\Theta_{\alpha\mu}^{\left(  2\right)  }$, is not conserved \cite{J}. \ On the
other hand, the action (\ref{A2}) is homogeneous in $\phi$ and its
derivatives, and is clearly invariant under the scale transformations
$x\rightarrow sx$ and $\phi\left(  x\right)  \rightarrow s^{\left(
4-n\right)  /3}\phi\left(  sx\right)  $. \ Hence the corresponding Noether
current must be conserved. \ This current is easily found, especially for
$n=4$, so let us restrict our attention to four spacetime dimensions in the
following. \ 

In that case the trace is obviously a total divergence:%
\begin{equation}
\Theta^{\left(  2\right)  }\equiv\delta_{\mu\nu}\Theta_{\mu\nu}^{\left(
2\right)  }=\partial_{\alpha}\left(  \phi_{\alpha}\phi_{\beta}\phi_{\beta
}\right)  \ .
\end{equation}
That is to say, for $n=4$ the virial is the trilinear $V_{\alpha}=\phi
_{\alpha}\phi_{\beta}\phi_{\beta}$. \ So a conserved scale current is given by
the combination,%
\begin{equation}
S_{\mu}=x_{\alpha}\Theta_{\alpha\mu}^{\left(  2\right)  }-\phi_{\alpha}%
\phi_{\alpha}\phi_{\mu}\ .
\end{equation}
Interestingly, this virial is not a divergence modulo a conserved current, so
this model is \emph{not} conformally invariant despite being scale invariant.
\ Be that as it may, it is not our principal concern here.

Our interest here is that the nonzero trace suggests an additional interaction
where $\phi$ couples directly to its own $\Theta^{(2)}$. \ This is similar to
coupling a conventional \emph{massive} scalar to the trace of its own
energy-momentum tensor \cite{FN}. \ In that previously considered example,
however, the consistent coupling of the field to its trace required an
iteration to all orders in the coupling. \ Upon summing the iteration and
making a field redefinition, the Nambu-Goldstone model emerged. \ But, for the
simplest galileon model in four spacetime dimensions, (\ref{A2}), a consistent
coupling of field and trace is much easier to implement. \ \emph{No iteration
is required.} \ The first-order coupling alone is consistent, after
integrating by parts and ignoring boundary contributions, so
that\footnote{Also note that $A_{2}$ follows from coupling $\phi$ to the trace
of the manifestly chargeless tensor $\left(  \partial_{\mu}\partial_{\nu
}-\delta_{\mu\nu}\partial_{\alpha}\partial_{\alpha}\right)  \phi_{\beta}%
\phi_{\beta}$.}%
\begin{equation}
-\tfrac{1}{4}\int\phi~\partial_{\alpha}\left(  \phi_{\alpha}\phi_{\beta}%
\phi_{\beta}\right)  ~d^{4}x=\tfrac{1}{4}\int\phi_{\alpha}\phi_{\alpha}%
\phi_{\beta}\phi_{\beta}~d^{4}x\ . \label{TraceTerm}%
\end{equation}
(Similar quadrilinear terms have appeared previously in \cite{DEV,DDE}, only
multiplied there by scalar curvature $R$ so that they would drop out in the
flat spacetime limit that we consider.) \ Consistency follows because
(\ref{TraceTerm}) gives an additional contribution to the energy-momentum
tensor which is \emph{traceless}, in 4D spacetime:%
\begin{equation}
\Theta_{\mu\nu}^{(3)}=\phi_{\mu}\phi_{\nu}\phi_{\alpha}\phi_{\alpha}-\tfrac
{1}{4}\delta_{\mu\nu}\phi_{\alpha}\phi_{\alpha}\phi_{\beta}\phi_{\beta
}\ ,\ \ \ \Theta^{(3)}=0\ .
\end{equation}
Of course, coupling $\phi$ to its own trace may impact the Vainstein mechanism
\cite{V} by changing the effective coupling of $\Theta^{\text{(matter)}}$ to
both backgrounds and fluctuations in $\phi$. \ We leave this as an exercise
for the reader.

\subsection{A model with additional quartic self-coupling}

Based on these elementary observations, we consider a model with action%
\begin{equation}
A=\int\left(  \tfrac{1}{2}\phi_{\alpha}\phi_{\alpha}-\tfrac{1}{2}\lambda
\phi_{\alpha}\phi_{\alpha}\phi_{\beta\beta}-\tfrac{1}{4}\kappa\phi_{\alpha
}\phi_{\alpha}\phi_{\beta}\phi_{\beta}\right)  ~d^{4}x\ , \label{A}%
\end{equation}
where for the Lagrangian $L$ we take a mixture of three terms: \ the standard
bilinear, the trilinear galileon, and its corresponding quadrilinear
trace-coupling. \ The quadrilinear is reminiscent of the Skyrme term in
nonlinear $\sigma$ models \cite{S} although here the topology would appear to
be always trivial. \ 

The second and third terms in $A$ are logically connected, as we have
indicated. \ But why include in $A$\ the standard bilinear term? \ The reasons
for including this term are to soften the behavior of solutions at large
distances, as will be evident below, and also to satisfy Derrick's criterion
for classical stability under the rescaling of $x$. \ Without the bilinear
term in $L$\ the energy within a spatial volume would be neutrally stable
under a uniform rescaling of $x$, and therefore able to disperse \cite{D,E}.

Similarly, for positive $\kappa$, the last term in $A$ ensures the energy
density of static solutions is always bounded below under a rescaling of the
field $\phi$, a feature that would not be true if $\kappa=0$ but $\lambda
\neq0$. \ So, we only consider $\kappa>0$ in the following. \ But before
discussing the complete $\Theta_{\mu\nu}$ for the model, we note that we
did\emph{\ not} include in $A$ a term coupling $\phi$ to the trace of the
energy-momentum due to the standard bilinear term, namely, $\int\phi
\Theta^{(1)}d^{4}x$, where%
\begin{equation}
\Theta_{\mu\nu}^{(1)}=\phi_{\mu}\phi_{\nu}-\tfrac{1}{2}\delta_{\mu\nu}%
\phi_{\alpha}\phi_{\alpha}\ ,\ \ \ \Theta^{(1)}=-\phi_{\alpha}\phi_{\alpha}\ .
\end{equation}
We have omitted such an additional term in $A$ solely as a matter of taste,
thereby ensuring that $L$ is invariant under constant shifts of the field.
\ Among other things, this greatly simplifies the task of finding solutions to
the equations of motion.

The field equation of motion for the model is $0=\delta A/\delta
\phi=-\mathcal{E}\left[  \phi\right]  $, where
\begin{equation}
\mathcal{E}\left[  \phi\right]  \equiv\phi_{\alpha\alpha}-\lambda\left(
\phi_{\alpha\alpha}\phi_{\beta\beta}-\phi_{\alpha\beta}\phi_{\alpha\beta
}\right)  -\kappa\left(  \phi_{\alpha}\phi_{\beta}\phi_{\beta}\right)
_{\alpha}\ .
\end{equation}
As expected, this field equation is second-order, albeit nonlinear. \ Also
note, under a rescaling of both $x$ and $\phi$, nonzero parameters $\lambda$
and $\kappa$ can be scaled out of the equation. \ Define%
\begin{equation}
\phi\left(  x\right)  =\frac{\lambda}{\kappa}~\psi\left(  \sqrt{\frac{\kappa
}{\lambda^{2}}}x\right)  \ . \label{PhiRescaled}%
\end{equation}
Then the field equation for $\psi\left(  z\right)  $ becomes%
\begin{equation}
\psi_{\alpha\alpha}-\left(  \psi_{\alpha\alpha}\psi_{\beta\beta}-\psi
_{\alpha\beta}\psi_{\alpha\beta}\right)  -\left(  \psi_{\alpha}\psi_{\beta
}\psi_{\beta}\right)  _{\alpha}=0\ , \label{PsiEofM}%
\end{equation}
where $\psi_{\alpha}=\partial\psi\left(  z\right)  /\partial z^{\alpha}$, etc.
\ In effect then, if both $\lambda$ and $\kappa$ do not vanish, it is only
necessary to solve the model's field equation for $\lambda=\kappa=1$.

\subsection{Static solutions}

For static, spherically symmetric solutions,\ $\phi=\phi\left(  r\right)  $,
the field equation of motion becomes%
\begin{equation}
0=\frac{1}{r^{2}}\frac{d}{dr}\left(  r^{2}\left(  \phi^{\prime}+\lambda
\frac{2}{r}\left(  \phi^{\prime}\right)  ^{2}+\kappa\left(  \phi^{\prime
}\right)  ^{3}\right)  \right)  \ .
\end{equation}
where $\phi^{\prime}=d\phi/dr$. This is immediately integrated once to obtain
a cubic equation,
\begin{equation}
r^{2}\phi^{\prime}+2\lambda r\left(  \phi^{\prime}\right)  ^{2}+\kappa
r^{2}\left(  \phi^{\prime}\right)  ^{3}=C\ , \label{SSSFieldEqn}%
\end{equation}
where $C$ is the constant of integration. \ Now, without loss of generality
(cf. (\ref{PhiRescaled}) and (\ref{PsiEofM})) we may choose $\lambda>0$.
\ Then, if $C=0$, either $\phi^{\prime}$ vanishes, or else there are two
solutions that are real only within a finite sphere of radius $r=\sqrt
{\lambda^{2}/\kappa}$. \ These two \textquotedblleft
interior\textquotedblright\ solutions are given\ exactly by%
\begin{equation}
\phi_{\pm}^{\prime}=-\frac{1}{r\kappa}\left(  \lambda\pm\sqrt{\lambda
^{2}-r^{2}\kappa}\right)  \text{ .} \label{C=0Solns}%
\end{equation}
Note that these solutions always have $\phi^{\prime}<0$ within the finite sphere.

Otherwise, if $C\neq0$, then examination of the cubic equation for small and
large $\left\vert \phi^{\prime}\right\vert $ determines the asymptotic
behavior of $\phi^{\prime}$\ for large and small $r$. \ In particular, there
is only one type of asymptotic behavior for large $r$:%
\begin{equation}
\phi^{\prime}\underset{r\rightarrow\infty}{\sim}\frac{C}{r^{2}}\text{ \ \ for
either sign of }C\text{\ .}%
\end{equation}
However, there are two types of behavior for large $\left\vert \phi^{\prime
}\right\vert $, corresponding to small $r$. \ Either%
\begin{equation}
r=\frac{-2\lambda}{\phi^{\prime}\kappa}\left(  1+O\left(  \frac{1}%
{\phi^{\prime}}\right)  \right)
\end{equation}
provided $\phi^{\prime}<0$, but with either sign of $C$; or else
\begin{equation}
r=\frac{1}{\phi^{\prime2}}\left(  \frac{C}{2\lambda}+O\left(  \frac{1}%
{\phi^{\prime}}\right)  \right)
\end{equation}
provided $C>0$, but with either sign of $\phi^{\prime}$. \ The corresponding
real solutions behave as%
\begin{gather}
\phi^{\prime}\underset{r\rightarrow0}{\sim}\frac{-2\lambda}{\kappa r}\text{
\ \ for either sign of }C\text{, or}\\
\phi^{\prime}\underset{r\rightarrow0}{\sim}\pm\sqrt{\frac{C}{2\lambda r}%
}\text{ \ \ provided }C>0\ .
\end{gather}
Comparison of\ the small $r$ behavior to the large $r$ asymptotics shows that
in half these cases the solutions would require zeroes to be real and
continuous for all $r$. \ But such zeroes do not occur. \ Instead, half of the
cases provide real solutions only over a finite interval of $r$, somewhat
similar to the $C=0$ solutions in (\ref{C=0Solns}), but not so easily
expressed, analytically. \ 

The solutions which are real for all $r>0$ boil down to two cases, with small
and large $r$ behavior given by either
\begin{equation}
\phi^{\prime}\underset{r\rightarrow0}{\sim}\sqrt{\frac{C}{2\lambda r}%
}\text{\ \ \ and \ \ }\phi^{\prime}\underset{r\rightarrow\infty}{\sim}\frac
{C}{r^{2}}\text{ \ \ for }C>0\text{,} \label{C>0Asymps}%
\end{equation}
or else%
\begin{equation}
\phi^{\prime}\underset{r\rightarrow0}{\sim}\frac{-2\lambda}{\kappa
r}\text{\ \ \ and \ \ }\phi^{\prime}\underset{r\rightarrow\infty}{\sim}%
\frac{C}{r^{2}}\text{ \ \ for }C<0\text{.} \label{C<0Asymps}%
\end{equation}
From further inspection of the cubic equation to determine the behavior of
$\phi^{\prime}$ for intermediate values of $r$, when $C>0$ it turns out that
$\phi^{\prime}$ is a single-valued, positive function for all $r>0$, joining
smoothly with the asymptotic behaviors given in (\ref{C>0Asymps}).
\ However,\ it also turns out there is an additional complication when $C<0$.
\ In this case there is a critical value $\left(  \kappa^{3/2}/\lambda
^{2}\right)  C_{\text{critical}}=-4\sqrt{3}/27\approx-0.2566$ such that, if
$C\leq C_{\text{critical}}$ then $\phi^{\prime}$ is a single-valued, negative
function for all $r>0$, while if $C_{\text{critical}}<C<0$ then $\phi^{\prime
}$ is triple-valued for an open interval in $r>0$. \ It is not completely
clear to us what physics underlies this multivalued-ness for some negative
$C$. \ But in any case, when $C<0$\ it is also true that $\phi^{\prime}$ joins
smoothly with the asymptotic behaviors given in (\ref{C<0Asymps}). \ All this
is illustrated in the two Figures to follow, for $\lambda=\kappa=1$.

A test particle coupled by $\phi\Theta^{\text{(matter)}}$ to any of these
galileon field configurations would see an effective potential which is not
$1/r$, for intermediate and small $r$. \ Therefore its orbit would show
deviations from the usual Kepler laws, including precession that is possibly
at variance with the predictions of conventional general relativity. \ It
would be interesting to search for such effects, say, by considering stars
orbiting around the galactic center \cite{MMG,MeyerEtAl}.%
\begin{center}
\includegraphics[
height=3.071in,
width=4.8152in
]%
{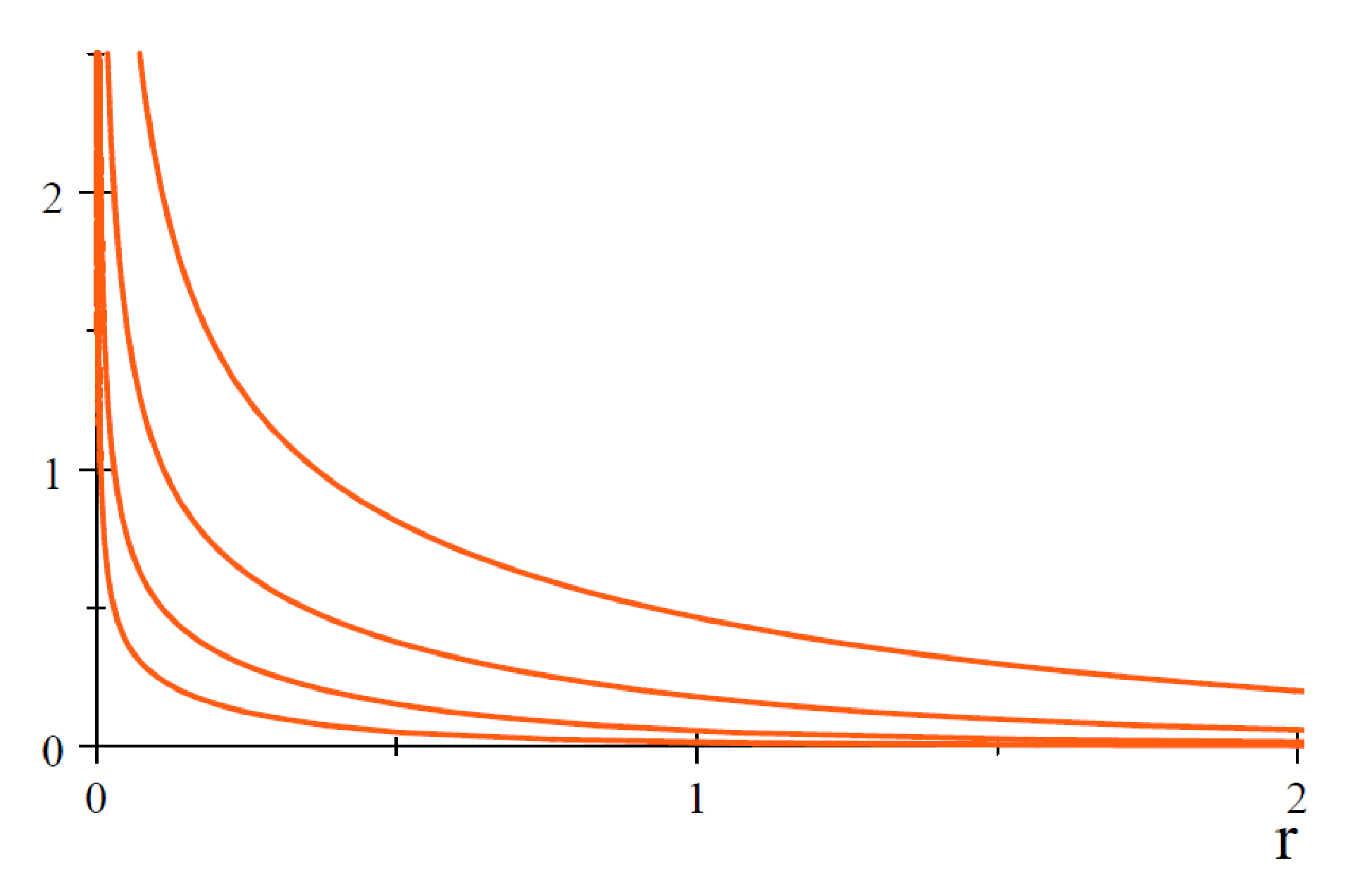}%
\\
$\psi^{\prime}\left(  r\right)  $ for $C=+1/4^{N}$, with $N=0,1,2,3$ for top
to bottom curves, respectively.
\end{center}
\begin{center}
\includegraphics[
height=2.9971in,
width=4.8219in
]%
{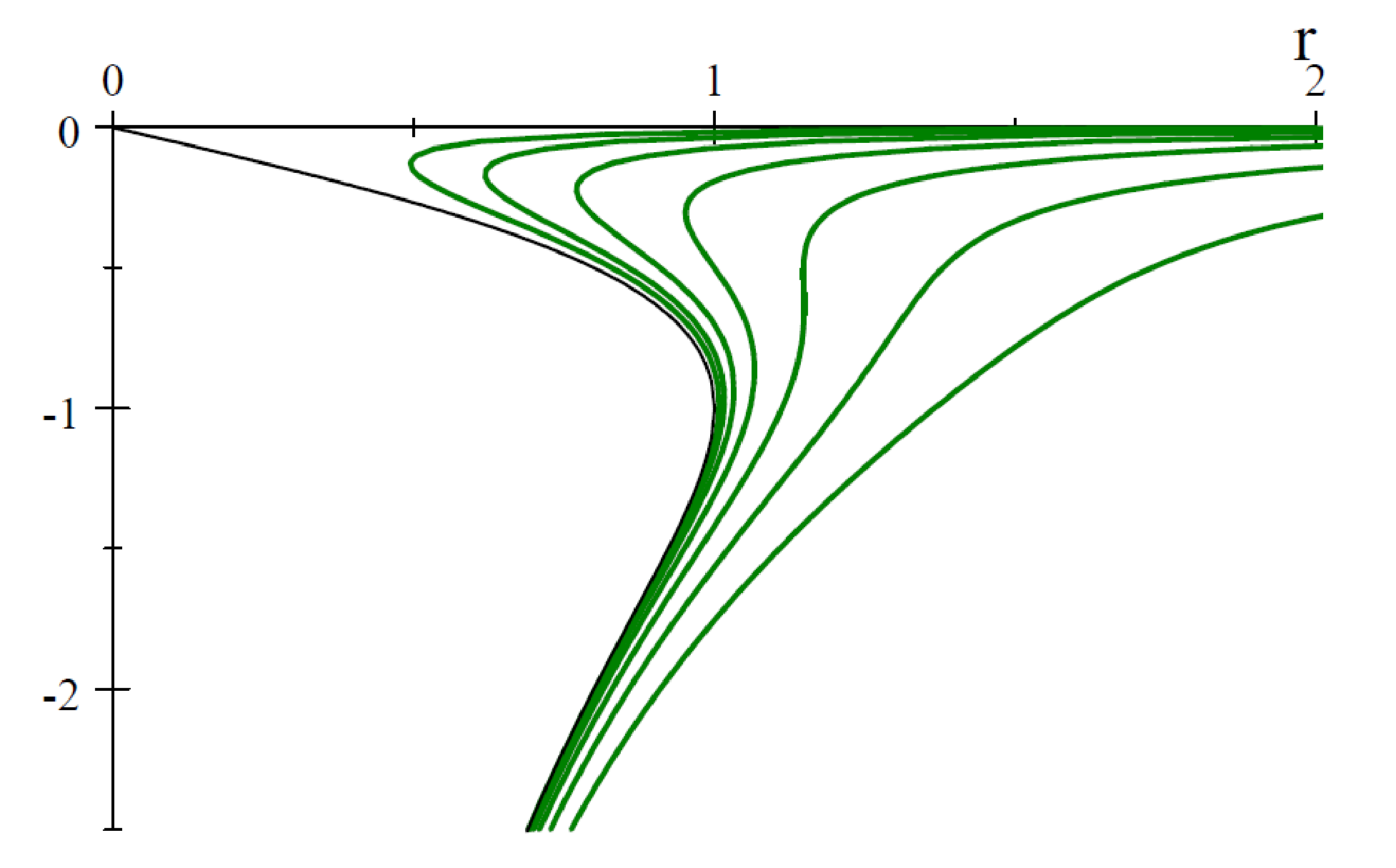}%
\\
$\psi^{\prime}\left(  r\right)  $ for $C=-1/2^{N}$, with $N=6,5,4,3,2,1,0$
from left to right, respectively. \ The thin black curve is a union of the two
$C=0$ solutions in (\ref{C=0Solns}).
\end{center}

For the solutions described by (\ref{C>0Asymps}) and (\ref{C<0Asymps}), the
total energy outside any large radius is obviously finite for both $C>0$ and
$C<0$. \ And if $C>0$, the total energy within a small sphere surrounding the
origin is also manifestly finite. \ But if $C<0$ the energy within that same
small sphere could be infinite \emph{unless} there is a cancellation between
the galileon term and the trace interaction term. \ Remarkably, this
cancellation does occur.\footnote{For $C<0$, to see cancellation between the
individually divergent galileon and trace interaction energies for small $r$
requires leading and next-to-leading terms in the expansion: $\ \phi^{\prime
}\underset{r\rightarrow0}{\sim}\tfrac{-2\lambda}{\kappa r}+\tfrac{\kappa
C}{4\lambda^{2}}+O\left(  r\right)  $.} \ So both $C>0$ and $C<0$ static
solutions for the model have finite total energy.

\subsection{Energy considerations again}

Complete information about the distribution of energy is provided by the
model's energy-momentum tensor,%
\begin{equation}
\Theta_{\mu\nu}=\Theta_{\mu\nu}^{(1)}-\lambda\Theta_{\mu\nu}^{\left(
2\right)  }-\kappa\Theta_{\mu\nu}^{(3)}\ .
\end{equation}
As expected, this is conserved, given the field equation $\mathcal{E}\left[
\phi\right]  =0$, since%
\begin{equation}
\partial_{\mu}\Theta_{\mu\nu}=\phi_{\nu}\mathcal{E}\left[  \phi\right]  \ .
\end{equation}
The energy density for \emph{static} solutions differs from the canonical
energy density for such solutions (namely, $-L$) by a total spatial divergence
that arises from the galileon term:
\begin{equation}
\Theta_{00}=-\left.  L\right\vert _{\text{static}}-\tfrac{1}{2}\lambda
\overrightarrow{\nabla}\cdot\left(  \left(  \nabla\phi\right)  ^{2}%
\overrightarrow{\nabla}\phi\right)  \ . \label{CvsSymmetric}%
\end{equation}
This divergence will not contribute to the total energy for fields such that
$\lim_{r\rightarrow\infty}\left(  \phi/\ln r\right)  $ exists. \ Assuming that
is the case, Derrick's scaling argument for static, finite energy
solutions\ of the equations of motion \cite{D} shows the energy is just twice
that due to the bilinear $\Theta_{00}^{\left(  1\right)  }$. \ Thus,
\begin{equation}
E=\int\Theta_{00}~d^{3}r=\int\left(  \overrightarrow{\nabla}\phi\right)
^{2}~d^{3}r\ .
\end{equation}

For the spherically symmetric static solutions of (\ref{SSSFieldEqn}), this
becomes an expression of the energy as a function of the parameters and the
constant of integration $C$:
\begin{equation}
E\left[  \lambda,\kappa,C\right]  =4\pi\int_{0}^{\infty}\left(  \phi^{\prime
}\right)  ^{2}~r^{2}dr\ .
\end{equation}
Again without loss of generality, consider $\lambda=\kappa=1$. \ Then for
either $C>0$ or for $C<C_{\text{critical}}<0$, change integration variables
from $r$ to $s\equiv\phi^{\prime}$ to find:\footnote{The multivalued behavior
of any solution for $C_{\text{critical}}<C<0$ makes the determination of the
total energy ambiguous, at best, for these cases. \ This is an unresolved
issue.}%
\begin{align}
E\left(  C\gtrless0\right)   &  =I\left(  \left\vert C\right\vert \right)
\mp\left(  \left\vert C\right\vert +\tfrac{1}{2}\pi\right)  \ ,\label{E(C)}\\
I\left(  C>0\right)   &  \equiv\tfrac{1}{2}\int_{0}^{\infty}\frac{P\left(
s,C\right)  ~ds}{\left(  s^{2}+1\right)  ^{4}\sqrt{s^{4}+s\left(
s^{2}+1\right)  C}}\ , \label{I(C)}%
\end{align}
where the numerator of the integrand is an eighth-order polynomial in $s$,
namely,$\ $%
\begin{equation}
P\left(  s,C\right)  =8s^{8}+12Cs^{7}+\left(  3C^{2}-8\right)  s^{6}%
+8Cs^{5}+7C^{2}s^{4}-4Cs^{3}+5C^{2}s^{2}+C^{2}\ .
\end{equation}
Thus, $I\left(  C\right)  $ is an elliptic integral. \ But rather than express
the final result in terms of standard functions, it suffices here just to plot
$E\left(  C\right)  $, in the Figure below. \ Note that $E$ increases
monotonically with $\left\vert C\right\vert $.

For other values of $\lambda$ and $\kappa$ with the constant of integration
$C$ specified as in (\ref{SSSFieldEqn}), the energy of the solution is given
in terms of the function defined by (\ref{E(C)},\ref{I(C)}):
\begin{equation}
E\left[  \lambda,\kappa,C\right]  =\left(  \lambda^{3}/\kappa^{5/2}\right)
~E\left(  \kappa^{3/2}C/\lambda^{2}\right)  \ .
\end{equation}
The energy curves indicate double degeneracy in $E$, for different values of
$\left\vert C\right\vert $, when $E\left[  \lambda,\kappa,C\right]
>\pi\lambda^{3}/\kappa^{5/2}$. \ Also, for a given $\left\vert C\right\vert $
the negative $C$ solutions are \emph{higher} in energy, with $E\left[
\lambda,\kappa,-\left\vert C\right\vert \right]  -E\left[  \lambda
,\kappa,\left\vert C\right\vert \right]  =\pi\lambda^{3}/\kappa^{5/2}%
+2\left\vert C\right\vert \lambda/\kappa$. \ Or at least this is true for all
$\left\vert C\right\vert \geq\left\vert C_{\text{critical}}\right\vert $ in
which case $E\left[  \lambda,\kappa,C\right]  \geq\frac{\lambda^{3}}%
{\kappa^{5/2}}E\left(  \frac{\kappa^{3/2}}{\lambda^{2}}C_{\text{critical}%
}\right)  \approx3.7396~\lambda^{3}/\kappa^{5/2}$.
\begin{center}
\includegraphics[
trim=0.000000in 0.000000in 0.376326in 0.266826in,
height=3.2071in,
width=4.8194in
]%
{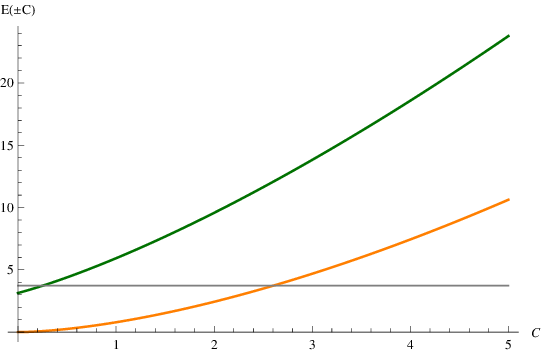}%
\\
$E\left(  \pm C\right)  $ versus $C\geq0$ as lower/upper curves (the
horizontal line is $E\left(  C_{\text{critical}}\right)  \approx3.7396$).
\end{center}

\subsection{Scalar geons and a shock-front conjecture}

Finite energy classical solutions of gravity-like theories bring to mind the
\textquotedblleft geons\textquotedblright\ proposed long ago by Wheeler
\cite{W}. \ These were envisioned in their purest form as distributions of
only gravitational energy held together solely by gravitational interaction.
$\ $Combinations of electromagnetic energy and gravity were also considered,
as were systems containing neutrinos. $\ $Wheeler argued that such
configurations would be \emph{relatively}\ stable, if they existed, but would
eventually dissipate due to a variety of both classical and quantum effects,
including\ light-light scattering, as well as production and absorption of
quanta. $\ $While plausible distributions were sketched, and decay rates were
estimated, \emph{exact} classical solutions were not found. $\ $

The same mechanisms would seem to apply to any hypothetical classical galileon
distributions such as those discussed here, the main difference being that
analytic spherically symmetric solutions might still be obtainable even if
conventional gravitational effects were included. \ Perhaps these
gravitational effects would not alter the qualitative features of the static
pure $\phi$ configurations given above. \ Should they really exist, presumably
these galileon geons could also be dissipated by various classical and quantum
effects. \ All this is far beyond our current abilities and the scope of this
paper, of course, but the general ideas suggest some interesting possibilities.

Whatever the cause, if the configuration's energy loss were gradual, as a
first step it might suffice to model the time-dependent system
quasi-statically, as a continuous flow from one static solution to another.
\ That is to say, perhaps a good approximation would be to take $C\left(
t\right)  $, with $\left\vert C\right\vert $ and $E\left(  C\right)  $
decreasing monotonically with time. \ For the positive $C$ case, this would be
more or less uneventful as the whole configuration would just slowly disappear
without any abrupt changes. \ But for the negative $C$ case, as $t$ increased
$C_{\text{critical}}$ would be reached, beyond which the solution would begin
to fold over, exhibiting the multivalued features shown in the Figure. \ But
this is just the usual picture for the formation of a shock front. \ These
particular galileon shocks would implode, converging towards the origin, as
shown \href{http://server.physics.miami.edu/~curtright/PsiWave.gif}{here}.
\ We believe this is a plausible scenario and a reasonable physical
interpretation of the model's multivalued solutions. \ Moreover, it would seem
to provide a signature for their existence.

As is clear from the Figure, the shock front would form when $d\phi^{\prime
}/dr=\infty$. \ For the $C<0$ static solutions of (\ref{SSSFieldEqn}) it is
not difficult to determine the locus of such singular points. \ It is given by
the intersection of the solutions, for various $C$, and the curve $\left(
1+3\kappa\phi^{\prime2}\right)  r=4\lambda\phi^{\prime}$. \ As usual for
singular points in the development of a shock, almost certainly there is some
physics missing from the equations. \ Since $\phi^{\prime\prime}$ is large,
the obvious modification would be to include higher derivative terms in the
action, which is tantamount to attempting an ultraviolet completion of the
model. \ This is an open question. \ Perhaps higher terms in the galileon
hierarchy would be natural candidates to be included.

\subsection{Comparison to the self-dual model}

To get a handle on such terms, and for purposes of comparison to the model in
(\ref{A}), consider briefly another model somewhat similar in form, but whose
Lagrangian consists only of terms taken from the galileon hierarchy, without
any coupling to $\Theta$. \ After rescaling the field and coordinates to
achieve a standard form, this alternate model may be defined by%
\begin{gather}
A_{\text{self-dual}}\left[  \psi\right]  =\int\left(  \tfrac{1}{2}\psi
_{\alpha}\psi_{\alpha}-\tfrac{1}{4}\psi_{\alpha}\psi_{\alpha}\psi_{\beta\beta
}\right. \nonumber\\
\left.  +\tfrac{1}{12}\psi_{\alpha}\psi_{\alpha}\left(  \psi_{\beta\beta}%
\psi_{\gamma\gamma}-\psi_{\beta\gamma}\psi_{\beta\gamma}\right)  \right)
d^{4}x\ . \label{ASelfDual}%
\end{gather}
The difference with (\ref{A})\ lies in the last term, which is quadrilinear in
the field, as before, but now has two fields with second derivatives. \ 

As the name suggests, this model is self-dual, in the following sense: \ The
action retains its form under a Legendre transformation \cite{FG2} (also see
\cite{G}) to a new field $\Psi$ and new coordinates $X$, as defined by:%
\begin{equation}
\psi\left(  x\right)  +\Psi\left(  X\right)  =x_{\alpha}X_{\alpha}\ .
\end{equation}
Thus $A_{\text{self-dual}}\left[  \psi\right]  =A_{\text{self-dual}}\left[
\Psi\right]  $, provided integrations by parts give no surface contributions.
\ This identity suggests that there are interesting properties for the
quantized model, such as its ultraviolet behavior, but that is outside the
scope of the present discussion. \ 

Here it suffices to compare the classical physics following from
(\ref{ASelfDual}) with that following from (\ref{A}). \ Upon integrating once
the classical equations of motion for static, spherically symmetric solutions
of the field equations for (\ref{ASelfDual}), the result is again a cubic
equation,%
\begin{equation}
r^{2}\psi^{\prime}+r\left(  \psi^{\prime}\right)  ^{2}+\tfrac{1}{3}\left(
\psi^{\prime}\right)  ^{3}=C\ , \label{SDSSSFieldEqn}%
\end{equation}
but the $\left(  \psi^{\prime}\right)  ^{3}$ term is no longer weighted by
$r^{2}$ as it was in (\ref{SSSFieldEqn}). \ Thus the small and large $r$
behaviors are now given by%
\begin{equation}
\psi^{\prime}\underset{r\rightarrow0}{\sim}\left(  3C\right)  ^{1/3}%
\text{\ \ \ and \ \ }\psi^{\prime}\underset{r\rightarrow\infty}{\sim}\frac
{C}{r^{2}}\text{,}%
\end{equation}
for either sign of the constant of integration, $C$. \ These static solutions
have finite total energy for either sign of $C$, as before, only now
$\psi^{\prime}$ is always bounded. \ Moreover, upon inspection of the behavior
of $\psi^{\prime}$ for intermediate $r$, and various $C$, unlike the previous
model the solutions are now always single-valued for either $C>0$ or $C<0$.
\ Thus there are no multivalued solutions like those shown in the previous
Figure for various $C<0$. \ However, each of the $C<0$ static solutions now
has a single point for which $d\psi^{\prime}/dr=\infty$, namely, $r=\left(
3\left\vert C\right\vert \right)  ^{1/3}$. \ So there is still a reason to
expect the existence of shock fronts for quasi-static time-dependent fields in
this alternate model. \ Finally, again for $C<0$, to have $\phi^{\prime}$ real
for all $r>0$, it is necessary to join together \textquotedblleft
interior\textquotedblright\ and \textquotedblleft exterior\textquotedblright%
\ solutions at $r=\left(  3\left\vert C\right\vert /2\right)  ^{1/3}$. \ These
features are illustrated in the following Figure.%
\begin{center}
\includegraphics[
height=6.308in,
width=6.6816in
]%
{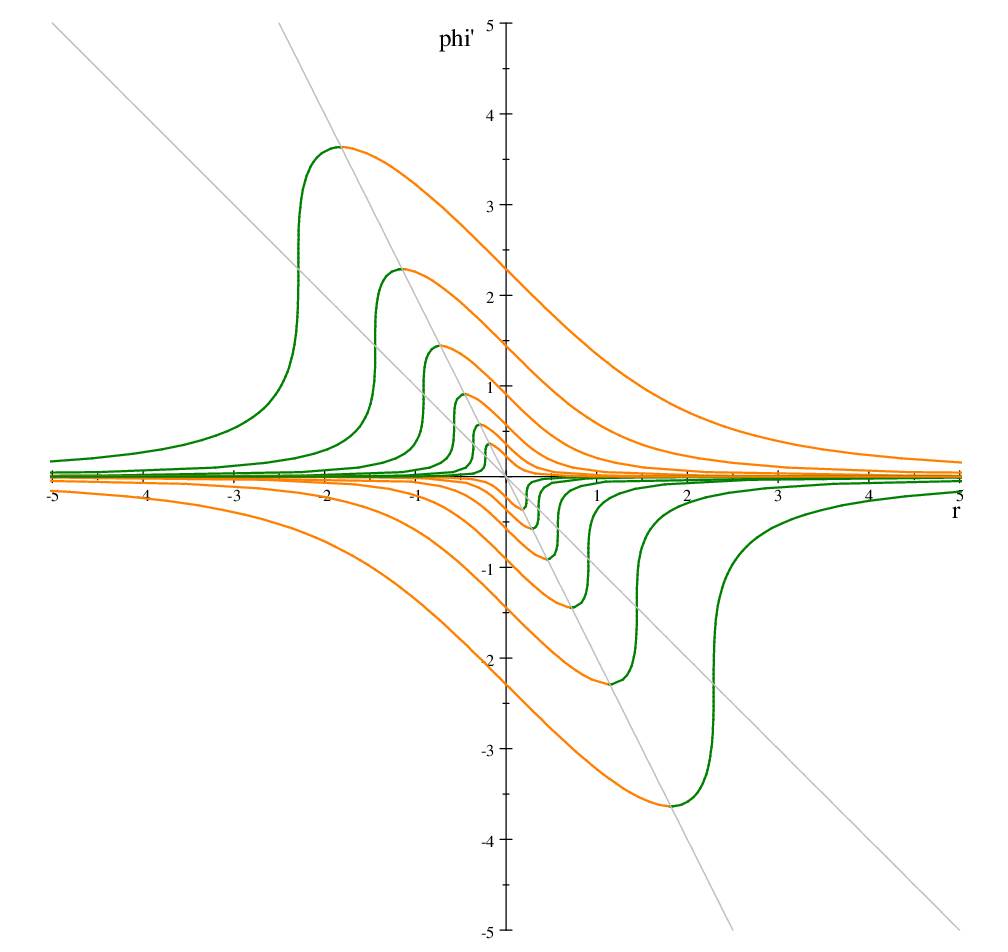}%
\\
Static solutions of the self-dual model for $C>0$ upper half-plane, and for
$C<0$, lower half plane. $\ $The solutions $r_{\pm}=\frac{\phi^{\prime}}%
{2}\left(  -1\pm\sqrt{\frac{4C}{\left(  \phi^{\prime}\right)  ^{3}}-\frac
{1}{3}}\right)  $ are shown in orange/green.
\end{center}
In these graphs, the solutions of (\ref{SDSSSFieldEqn}) are shown for both
physical $r>0$ and unphysical $r<0$ to display some symmetry relations between
the $C>0$ cases and the interior and exterior solutions for $C<0$. \ The
straight lines, in gray, are the loci of points where the solutions have zero
and infinite slopes, for different values of $C$.

It remains to investigate the stability of these spherically symmetric
solutions under perturbations, especially to check for the existence of
superluminal modes, along the lines of \cite{GHT}. \ Evidently, superluminal
modes are a possible feature for models of this type.\newpage

\section{General relativistic effects}

\textit{In this section, the simple trace-coupled Galileon model of the
previous section is coupled minimally to gravitation (GR) and shown to
admit\ spherically symmetric static solutions with naked spacetime curvature
singularities.}

In the previous section, based on \cite{CF},\ the effects of coupling a
Galileon to its own energy-momentum trace were considered in the flat
spacetime limit. \ Here, general relativistic effects are taken into
consideration and additional features of this same model are explored in
curved spacetime \cite{DDE,DEV}. \ Such features have been explored in the
literature (see \cite{C}, and for a related class of models, \cite{CharmEtAl.}%
). \ The main point to be emphasized here is that there can be solutions with
\emph{naked singularities} when the energy in the scalar field is finite and
not too large, and for which the effective mass of the system is positive.
\ Thus for the simple model at hand there is an open set of \emph{physically
acceptable} scalar field data for which curvature singularities are \emph{not}
hidden inside event horizons \cite{Rama,Rinaldi}. \ This would seem to have
important implications for the cosmic censorship conjecture
\cite{Penrose,Wald,Singh}. \ It is worthwhile to note that, in general, naked
singularities have observable consequences that differ from those due to black
holes \cite{VirbhadraEtAl.}.

\subsection{Minimal coupling to gravity}

The scalar field part of the action in curved space is
\begin{equation}
A=\frac{1}{2}\int g^{\alpha\beta}\phi_{\alpha}\phi_{\beta}\left(  1-\frac
{1}{\sqrt{-g}}~\partial_{\mu}\left(  \sqrt{-g}g^{\mu\nu}\phi_{\nu}\right)
-\frac{1}{2}~g^{\mu\nu}\phi_{\mu}\phi_{\nu}\right)  \sqrt{-g}~d^{4}x\ .
\label{Action}%
\end{equation}
This gives a symmetric energy-momentum tensor $\Theta_{\alpha\beta}$ for
$\phi$ upon variation of the metric. \
\begin{equation}
\delta A=\tfrac{1}{2}\int\sqrt{-g}~\Theta_{\alpha\beta}~\delta g^{\alpha\beta
}~d^{4}x\ ,
\end{equation}%
\begin{align}
\Theta_{\alpha\beta}  &  =\phi_{\alpha}\phi_{\beta}\left(  1-g^{\mu\nu}%
\phi_{\mu}\phi_{\nu}\right)  -\tfrac{1}{2}g_{\alpha\beta}~g^{\mu\nu}\phi_{\mu
}\phi_{\nu}\left(  1-\tfrac{1}{2}g^{\rho\sigma}\phi_{\rho}\phi_{\sigma}\right)
\nonumber\\
&  -\phi_{\alpha}\phi_{\beta}\tfrac{1}{\sqrt{-g}}\partial_{\mu}\left(
\sqrt{-g}g^{\mu\nu}\phi_{\nu}\right)  +\tfrac{1}{2}\partial_{\alpha}\left(
g^{\mu\nu}\phi_{\mu}\phi_{\nu}\right)  \phi_{\beta}+\tfrac{1}{2}%
\partial_{\beta}\left(  g^{\mu\nu}\phi_{\mu}\phi_{\nu}\right)  \phi_{\alpha
}-\tfrac{1}{2}g_{\alpha\beta}\partial_{\rho}\left(  g^{\mu\nu}\phi_{\mu}%
\phi_{\nu}\right)  g^{\rho\sigma}\phi_{\sigma}\ . \label{EnMomTensor}%
\end{align}
It also gives the field equation for $\phi$ upon variation of the scalar
field, $\mathcal{E}\left[  \phi\right]  =0$, where
\begin{gather}
\delta A=-\int\sqrt{-g}~\mathcal{E}\left[  \phi\right]  ~\delta\phi
~d^{4}x\ ,\\
\mathcal{E}\left[  \phi\right]  =\partial_{\alpha}\left[  g^{\alpha\beta}%
\phi_{\beta}\sqrt{-g}-g^{\alpha\beta}\phi_{\beta}~g^{\mu\nu}\phi_{\mu}%
\phi_{\nu}\sqrt{-g}-g^{\alpha\beta}\phi_{\beta}\partial_{\mu}\left(  \sqrt
{-g}g^{\mu\nu}\phi_{\nu}\right)  +\tfrac{1}{2}\sqrt{-g}g^{\alpha\beta}%
\partial_{\beta}\left(  g^{\mu\nu}\phi_{\mu}\phi_{\nu}\right)  \right]  \ .
\end{gather}
Since $\mathcal{E}\left[  \phi\right]  $ is a total divergence, it easily
admits a first integral for static, spherically symmetric configurations.
\ Consider \emph{only} those situations in the following.

\subsection{Static spherical solutions}

For such configurations the metric in generalized Schwarzschild coordinates is
\cite{Tolman}%
\begin{equation}
\left(  ds\right)  ^{2}=e^{N\left(  r\right)  }\left(  dt\right)
^{2}-e^{L\left(  r\right)  }\left(  dr\right)  ^{2}-r^{2}\left(
d\theta\right)  ^{2}-r^{2}\sin^{2}\theta\left(  d\varphi\right)  ^{2}\ .
\end{equation}
Thus for static, spherically symmetric $\phi$, with covariantly conserved
energy-momentum tensor (\ref{EnMomTensor}), Einstein's equations reduce to
just a pair of coupled $1$st-order nonlinear equations:%
\begin{align}
r^{2}\Theta_{t}^{\ t}  &  =e^{-L}\left(  rL^{\prime}-1\right)
+1\ ,\label{Albert1}\\
r^{2}\Theta_{r}^{\ r}  &  =e^{-L}\left(  -rN^{\prime}-1\right)  +1\ .
\label{Albert2}%
\end{align}
These are to be combined with the first integral of the $\phi$ field equation
in this situation. \ Defining
\begin{equation}
\eta\left(  r\right)  \equiv e^{-L\left(  r\right)  /2}\ ,\ \ \ \varpi\left(
r\right)  \equiv\eta\left(  r\right)  \phi^{\prime}\left(  r\right)  \ ,
\label{TwoDefns}%
\end{equation}
that first integral becomes
\begin{equation}
\frac{Ce^{-N/2}}{r^{2}}=\varpi\left(  1+\varpi^{2}\right)  +\frac{1}{2}\left(
N^{\prime}+\frac{4}{r}\right)  \eta\varpi^{2}\ , \label{1stIntegral}%
\end{equation}
where for asymptotically flat spacetime the constant $C$ is given by
$\lim_{r\rightarrow\infty}r^{2}\phi^{\prime}\left(  r\right)  =C$. \ Then upon
using
\begin{align}
\Theta_{t}^{\ t}  &  =\Theta_{\theta}^{\ \theta}=\Theta_{\varphi}^{\ \varphi
}=\tfrac{1}{2}\varpi^{2}\left(  1+\tfrac{1}{2}\varpi^{2}\right)  -\eta
\varpi^{2}\varpi^{\prime}\ ,\label{EM1}\\
\Theta_{r}^{\ r}  &  =-\tfrac{1}{2}\varpi^{2}\left(  1+\tfrac{3}{2}\varpi
^{2}\right)  -\tfrac{1}{2}\eta\varpi^{3}\left(  N^{\prime}+\tfrac{4}%
{r}\right)  \ , \label{EM2}%
\end{align}
the remaining steps to follow are clear.

First, for $C\neq0$, one can eliminate $N^{\prime}$ from (\ref{Albert2}) and
(\ref{1stIntegral}) to obtain an exact expression for $N$ in terms of $\eta$,
$\varpi$, and $C$:%
\begin{equation}
e^{N/2}=\frac{8C}{r\varpi}\frac{\eta-\frac{1}{2}r\varpi^{3}}{\left(
4\varpi-2r^{2}\varpi^{3}-r^{2}\varpi^{5}+8r\eta+12\varpi\eta^{2}+8r\varpi
^{2}\eta\right)  }\ . \label{ResultsOf1stIntegral}%
\end{equation}
If the numerator of this last expression vanishes there is an \emph{event
horizon}, otherwise not. \ When $\eta=\frac{1}{2}r\varpi^{3}$ the denominator
of (\ref{ResultsOf1stIntegral}) is positive definite. \ 

Next, in addition to (\ref{Albert1}) one can now eliminate $N$ from either
(\ref{Albert2}) or (\ref{1stIntegral}) to obtain two coupled first-order
nonlinear equations for $\eta$ and $\varpi$. \ These can be integrated, at
least numerically. \ Or they can be used to determine analytically the large
and small $r$ behaviors, hence to see if the energy and curvature are finite.
\ For example, again for asymptotically flat spacetime, it follows that%
\begin{align}
&  e^{L/2}\underset{r\rightarrow\infty}{\sim}1+\frac{M}{r}+\frac{1}{4}\left(
6M^{2}-C^{2}\right)  \frac{1}{r^{2}}+\frac{1}{2}M\left(  5M^{2}-2C^{2}\right)
\frac{1}{r^{3}}+O\left(  \frac{1}{r^{4}}\right)  \ ,\label{AsympRadial}\\
&  e^{N/2}\underset{r\rightarrow\infty}{\sim}1-\frac{M}{r}-\frac{1}{2}%
M^{2}\frac{1}{r^{2}}+\frac{1}{12}M\left(  C^{2}-6M^{2}\right)  \frac{1}{r^{3}%
}+O\left(  \frac{1}{r^{4}}\right)  \ ,\label{AsympTime}\\
&  \varpi\underset{r\rightarrow\infty}{\sim}\frac{C}{r^{2}}\left(  1+\frac
{M}{r}+\frac{3}{2}M^{2}\frac{1}{r^{2}}\right)  +O\left(  \frac{1}{r^{5}%
}\right)  \ , \label{AsympCharge}%
\end{align}
for constant $C$ and $M$.

As of this writing the details of the two remaining first-order ordinary
differential equations are not pretty, but the equations are numerically
tractable. \ In terms of the variables defined in (\ref{TwoDefns}), in light
of (\ref{ResultsOf1stIntegral}), Einstein's equation (\ref{Albert2}) becomes%
\begin{equation}
F\left(  r,\varpi,\eta\right)  r\frac{d}{dr}\varpi+G\left(  r,\varpi
,\eta\right)  r\frac{d}{dr}\eta=H\left(  r,\varpi,\eta\right)  \ ,
\label{Poly2}%
\end{equation}%
\begin{align}
F\left(  r,\varpi,\eta\right)   &  =-4\eta\left[  \left.  2r^{3}\varpi
^{6}+3r^{3}\varpi^{8}+16\varpi\eta+4r\varpi^{4}\right.  _{\ }\right.
\nonumber\\
&  \left.  _{\ }\left.  +16r\eta^{2}+48\varpi\eta^{3}+48r\varpi^{2}\eta
^{2}+12r\varpi^{4}\eta^{2}-12r^{2}\varpi^{5}\eta\right.  \right]  \ ,\\
G\left(  r,\varpi,\eta\right)   &  =8\eta\varpi^{2}\left[  \left.
2r^{2}\varpi^{2}+3r^{2}\varpi^{4}-12\eta^{2}+12r\varpi^{3}\eta+4\right.
_{\ }\right]  \ ,\\
H\left(  r,\varpi,\eta\right)   &  =\varpi\left[  \left.  8\eta\varpi\left(
4r\varpi^{3}-4\eta+2r^{2}\varpi^{2}\eta+3r^{2}\varpi^{4}\eta+12r\varpi^{3}%
\eta^{2}-12\eta^{3}\right)  \right.  _{\ }\right. \nonumber\\
&  \left.  _{\ }\left.  +\left(  4+3r^{2}\varpi^{4}+2r^{2}\varpi^{2}%
+12\eta^{2}\right)  \left(  4\varpi-r^{2}\varpi^{5}-2r^{2}\varpi^{3}%
+8r\varpi^{2}\eta+8r\eta+12\varpi\eta^{2}\right)  \right.  \right]  \ ,
\end{align}
while Einstein's equation (\ref{Albert1})\ becomes%
\begin{gather}
I\left(  r,\varpi,\eta\right)  r\frac{d}{dr}\varpi+J\left(  r,\varpi
,\eta\right)  r\frac{d}{dr}\eta=K\left(  r,\varpi,\eta\right)
\ ,\label{Poly1}\\
I\left(  r,\varpi,\eta\right)  =r\eta\varpi^{2}\ ,\ \ \ J\left(  r,\varpi
,\eta\right)  =-2\eta\ ,\\
K\left(  r,\varpi,\eta\right)  =\tfrac{1}{2}r^{2}\varpi^{2}\left(  1+\tfrac
{1}{2}\varpi^{2}\right)  +\eta^{2}-1\ .
\end{gather}

\subsection{Numerical results}

As a representative example with $\varpi>0$, (\ref{Poly1}) and (\ref{Poly2})
were integrated numerically to obtain the results shown in the Figure, for
data initialized as$\ \left.  \varpi\right\vert _{r=1}=0.5$ and $\left.
\eta\right\vert _{r=1}=1$. \ Evidently it is true that $\eta\left(  r\right)
\neq\frac{1}{2}r\varpi^{3}\left(  r\right)  $ for this case, so $e^{N\left(
r\right)  }$ does not vanish for any $r>0$ and there is no event horizon. \ %

\begin{center}
\includegraphics[
height=4.0133in,
width=6.0265in
]%
{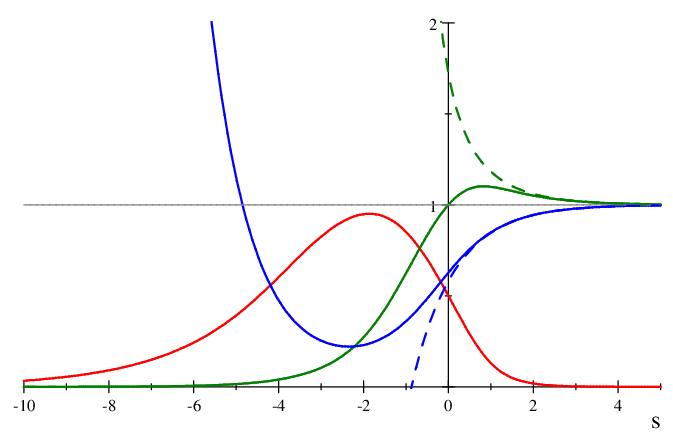}%
\\
For initial values $\left.  {\protect\footnotesize \varpi(s)}\right\vert
_{s=0}=0.5$ and$\ \left.  {\protect\footnotesize \eta(s)}\right\vert
_{s=0}=1.0$, $d\phi/dr=\varpi/\eta$ is shown in red,
$e^{{\protect\footnotesize L}}=1/\eta^{{\protect\footnotesize 2}}$ in green,
and $e^{{\protect\footnotesize N}}$ in blue, where
$r=e^{{\protect\footnotesize s}}$. \ For comparison, Schwarzschild
$e^{{\protect\footnotesize L}}$ and $e^{{\protect\footnotesize N}}$ are also
shown as resp. green and blue dashed curves for the same $M\approx0.21$.
\end{center}
\textquotedblleft\textit{\ ... an exotic type of matter with which human
science is entirely unfamiliar is required for such a geometry to
exist.}\textquotedblright\ --- B K Tippett \cite{Tippett}%

\begin{center}
\includegraphics[
height=4.0133in,
width=6.0265in
]%
{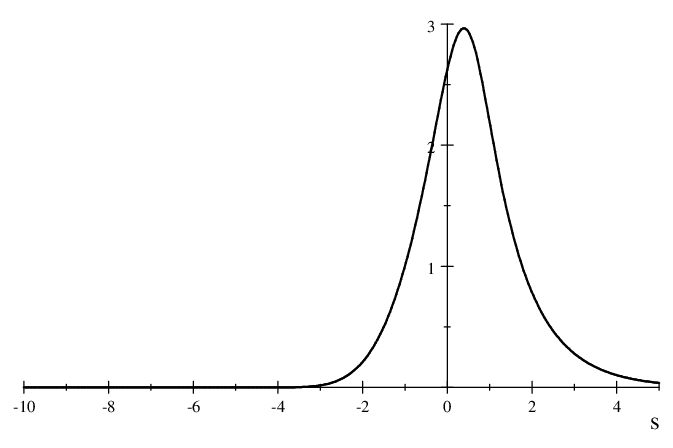}%
\\
$e^{s}\mathcal{H}\left(  e^{s}\right)  $ for $\left.  \varpi\left(  s\right)
\right\vert _{s=0}=0.5$ and$\ \left.  \eta\left(  s\right)  \right\vert
_{s=0}=1.0$, where $r=e^{{\protect\footnotesize s}}$.
\end{center}
\newpage

However, there is a geometric singularity at $r=0$ with divergent scalar
curvature: $\ \lim_{r\rightarrow0}r^{3/2}R=const$. \ Since $R=-\Theta_{\mu
}^{\ \mu}$, and $\lim_{r\rightarrow0}\varpi$ is finite, this divergence in $R$
comes from the last term in (\ref{EM2}), which in turn comes from the second
term in $A$, i.e. the covariant $\partial\phi\partial\phi\partial^{2}\phi$ in
(\ref{Action}). \ In fact, it it not difficult to establish analytically for a
class of solutions of the model, for which the example in the Figure is
representative, the following limiting behavior holds.%
\begin{equation}
\lim_{r\rightarrow0}\left(  e^{L/2}/\sqrt{r}\right)  =\ell\ ,\ \ \ \lim
_{r\rightarrow0}\left(  \sqrt{r}e^{N/2}\right)  =n\ ,\ \ \ \lim_{r\rightarrow
0}\varpi=p\ ,\ \ \ \lim_{r\rightarrow0}\left(  \phi^{\prime}/\sqrt{r}\right)
=p\ell\ ,
\end{equation}
where $\ell$, $n$, and $p$ are constants related to the constant $C$ in
(\ref{1stIntegral}):%
\begin{equation}
2C=3np^{2}/\ell\ .
\end{equation}
It follows that for solutions in this class,%
\begin{equation}
\lim_{r\rightarrow0}r^{3/2}R=pC/n\ .
\end{equation}
For the example shown in the Figure: \ $\ell\approx1.5$, $n\approx0.086$,
$p\approx3.3$, $C\approx0.94$, and $pC/n\approx36$.

For the same $\left.  \eta\right\vert _{r=1}=1$, further numerical results
show there are also curvature singularities without horizons for smaller
$\left.  \varpi\right\vert _{r=1}>0$, but event horizons are present for
larger scalar fields (roughly when $\left.  \varpi\right\vert _{r=1}>2/3$).
\ A more precise and complete characterization of the data set $\left\{
\left.  \varpi\right\vert _{r=1},\left.  \eta\right\vert _{r=1}\right\}  $ for
which there are naked singularities is in progress, but it is already evident
from the preceding remarks that the set has nonzero measure. \ 

The energy contained in \emph{only} the scalar field in the curved spacetime
is given by%
\begin{gather}
E_{\text{Galileon}}=\int_{0}^{\infty}\mathcal{H}\left(  r\right)
dr=\int_{-\infty}^{\infty}e^{s}\mathcal{H}\left(  e^{s}\right)  ds\ ,\\
\mathcal{H}\left(  r\right)  \equiv4\pi r^{2}e^{L/2}e^{N/2}\Theta_{t}%
^{\ t}=2\pi e^{2s}e^{L/2}e^{N/2}\varpi^{2}\left(  s\right)  \left(
1+\tfrac{1}{2}\varpi^{2}\left(  s\right)  \right)  -4\pi e^{s}e^{N/2}%
\varpi^{2}\left(  s\right)  \tfrac{d}{ds}\varpi\left(  s\right)  \ .
\end{gather}
For the above numerical example, the integrand $e^{s}\mathcal{H}\left(
e^{s}\right)  $ is shown in the Figure. \ Evidently, $E_{\text{Galileon}}$\ is
finite in this case. \ It is also clear from the Figures that the Galileon
field has significant effects on the geometry in the vicinity of the peak of
its radial energy density. \ There the metric coefficients are greatly
distorted from the familiar Schwarzschild values, and as a consequence, the
horizon is eliminated.\newpage

\subsection*{Other numerical examples}

Here are additional plots for $\left.  {\footnotesize \eta(s)}\right\vert
_{s=0}=1.0$ and various initial values $\left.  {\footnotesize \varpi
(s)}\right\vert _{s=0}$. \ As before, $d\phi/dr=\varpi/\eta$ is shown in red,
$e^{{\footnotesize L}}=1/\eta^{{\footnotesize 2}}$ in green, and
$e^{{\footnotesize N}}$ in blue, where $r=e^{{\footnotesize s}}$. \ For
comparison, Schwarzschild $e^{{\footnotesize L}}$ and $e^{{\footnotesize N}}$
are also shown as resp. green and blue dashed curves for the same $M$, as
given in the Figure labels.

\noindent\hspace{-0.63in}%
{\parbox[b]{3.8555in}{\begin{center}
\includegraphics[
height=2.5687in,
width=3.8555in
]%
{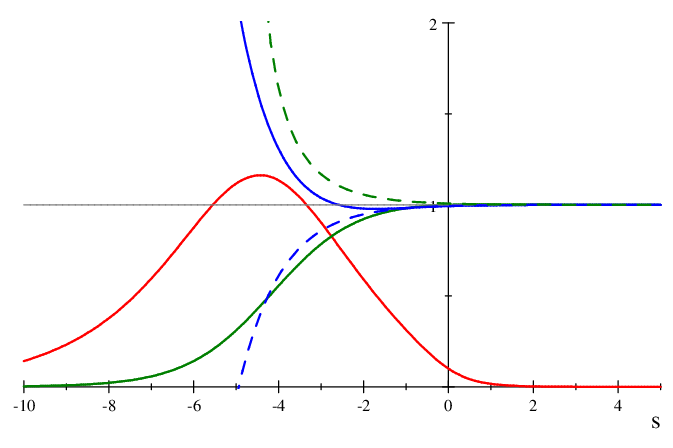}%
\\
Initial values $\left.  {\protect\footnotesize \varpi(s)}\right\vert
_{s=0}=0.100$ and$\ \left.  {\protect\footnotesize \eta(s)}\right\vert
_{s=0}=1.00$ corresponding to $M=0.00358$ and $C=0.121$%
\end{center}}}
{\parbox[b]{3.8555in}{\begin{center}
\includegraphics[
height=2.5679in,
width=3.8555in
]%
{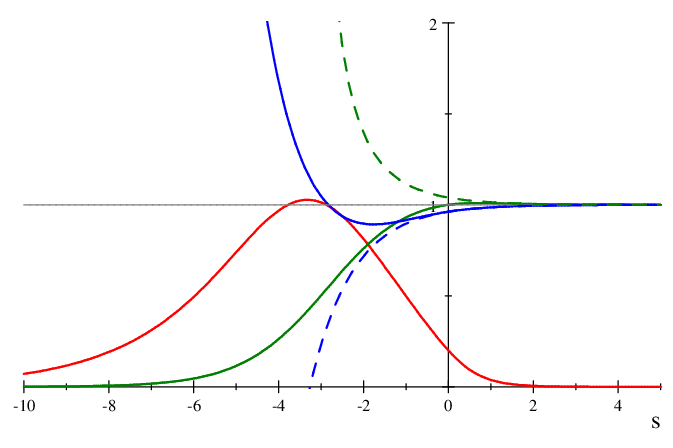}%
\\
Initial values $\left.  {\protect\footnotesize \varpi(s)}\right\vert
_{s=0}=0.200$ and$\ \left.  {\protect\footnotesize \eta(s)}\right\vert
_{s=0}=1.00$ corresponding to $M=0.0191$ and $C=0.283$%
\end{center}}}

\noindent\hspace{-0.63in}%
{\parbox[b]{3.8555in}{\begin{center}
\includegraphics[
height=2.5687in,
width=3.8555in
]%
{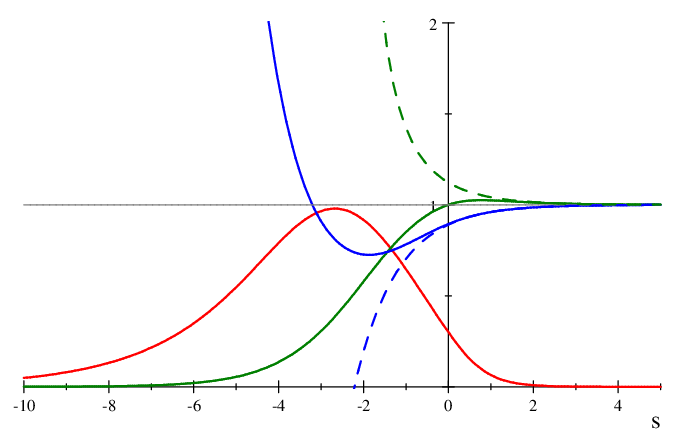}%
\\
Initial values $\left.  {\protect\footnotesize \varpi(s)}\right\vert
_{s=0}=0.300$ and$\ \left.  {\protect\footnotesize \eta(s)}\right\vert
_{s=0}=1.00$ corresponding to $M=0.0546$ and $C=0.484$%
\end{center}}}
{\parbox[b]{3.8555in}{\begin{center}
\includegraphics[
height=2.5687in,
width=3.8555in
]%
{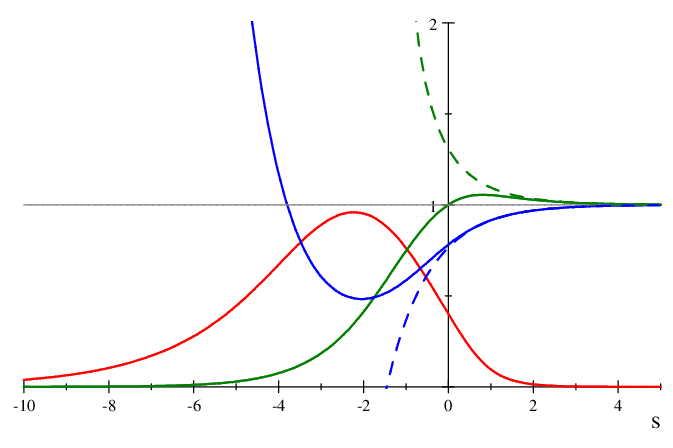}%
\\
Initial values $\left.  {\protect\footnotesize \varpi(s)}\right\vert
_{s=0}=0.400$ and$\ \left.  {\protect\footnotesize \eta(s)}\right\vert
_{s=0}=1.00$ corresponding to $M=0.117$ and $C=0.710$%
\end{center}}}

\noindent\hspace{-0.63in}%
{\parbox[b]{3.8555in}{\begin{center}
\includegraphics[
height=2.5762in,
width=3.8555in
]%
{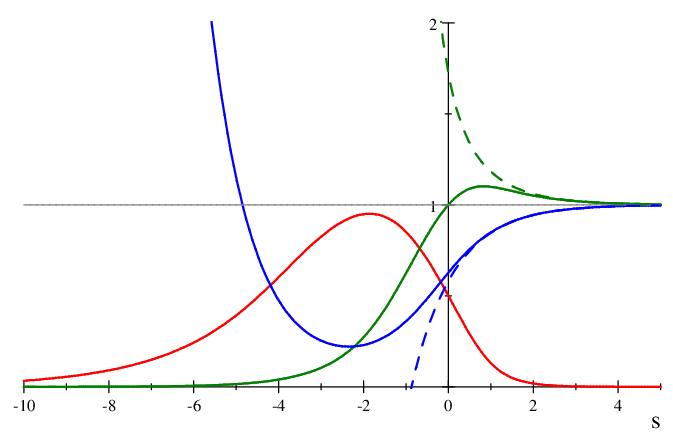}%
\\
Initial values $\left.  {\protect\footnotesize \varpi(s)}\right\vert
_{s=0}=0.500$ and$\ \left.  {\protect\footnotesize \eta(s)}\right\vert
_{s=0}=1.00$ corresponding to $M=0.209$ and $C=0.936$%
\end{center}}}
{\parbox[b]{3.8555in}{\begin{center}
\includegraphics[
height=2.5695in,
width=3.8555in
]%
{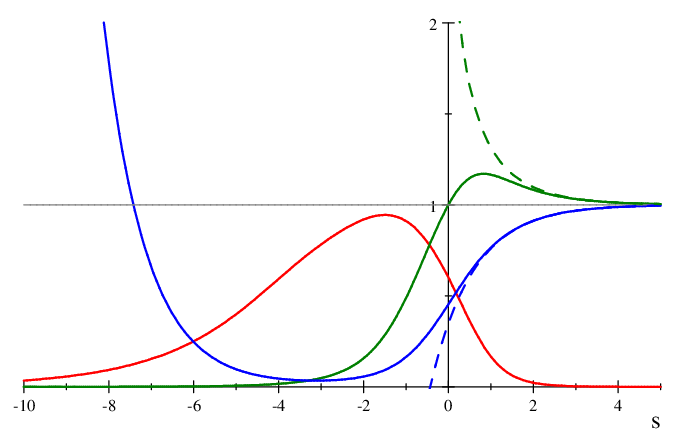}%
\\
Initial values $\left.  {\protect\footnotesize \varpi(s)}\right\vert
_{s=0}=0.600$ and$\ \left.  {\protect\footnotesize \eta(s)}\right\vert
_{s=0}=1.00$ corresponding to $M=0.326$ and $C=1.13$%
\end{center}}}

\noindent\hspace{-0.63in}%
{\parbox[b]{3.8555in}{\begin{center}
\includegraphics[
height=2.5695in,
width=3.8555in
]%
{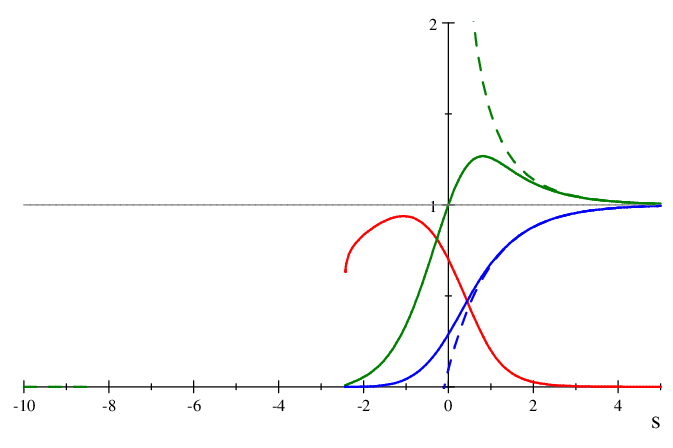}%
\\
Initial values $\left.  {\protect\footnotesize \varpi(s)}\right\vert
_{s=0}=0.700$ and$\ \left.  {\protect\footnotesize \eta(s)}\right\vert
_{s=0}=1.00$ corresponding to $M=0.453$ and $C=1.26$%
\end{center}}}
{\parbox[b]{3.8514in}{\begin{center}
\includegraphics[
height=2.5728in,
width=3.8514in
]%
{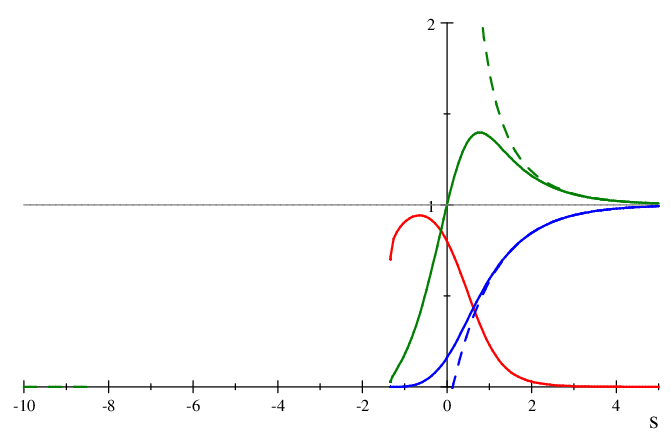}%
\\
Initial values $\left.  {\protect\footnotesize \varpi(s)}\right\vert
_{s=0}=0.800$ and$\ \left.  {\protect\footnotesize \eta(s)}\right\vert
_{s=0}=1.00$ corresponding to $M=0.573$ and $C=1.32$%
\end{center}}}

\noindent For each of the last two plots, the numerical integration of the
coupled galileon-GR equations has encountered a mathematical (as opposed to
physical) singularity and terminated, resp. at $r\approx e^{-2.5}=0.082$ and
$r\approx e^{-1.3}=0.27$, as is indicative of an horizon for which $e^{N}=0$.
\ This feature persists for initial data with larger values of $\left.
\varpi(s)\right\vert _{s=0}$, when $\left.  \eta(s)\right\vert _{s=0}%
=1$.\newpage Here are two more cases, just below and just above the point
where horizons are formed. \ Again, for the second of these plots, the
numerical integration of the coupled galileon-GR equations has encountered a
mathematical singularity, and terminated at the point where $e^{N\left(
r\right)  }$ (blue curve) vanishes.%
\begin{center}
\includegraphics[
height=2.9896in,
width=4.4765in
]%
{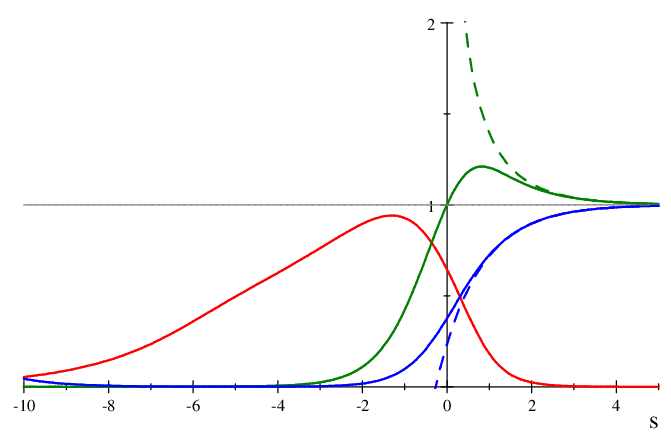}%
\\
Initial values $\left.  \varpi(s)\right\vert _{s=0}=0.645$ and$\ \left.
\eta(s)\right\vert _{s=0}=1.00$ corresponding to $M=0.383$ and $C=1.199$%
\end{center}
\begin{center}
\includegraphics[
height=2.9896in,
width=4.4765in
]%
{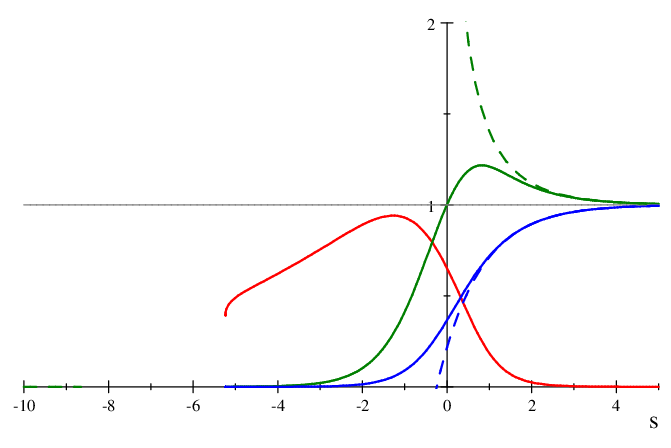}%
\\
Initial values $\left.  \varpi(s)\right\vert _{s=0}=0.652$ and$\ \left.
\eta(s)\right\vert _{s=0}=1.00$ corresponding to $M=0.392$ and $C=1.209$%
\end{center}
A useful test for an horizon is provided by the numerator of $e^{N}$ in
(\ref{ResultsOf1stIntegral}). \ Define the discriminant%
\begin{equation}
disc\left(  r\right)  =1-\frac{r~\varpi\left(  r\right)  ^{3}}{2~\eta\left(
r\right)  }\ .
\end{equation}
Should this vanish at some radius for which $\eta\left(  r\right)  $ is
finite, then at that radius $e^{N\left(  r\right)  }=0$, thereby indicating an
horizon at that radius.

The \emph{critical case}, separating solutions with naked singularities from
those with event horizons, has the small $r$ limiting behavior $\eta\left(
r\right)  \underset{r\rightarrow0}{\sim}r~\varpi^{3}\left(  r\right)  $, such
that the discriminant $disc\underset{r\rightarrow0}{\sim}\frac{1}{2}$ as
illustrated here for specific data. \
\begin{center}
\includegraphics[
height=2.3869in,
width=3.6206in
]%
{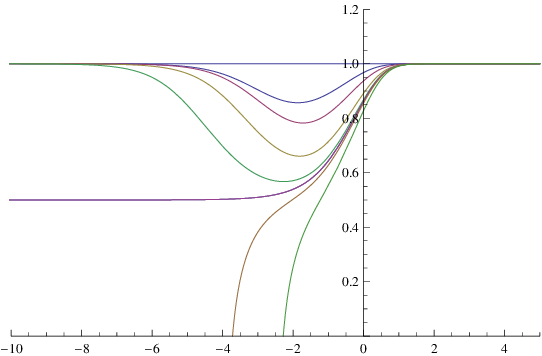}%
\\
The discriminant $disc=1-\frac{1}{2}r\varpi^{3}/\eta$ versus $s=\ln r$ for
various $\left.  \varpi\right\vert _{r=1}$ (namely, 0.4, 0.5, 0.6, 0.64,
critical, 0.66, and 0.7) with $\left.  \eta\right\vert _{r=1}=1$. \ The
critical initial value for the separatrix, for which
$disc\protect\underset{r\rightarrow0}{\sim}1/2$, is $\left.  \varpi\right\vert
_{r=1}=0.65002917\cdots$.
\end{center}
For initial data giving rise to naked singularities, $disc>1/2$ (cf. the upper
curves in the Figure above), while for data leading to horizons, $e^{N}$
vanishes at the horizon radius, and therefore at that radius $disc=0$ (cf. the
lower two curves in the Figure). \ When the limiting critical behavior
$\eta\left(  r\right)  \underset{r\rightarrow0}{\sim}r~\varpi^{3}\left(
r\right)  $ is inserted into the differential equations (\ref{Poly1}) and
(\ref{Poly2}) we find the power law behavior:
\begin{equation}
\eta_{\text{critical}}\left(  r\right)  \underset{r\rightarrow0}{\sim}%
c^{3}r^{-4/5}\ ,\ \ \ \varpi_{\text{critical}}\left(  r\right)
\underset{r\rightarrow0}{\sim}cr^{-3/5}\ ,\ \ \ \phi_{\text{critical}}%
^{\prime}\left(  r\right)  \underset{r\rightarrow0}{\sim}\frac{r^{1/5}}{c^{2}%
}\ . \label{critical}%
\end{equation}
Moreover, critical cases are easily determined numerically for various initial
data, $\left\{  \left.  \varpi(s)\right\vert _{s=0},\ \left.  \eta
(s)\right\vert _{s=0}\right\}  $, thereby allowing determination of a curve
that separates the open set of initial data that exhibits naked singularities
from the set that exhibits event horizons. \ 

\subsection{Censored and naked phases}

The situation for a portion of the initial data plane is as follows. \
\begin{center}
\includegraphics[
height=3.0602in,
width=4.6011in
]%
{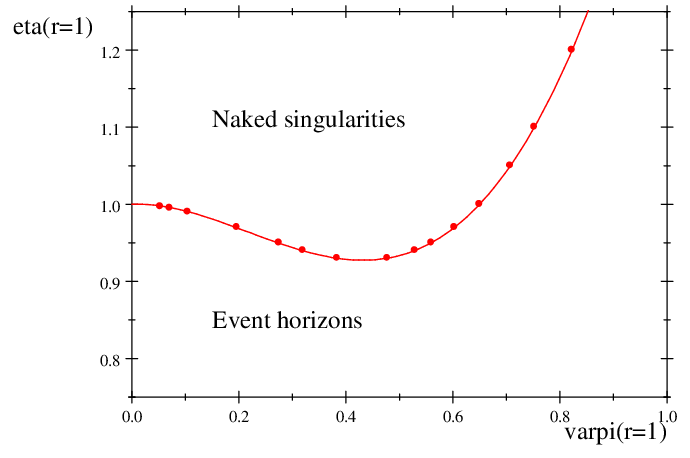}%
\\
$\left(  \left.  \varpi\right\vert _{r=1},\left.  \eta\right\vert
_{r=1}\right)  $ boundary separating initial data that exhibit naked
singularities from data that exhibit horizons. \ The curve is a fourth-order
polynomial fit to the numerically computed critical points (dots), namely,
$\eta_{\text{fit}}\left(  \varpi\right)  =1+0.0255538\varpi-1.34405\varpi
^{2}+2.20589\varpi^{3}-0.304933\varpi^{4}$.
\end{center}
This shows naked singularities for the model exist for an initial data set of
non-zero measure, and are actually encountered for a significant portion of
the initial data plane. \ \newpage

A similar demarcation between naked/censored solutions can be presented in
terms of asymptotic $r\rightarrow\infty$ data instead of initial $r=1$ data.
\ With $M$ and $C$ defined as in (\ref{AsympRadial}), (\ref{AsympTime}), and
(\ref{AsympCharge}), we find the following curve separating the two types of
solutions. \ Solutions for points above the red curve have naked
singularities, while solutions for points below that curve have event
horizons.%
\begin{center}
\includegraphics[
height=2.978in,
width=4.489in
]%
{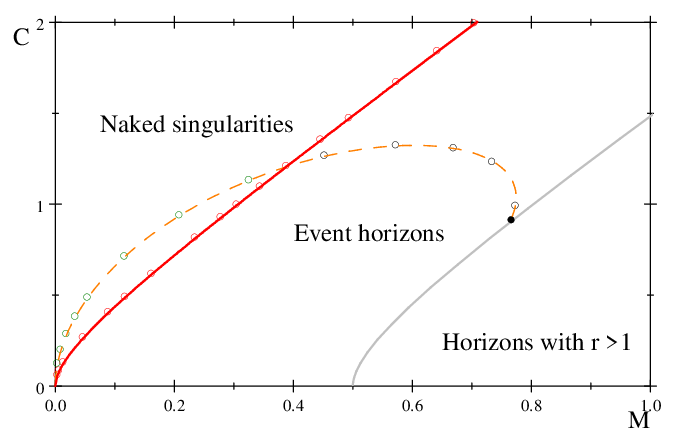}%
\\
Computed points (red circles) and an interpolating curve (solid red)
separating the $r\rightarrow\infty$ asymptotic data for solutions with naked
singularities from that for solutions with event horizons.
\end{center}

By imposing the same $\left.  \eta(s)\right\vert _{s=0}$ initial condition for
various values of $\left.  \varpi(s)\right\vert _{s=0}$, the numerical data
also shows that the corresponding $C\left(  M\right)  $ has a local maximum,
and hence $M\left(  C\right)  $ becomes double-valued near that point. \ For
example, when $\left.  \eta(s)\right\vert _{s=0}=1$ the local maximum for
$C\left(  M\right)  $ is near $\left.  \varpi(s)\right\vert _{s=0}\approx
0.78$. \ By examining larger $\left.  \varpi(s)\right\vert _{s=0}$ for the
same $\left.  \eta(s)\right\vert _{s=0}$, it is apparent that $C\left(
M\right)  $ can also be double-valued. \ All this is evident in a parametric
plot of the corresponding $\left(  C,M\right)  $ points on the data plane.
\ For example, for $\left.  \eta(s)\right\vert _{s=0}=1$ and various $\left.
\varpi(s)\right\vert _{s=0}\in\left[  0.1,~1.259\,92=\sqrt[3]{2}\right]  $, we
find the naked (green circle) and censored (black circle) data as included in
the last Figure, with a fitted interpolating curve (orange dashes) connecting
the computed points. \ In this numerical analysis, care should be taken not to
have $\left.  \varpi(s)\right\vert _{s=0}$ larger than $\sqrt[3]{2\left.
\eta(s)\right\vert _{s=0}}$ because otherwise this would place data
initialized at $r=1$ \emph{within} the horizon. \ The horizon is exactly at
the radius $r=1$ when $\left.  \varpi(s)\right\vert _{s=0}=$ $\sqrt[3]%
{2\left.  \eta(s)\right\vert _{s=0}}$. \ The gray curve in the last Figure is
the image of $\left.  \varpi(s)\right\vert _{s=0}=$ $\sqrt[3]{2\left.
\eta(s)\right\vert _{s=0}}$ on the $\left(  M,C\right)  $ plane. \ Points
below this gray curve can be investigated numerically using Schwarzshild
coordinates but only if the initial data is specified for $r>1$, i.e.
\emph{outside} the horizon. \ (Also note the portion of the initial data plane
shown in the previous Figure lies entirely above the curve $\left.
\eta(s)\right\vert _{s=0}=\frac{1}{2}\left.  \varpi^{3}(s)\right\vert _{s=0}$,
so all initial data points in that Figure lie outside of any horizons.)

\section{Conclusions}

In conclusion, as previously emphasized by many authors it would be
interesting to search for evidence of galileons at all distance scales,
including galactic and sub-galactic, as well as cosmological. \ Perhaps a
combination of trace couplings and various galileon terms, such as those in
(\ref{A}) and (\ref{ASelfDual}) extended to included GR effects, will
ultimately lead to a realistic physical model. \ In particular, it is
important to investigate the stability of galileon solutions and to consider
the dynamical evolution of generic galileon and other matter field initial
data, along the lines of \cite{Choptuik1993,Choptuik2003},\ to determine under
what physical conditions naked singularities are actually formed.

\paragraph{Acknowledgements:}

We thank S Deser and C Zachos for constructive comments. We also thank S K
Rama, N Rinaldi, K S Virbhadra, and R Wald for discussions of naked
singularities. \ This research was supported by a University of Miami Cooper
Fellowship, and by NSF Awards PHY-0855386 and PHY-1214521.

\end{document}